\documentclass[12pt]{elsarticle}                                          
\usepackage{axodraw4j}
\usepackage{hyperref}
\usepackage{graphicx}                                                         
\usepackage{a41}                                                                
\usepackage{xcolor}                     
\usepackage{slashed}
\usepackage[rflt]{floatflt}
\usepackage{float}
\usepackage{hyperref}
\hypersetup{colorlinks=true, 
linkcolor=blue, filecolor=blue, urlcolor=mygreen}
\usepackage{breakurl} 
\usepackage{times}     %  
\usepackage{array}
\usepackage[english]{babel}
\usepackage[latin1]{inputenc}
\usepackage[T1]{fontenc}
\usepackage{ae}
\usepackage{url}
\usepackage{amsmath, amsthm, amssymb}
\usepackage{slashed}

\usepackage{rotating}
\usepackage{graphicx}
\usepackage{comment}
\newcounter{mmacnt}
\def\restartmma{\setcounter{mmacnt}{0}}
\restartmma \catcode`|=\active
\def|#1|{\mathrm{#1}}
\catcode`|=12
\newenvironment{mma}{
\par\smallskip
\catcode`|=\active
\parskip=0pt\parindent=0pt % locally
\small
\def\In##1\\{%
\def\linebreak{\hfill\break\null\qquad}%
\refstepcounter{mmacnt}
\hangindent=2.5em\hangafter=0
\leavevmode
\llap{\tiny\sffamily In[\arabic{mmacnt}]:=\kern.5em}%
\mathversion{bold}\footnotesiVe$
\displaystyle##1$\normalsiVe
\mathversion{normal}\par
 }%
\def\Print##1\\{%
\def\linebreak{\hfill\break}%
\hangindent=2.5em\hangafter=0
\leavevmode ##1\par}%
\def\Out##1\\{%
\def\linebreak{$\hfill\break\null\hfill$}%
\kern\abovedisplayskip\par
\hangindent=2.5em\hangafter=0
\leavevmode
\llap{\tiny\sffamily Out[\arabic{mmacnt}]=\kern.5em}
\footnotesiVe$\displaystyle##1$
\normalsiVe\hfill\null\par
\kern\belowdisplayskip
}%
\def\Warning##1##2\\{%
\def\linebreak{\hfill\break}%
\hangindent=2.5em\hangafter=0
\leavevmode
{\scriptsiVe##1 : ##2}\par}%
}{%
\par\smallskip
}

\usepackage{color}
\newenvironment{fshaded}{%
\MakeFramed {\FrameRestore}
}%
{\endMakeFramed}

\makeatletter
\def\ps@pprintTitle{%
\let\@oddhead\@empty
\let\@evenhead\@empty
\def\@oddfoot{\reset@font\hfil\thepage\hfil}
\let\@evenfoot\@oddfoot
}
\makeatother
\newcommand{\n}{\nonumber}

\usepackage{tcolorbox}
\usepackage{slashed}
\usepackage{comment}
\usepackage{amsmath, tikz, amsfonts, 
amssymb, xfrac, graphicx, wrapfig, 
slashed, physics,simplewick, 
simpler-wick, 
booktabs, bm, xcolor, enumerate, 
mathtools, multirow,subcaption}
\begin{document}  %%%%
%%%%%%%%%%%%%%%%%%%%%%
\begin{frontmatter}%%%
%%%%%%%%%%%%%%%%%%%%%%%%%%%%%%%%%%%%%%%%%%%%%%%%%%%%
\title{{\bf One-loop contributions 
for $h\rightarrow \ell \bar{\ell}\gamma$ 
and $e^-e^+\rightarrow h\gamma$ 
in $U(1)_{B-L}$ extension of the standard model
}}
%%%%%%%%%%%%%%%%%%%%%%%%%%%%%%%%%%%%%%%%%%%%%%%%%%%%
\author[1,2]{Dzung Tri Tran}
\author[3]{Thanh Huy Nguyen}
\author[1,2]{Khiem Hong Phan}
\ead{phanhongkhiem@duytan.edu.vn}
%%%%%%%%%%%
\address[1]{\it Institute of Fundamental 
and Applied Sciences, Duy Tan University, 
Ho Chi Minh City $700000$, Vietnam}
\address[2]{Faculty of Natural Sciences, 
Duy Tan University, Da Nang City $550000$, 
Vietnam}
\address[3]{\it 
VNUHCM-University of Science, 
$227$ Nguyen Van Cu, District $5$, 
Ho Chi Minh City, Vietnam}
\pagestyle{myheadings}
\markright{}
%%%%%%%%%%%%%%%%%%%%%%%%%%%%%%%%%%%%%%%%%%
\begin{abstract} %%%
%%%%%%%%%%%%%%%%%%%%
We present one-loop contributing
for $h\rightarrow \ell \bar{\ell}\gamma$
with $\ell =\nu_{e,\mu, \tau}, e, \mu$ 
and $e^-e^+\rightarrow h\gamma$ 
in $U(1)_{B-L}$ extension 
of the standard models. 
In phenomenological results, the signal 
strengths for $h\rightarrow 
\ell \bar{\ell}\gamma$ at 
Large Hadron Collider and 
for $e^-e^+\rightarrow h\gamma$
at future Lepton Colliders
are analyzed in physical 
parameter space for both 
vector and chiral $B-L$ models. 
We find that the contributions 
from neutral gauge boson $Z'$ 
to the signal strengths are 
rather small. 
Consequently, the effects are 
hard to probe at future colliders.
While the impacts of charged Higgs, 
CP-odd Higgs in the chiral $B-L$ model 
on the signal strengths
are significant and 
can be measured with the help 
of the initial polarization 
beams at 
future lepton colliders. 
\end{abstract}
%%%%%%%%%%%%%%%%%%%%%%%%%%%%%%%%%%
\end{frontmatter}
%%%%%%%%%%%%%%%%%%%%%%%%%%
\section{Introduction}%%%%
%%%%%%%%%%%%%%%%%%%%%%%%%%
The precise measurements for 
the decay rates and the production
cross-sections of standard-model-like 
Higgs boson (SM-like Higgs boson $h$) 
are main targets at the High-Luminosity 
Large Hadron Collider 
(HL-LHC)~\cite{Liss:2013hbb,CMS:2013xfa}
as well as future Lepton Colliders 
(LC)~\cite{Baer:2013cma}. 
From the measured data, we can  
probe accurately the properties of the
SM-like Higgs boson. In this perspective,
the nature of the scalar Higgs potential
hence can be discovered. Other words, 
we can understand deeply the electroweak 
spontaneous symmetry breaking (EWSB). 
The updated measurements for the Higgs 
boson production cross-sections 
and decay rates at ATLAS and CMS can be 
found in 
Refs.~\cite{ATLAS:2016neq,ATLAS:2019nkf} 
and references therein. Moreover,  
loop-induced processes 
$h\rightarrow \gamma\gamma$
and $h\rightarrow Z \gamma$ 
have been measured at the 
LHC~\cite{CMS:2014fzn, ATLAS:2015egz, 
CMS:2021kom, D0:2008swt,  
CMS:2013rmy, ATLAS:2017zdf,ATLAS:2020qcv}. 
Morerecently, 
the decay processes 
$h\rightarrow \ell \bar{\ell}\gamma$ 
with $\ell =\nu_{e,\mu, \tau}, e, \mu$ also
have been greatly paid 
attention at the LHC~\cite{CMS:2015tzs,CMS:2017dyb,
CMS:2018myz, ATLAS:2021wwb}.
Together with $h\rightarrow \gamma\gamma, 
Z \gamma$, the decay processes 
$h\rightarrow \ell \bar{\ell} \gamma$ 
also provide an important information for 
testing the standard model 
(SM)~\cite{Chen:2012ju, Gainer:2011aa,
Korchin:2014kha} and constraining parameters 
in many of beyond the standard models 
(BSM)~\cite{Korchin:2014kha}.

One-loop contributions for the decay processes 
$h\rightarrow \ell \bar{\ell}\gamma$ have 
performed in Refs.~\cite{Abbasabadi:2000pb, 
Dicus:2013ycd, Sun:2013rqa, Passarino:2013nka, 
Dicus:2013lta, Kachanovich:2020xyg, 
Kachanovich:2021pvx, Ahmed:2023vyl} in the SM. 
In the framework of 
Two Higgs Doublet Models (THDM), one-loop decay 
processes $h\rightarrow \ell \bar{\ell}\gamma$ have 
evaluated in Refs.~\cite{Li:1998rp, 
Sasaki:2017fvk}.
Recently, one-loop expressions for 
$h\rightarrow \ell \bar{\ell}\gamma$ in general 
BSM frameworks have presented in 
Refs.~\cite{Phan:2021xwc,VanOn:2021myp}.
We have derived alternative
presentations for one-loop contributing
of $h\rightarrow \ell \bar{\ell}\gamma$
in Higgs extension of the standard models
(HESM)~\cite{Hue:2023tdz}. 
In the work~\cite{Hue:2023tdz}, 
the computations have 
handled in
the 't Hooft-Feynman (HF) gauge and 
confirm the results in our previous 
studies~\cite{Phan:2021xwc,VanOn:2021myp}
in the unitary gauge. 
We note that all contributions of singly 
and doubly charged Higgses
propagating in the loop diagrams 
have taken into account in the 
article~\cite{Hue:2023tdz}.
It is well-known that there 
also exist of 
many new gauge bosons exchanging 
in the loop diagrams of the decay 
processes $h\rightarrow \ell \bar{\ell}
\gamma$ in other BSMs. 
For examples, the simplest $U(1)_{B-L}$ 
extension for the SM~\cite{Mandal:2023oyh,
Basso:2008iv,Basso:2009hf,
Basso:2010jt,Basso:2010pe,Basso:2010jm,Basso:2010hk,
Basso:2010yz,Basso:2010si,Basso:2011na}, 
there exists of neutral $Z'$ gauge boson.
In the left-right models (LR) constructed from 
the $SU(2)_L\times SU(2)_R\times 
U(1)_Y$~\cite{Pati:1974yy,
Mohapatra:1974gc, Senjanovic:1975rk}, 
the 3-3-1 models  
($SU(3)_L\times U(1)_X$)~\cite{Singer:1980sw, 
Valle:1983dk, Pisano:1991ee, Frampton:1992wt, 
Diaz:2004fs,Fonseca:2016tbn, Foot:1994ym},  
the $3$-$4$-$1$ models ($SU(4)_L\times U(1)_X$)
\cite{Foot:1994ym, Sanchez:2004uf, Ponce:2006vw, 
Riazuddin:2008yx, Jaramillo:2011qu, Long:2016lmj}, 
etc, include new charged gauge bosons as well as
new neutral gauge bosons.
Therefore, one-loop formulas for decay rates 
of $h\rightarrow \ell \bar{\ell}\gamma$ 
in the above-mentioned models 
are great of interest. 
The calculations for
$h\rightarrow \ell \bar{\ell}\gamma$
in the HF gauge have several advantages
as discussed in~\cite{Hue:2023tdz}. 
It is difficult to derive a general one-loop
formulas for $h\rightarrow \ell \bar{\ell}\gamma$
in the HF gauge for the arbitrary BSM as we 
have performed in~\cite{VanOn:2021myp}. 
Because the couplings of new particles
to Goldstone bosons, Ghost 
partiles depend on the models under cosnideration. 
Due to gauge invariance, the contributions from 
Goldstone bosons, Ghost particles propagating 
in the loop will be cancelled out at the 
final results. 
The cancellations within  the vector boson 
loop with their Goldstone bosons, 
Ghost particles exchanging in the 
loop may not be the same in 
the different beyond the standard models.
For this reason, in 
scope of this article we compute 
one-loop contributing
for $h\rightarrow \ell \bar{\ell}\gamma$
with $\ell =\nu_{e,\mu, \tau}, e, \mu$
and $e^-e^+\rightarrow h\gamma$ 
in $U(1)_{B-L}$ extension 
of the standard model in the HF gauge. 
In comparison with
the work in~\cite{VanOn:2021myp},
we present alternative expressions for 
one-loop contributing for 
$h\rightarrow \ell \bar{\ell}\gamma$
in the HF gauge within 
the $U(1)_{B-L}$ models. This work is also
extended
the results in~\cite{Hue:2023tdz}
with adding the new contributions of
neutral gauge boson $Z'$ as well as 
heavy neutrinos in the models.

Another aspect, the couplings of Higgs 
boson with $Z\gamma, \gamma\gamma$ can 
be probed by measuring 
cross-sections of Higgs boson production 
associated with a photon at future Lepton 
Colliders. It is important to note that 
the tree-level cross-section for this 
process is proportional to electron mass and 
the process follows electromagnetic 
gauge symmetry. Therefore, 
it is significant contributions at first 
from one-loop level.
% it contributes at first from one-loop level. 
As a result, 
cross-section is rather small. Because the 
LC compared 
with the LHC have a cleaner background. 
The new physic signals hence are easily 
extracted from the background. 
With the high-luminosity designed at 
future LC~\cite{Baer:2013cma},
the signal of  Higgs boson production 
associated with a photon can be probed. 
There have been available many computations 
for one-loop corrections to 
$e^-e^+ \rightarrow h \gamma$
at future LC 
in the SM~\cite{Abbasabadi:1995rc,Djouadi:1996ws,
Abbasabadi:1997zr},
in many frameworks of HESM~\cite{Arhrib:2014pva,
Rahili:2019ixf,Kanemura:2018esc},  
in Minimal Supersymmetric Standard 
Model~\cite{Demirci:2019ush}, etc. 
In this work, we illustrate that 
cross-sections of $e^-e^+ \rightarrow 
h\gamma$ can be derived by using 
one-loop form factors in 
$h\rightarrow \ell \bar{\ell}\gamma$. 
The results are 
also valid in the SM, 
within the $U(1)_{B-L}$ extension 
of the standard model as well as
can be extended for other BSMs.

In phenomenological studies, 
we present first
results of analyzing the signal 
strengths for 
$h\rightarrow \ell \bar{\ell}\gamma$ 
at Large Hadron Collider and for 
$e^-e^+\rightarrow h\gamma$
at future LC
in physical parameter space 
for both vector
and chiral $B-L$ models. 
For decay processes 
$h\rightarrow \ell \bar{\ell}\gamma$, 
we focus on the possibility to probe 
heavy neutral gauge boson $Z'$
as well as charged Higgs, CP-odd Higgs
at the LHC. 
For this reason, we are interested in 
the case of heavy neutrinos scenario 
($M_{N_i}
\sim \textrm{TeV}$). For the production 
process $e^-e^+\rightarrow h\gamma$
at future LC, 
the signal strengths are examined
by including the initial beam 
polarization effects. The 
contributions of neutral gauge boson $Z'$, 
charged Higgs, CP-odd Higgs 
as well as additional
neutrinos are investigated 
in further detail at future LC.

Our work is organized as follows.
In section $2$, the $U(1)_{B-L}$ extension
of the standard models 
are reviewed in more detail. In section $3$, 
we present the concrete evaluations 
for one-loop form factors for 
the decay channels $h\rightarrow 
\ell \bar{\ell}\gamma$ 
and for the process
$e^-e^+ \rightarrow h\gamma$. 
Phenomenological results for 
the processes are discussed in 
section $4$. 
Conclusions and outlook are 
devoted in section $5$. In the appendices, 
we calculate all couplings
related to the processes under 
consideration.
%%%%%%%%%%%%%%%%%%%%%%%%%%%%%%%%%%
\section{Review of $U(1)_{B-L}$ %% 
extension of the standard model}%%
%%%%%%%%%%%%%%%%%%%%%%%%%%%%%%%%%%
In this section, 
following~\cite{Mandal:2023oyh} 
we review in detail
the $U(1)_{B-L}$ model, one of the simplest 
extensions of the standard model. The model 
is enlarged the SM gauge group with a  
new gauge symmetry $U(1)_{B-L}$. 
The $U(1)_{B-L}$ extension model then 
follows gauge group 
$SU(3)_C\otimes SU(2)_L\otimes 
U(1)_Y\otimes U(1)_{B-L}$.
As a result, the Yang-Mills 
Lagrangian including 
mixing of two $U(1)$ abelian 
gauge groups is modified as follows:
%%%%%%%%%%%%%%%%%%%%%%%%%%%%%%%%%%
\begin{eqnarray}
\mathcal{L}_{YM} &=&
-\frac{1}{4}G^{a,\mu\nu}G_{a,\mu\nu}
-\frac{1}{4}W^{a,\mu\nu}W_{a, \mu\nu}
-\frac{1}{4}B^{\mu\nu}B_{\mu\nu}
-\frac{1}{4}X^{\mu\nu}X_{\mu\nu}
-\frac{\kappa}{2}B^{\mu\nu}X_{\mu\nu}
\end{eqnarray}
where $G^{a,\mu\nu},~W^{a,\mu\nu}$ 
are corresponding to 
the field strengths 
of the gauge groups 
$SU(3)_C, SU(2)_L$. 
While $B_{\mu\nu}$ 
and $X_{\mu\nu}$ are 
the field strengths 
of the gauge groups $U(1)_Y$ 
and $U(1)_{B-L}$, respectively. 
Considering the mixing term 
(kinematic mixing case), we have 
to perform the following rotation 
for two gauge bosons in groups 
of $U(1)_Y$, 
$U(1)_{B-L}$ as follows:
%%%%%%%%%%%%%%%%%%%%%%%%%%%%%%%%%%
\begin{eqnarray}
\left(\begin{array}{c}
\tilde{B}_{\mu}  \\
\tilde{X}_{\mu} 
\end{array}\right)
=
\left(\begin{array}{cc}
1 & \kappa \\
0 & \sqrt{1-\kappa^2}
\end{array}\right)
\left(\begin{array}{c}
B_\mu  \\
X_\mu 
\end{array}\right). 
\label{mixingBL}
\end{eqnarray}
%%%%%%%%%%%%%%%%%%%%%%%%%%%%%%%%%%
Where $\kappa$ is the mixing parameter
between the gauge groups $U(1)_Y$ and 
$U(1)_{B-L}$. We note that 
$g, g'$ ($g_X$) are 
the couplings of $SU(2)_L, 
U(1)_Y$, ($U(1)_{B-L}$) gauge groups, 
respectively. In this paper, we note
the hypercharge $Y$ for $U(1)_Y$
and $Y_X$ for $U(1)_{B-L}$ gauge 
groups. The model is classified into 
two types 
of models such as vector 
$B-L$ model and chiral $B-L$
model~\cite{Mandal:2023oyh}. 
These models are studied in further 
detail in the following subsections.
%%%%%%%%%%%%%%%%%%%%%%%%%%%%%%%%%%
\subsection{Vector $B-L$ model} %%
%%%%%%%%%%%%%%%%%%%%%%%%%%%%%%%%%%
In the vector $B-L$ model, we have 
additional three right handed neutrinos 
(RHNs, $N_i$ for $i=\overline{1, 2, 3}$) 
and a new complex scalar singlet field $\chi$ 
which is required 
for the spontaneous symmetry breaking
(SSB) of the $U(1)_{B-L}$ gauge. In this 
version, 
the matter fields, scalar fields 
are listed with their quantum numbers
in the following Table~\ref{MatterBL}.
%%%%%%%%%%%%%%%%%%%%%%%%%%%%%%%%%%%
\begin{table}[H]
\centering
\begin{tabular}{|c|c|c|}
\hline \hline \centering
~~~Fields~~~ & $SU(3)_{c} \otimes 
SU(2)_{L} \otimes U(1)_{Y} \otimes 
U(1)_{B-L}$  \\ 
\hline \hline
$L_{L}$  & ($1,2,-\frac{1}{2},-1$) \\ 
\hline 
$Q_{L}$  & ($3,2,\frac{1}{6},\frac{1}{3}$)  \\
\hline 
$e_{R}$  & ($1, 1, -1, -1$) \\ 
\hline 
\hline
$\nu_{R}$  & ($1,1,0,-1$) \\ 
\hline
\hline 
$u_{R}$ & ($3,1,\frac{2}{3},\frac{1}{3}$) \\ 
\hline 
$d_{R}$  & ($3,1,-\frac{1}{3},\frac{1}{3}$) \\ 
\hline
\hline 
$\Phi$  & ($1,2,\frac{1}{2}, 0$) \\ 
\hline 
$\chi$  & ($1,1,0,2$) \\ 
\hline 
\end{tabular}  
\caption{\label{MatterBL} Table of 
matter fields, scalar bosons
with their charge quantum numbers of 
the vector $B-L$ model.
We omit the index of generations of 
matter particles in this Table. In 
this paper, we note 
$N_i$ as three RHNs for later uses.}
\end{table}
%%%%%%%%%%%%%%%%%%%%%%%%%%%%%%%%%%%
The Higgs sector is given by 
%%%%%%%%%%%%%%%%%%%%%%%%%%%%%%%%%%
\begin{eqnarray}
\mathcal{L}_{H}
&=& \mathcal{L}_{K} 
- \mathcal{V}(\Phi,\chi).
\end{eqnarray}
Where the kinematic part 
reads as
\begin{eqnarray}
\mathcal{L}_{K}
&=&
(D_\mu{\Phi})^\dagger{D_{\mu}\Phi} 
+ 
(D^\mu{\chi})^\dagger{D^{\mu}\chi}
\end{eqnarray}
and the scalar potential 
$\mathcal{V}(\Phi,\chi)$ 
is given by
%%%%%%%%%%%%%%%%%%%%%%%%%%%%%%%%%%
\begin{eqnarray}
\mathcal{V}(\Phi,\chi) 
= m_{\chi}^2(\chi^*\chi)
+ \frac{1}{2}\lambda_\chi(\chi^*\chi)^2
+ m_{\Phi}^2(\Phi^\dagger\Phi)
+ \frac{1}{2}\lambda_\Phi(\Phi^\dagger\Phi)^2
+ \lambda_{\Phi\chi}(\chi^*\chi)(\Phi^\dagger\Phi).
\end{eqnarray}
%%%%%%%%%%%%%%%%%%%%%%%%%%%%%%%%%%
The fields $\Phi$ and $\chi$ are 
parameterized for spontaneous 
symmetry breaking as follows:
%%%%%%%%%%%%%%%%%%%%%%%%%%%%%%%%%%
\begin{eqnarray}
\Phi = 
\frac{1}{\sqrt{2}}
\left(
\begin{array}{c}
\sqrt{2}G^{\pm}  \\
v_\Phi+R_1+iI_1 
\end{array}
\right), 
\quad 
\chi=\frac{1}{\sqrt{2}}(v_{\chi}+R_2+iI_2).
\end{eqnarray}
%%%%%%%%%%%%%%%%%%%%%%%%%%%%%%%%%%
Here, the Goldstone 
bosons $G^{\pm}$ are giving the masses for  
$W^{\pm}$ bosons. While $I_1, I_2$ fields
are mixed together being neutral 
Goldstone bosons $G^0_{1,2}$. They play 
a role
for giving masses
for gauge bosons $Z$ and $Z'$, respectively. 
The minimization for the scalar potential
leads to a system of equations as follows:
\begin{eqnarray}
m_{\chi}^2+ \frac{\lambda_{\chi}}{2} 
v_{\chi}^2 + \frac{\lambda_{\Phi\chi}}{2}
v_{\Phi}^2  =0,\\
m_{\Phi}^2+ \frac{\lambda_{\Phi}}{2} 
v_{\Phi}^2 + \frac{\lambda_{\Phi\chi}}{2}
v_{\chi}^2  =0.
\end{eqnarray}
From the minimization conditions 
for $\mathcal{V}(\Phi,\chi)$,
one can present $m_{\chi}^2, 
m_{\Phi}^2$ in terms of 
the remaining parameters in 
the potential. As a result, 
the mass matrix for neutral 
scalar components
are collected in the 
basis of $(R_1, R_2)$ 
as follows:
%%%%%%%%%%%%%%%%%%%%%%%%%%%%%%%%%%%%%%%%
\begin{eqnarray}
\mathcal{M}_S^2 = 
\begin{pmatrix}
v_{\Phi}^{2}\lambda_{\Phi} 
& v_{\Phi} v_{\chi} \lambda_{\Phi \chi}  \\
v_{\Phi} v_{\chi} \lambda_{\Phi \chi}
& v^{2}_{\chi} \lambda_{\chi}
\end{pmatrix}.
\label{Rmatrix}
\end{eqnarray}
In order to diagonalize the matrix 
Eq.~\ref{Rmatrix} for getting the Higgs 
physical masses of neutral Higgses, 
one first performs the rotation which 
shows the relation of
$(h, H)$ and $(R_1, R_2)$. 
The rotation matrix 
takes the form of
%%%%%%%%%%%%%%%%%%%%%%%%%%%%%%%%%%
\begin{eqnarray}
\left(\begin{array}{c}
h  \\
H 
\end{array}\right)
=
\left(
\begin{array}{cc}
c_\theta & -s_\theta \\
s_\theta & c_\theta
\end{array}
\right)
\left(
\begin{array}{c}
R_1  \\
R_2 
\end{array}
\right)
\end{eqnarray}
%%%%%%%%%%%%%%%%%%%%%%%%%%%%%%%%%%
with mixing angle
%%%%%%%%%%%%%%%%%%%%%%%%%%%%%%%%%%
\begin{eqnarray}
t_{2\theta} =  \frac{2\lambda_{\Phi\chi} v_{\Phi}
v_{\chi}}{\lambda_{\chi}v_{\chi}^2 - 
\lambda_{\Phi}v_{\Phi}^2}.
\end{eqnarray}
The masses of CP-even Higgs scalars are then 
taken the form of
%%%%%%%%%%%%%%%%%%%%%%%%%%%%%%%%%%
\begin{eqnarray}
M_h^2
&=&
\frac{1}{2}[v_\Phi^2\lambda_\Phi+v_\chi^2\lambda_\chi
-
\sqrt{(v_\Phi^2\lambda_\Phi-v_\chi^2\lambda_\chi)^2
+4v_\Phi^2{v_{\chi}^2}\lambda_{\Phi\chi}^2}], \\
% \end{eqnarray}
% %%%%%%%%%%%%%%%%%%%%%%%%%%%%%%%%%%
% \begin{eqnarray}
M_H^2 
&=& 
\frac{1}{2}[v_\Phi^2\lambda_\Phi 
+ v_\chi^2\lambda_\chi
+
\sqrt{(v_\Phi^2\lambda_\Phi-v_\chi^2\lambda_\chi)^2 
+ 4v_\Phi^2{v_{\chi}^2}\lambda_{\Phi\chi}^2}].
\end{eqnarray}
%%%%%%%%%%%%%%%%%%%%%%%%%%%%%%%%%%
In this work, we assume that 
$M_h^2\leq M_H^2$, $h$ being the SM-like 
Higgs boson with $M_h \sim 125$ GeV. 
%%%%%%%%%%%%%%%%%%%%%%%%%%%%%%%%%%

%%%%%%%%%%%%%%%%%%%%%%%%%%%%%%%%%%%%%%%%%%
From the kinematic term 
of scalar sector, we collect the mass matrix of
the neutral gauge bosons. This term is given by
%%%%%%%%%%%%%%%%%%%%%%%%%%%%%
\begin{align}
\mathcal{L}_{K} \supset 
\frac{1}{2} V_0^T M_G^2 V_0 ,
\end{align}
%%%%%%%%%%%%%%%%%%%%%%%%%%
where
\begin{eqnarray}
V_0^T = 
\begin{pmatrix}
B_\mu & W_{3\mu} & X_\mu 
\end{pmatrix} \text{   and    }
M_G^2=
\begin{pmatrix}
\frac{1}{4}g'^{2} v_\Phi^2 &  -\frac{1}{4} g g' v_\Phi^2   &   0 \\
-\frac{1}{4} g g' v_\Phi^2  &   \frac{1}{4} g^2 v_\Phi^2  &  0  \\
0  &  0  &   g_1^{'2} v_\chi^2  
\end{pmatrix}.
\end{eqnarray}
%%%%%%%%%%%%%%%%%%%%%%%%%%%%%%%%%%%%%%%%%%%
When we consider the mixing of two
$U(1)_Y, U(1)_{B-L}$ 
gauges, one first uses the rotation 
matrix in Eq.~\ref{mixingBL}
for obtaining the basis 
$(\tilde{B}_\mu\,\,\tilde{X}_\mu)$.
In order to find physical masses of gauge bosons, 
we have to diagonalize the mass matrix in 
the basis $\tilde{V}_0^T=(\tilde{B}_\mu\,
\,W_{3\mu}\,\,\tilde{X}_\mu)$. First, the mass matrix 
of the neutral gauge bosons in the kinetic term 
diagonalized basis 
$\tilde{V}_0^T=(\tilde{B}_\mu\,\,W_{3\mu}\,
\,\tilde{X}_\mu)$ can be collected as follows:
%%%%%%%%%%%%%%%%%%%%%%%%%
\begin{eqnarray}
\mathcal{L}_{K} \supset 
\frac{1}{2} \tilde{V}_0^T S^T M_G^2  S \tilde{V}_0 
= \frac{1}{2} \tilde{V}_0^T  
\tilde{M}_G^2  \tilde{V}_0,
\end{eqnarray}
%%%%%%%%%%%%%%%%%%%%%%%%%
where
\begin{align}
S=\begin{pmatrix} 
1  &  0  & -\frac{\kappa}{\sqrt{1-\kappa^2}}  \\
0  & 1   & 0 \\
0  &  0  & \frac{1}{\sqrt{1-\kappa^2}}
\end{pmatrix}, \; 
\tilde{M}_G^2
= S^T  M_G^2  S = \begin{pmatrix}
\frac{1}{4}g'^{2} v_\Phi^2 &  
-\frac{1}{4} g g' v_\Phi^2  &  
\frac{1}{4} g' \tilde{g}_t v_\Phi^2 \\
-\frac{1}{4} g g' v_\Phi^2  &   
\frac{1}{4} g^2 v_\Phi^2  & 
-\frac{1}{4} g  \tilde{g}_t v_\Phi^2  \\
\frac{1}{4} g'  \tilde{g}_t v_\Phi^2  
&  -\frac{1}{4} g  \tilde{g}_t v_\Phi^2  
&  \frac{1}{4}  \tilde{g}_t^2 v_\Phi^2
+ g_1^{''2} v_\chi^2  
\end{pmatrix},
\end{align}
%%%%%%%%%%%%
with the new coupling 
$\tilde{g}_t 
= -\frac{g'\kappa}{\sqrt{1-\kappa^2}}$. 
By changing  the basis from 
$\tilde{B}^{\mu}$, $W_{3}^{\mu}, \tilde{X}^{\mu}$ 
to the one in $A^{\mu}$, $Z^{\mu}, Z^{'\mu}$,
we have the mass eigenstates 
for physical gauge bosons. 
The relation is 
shown in the following
rotation matrix as 
%%%%%%%%%%%%%%%%%%%%%%%%%%%%%%%%%%
\begin{eqnarray}
\left(
\begin{array}{c}
\tilde{B}_{\mu}  \\
W_3^\mu \\
\tilde{X}_{\mu} 
\end{array}
\right)
=
\left(
\begin{array}{ccc}
c_W & -s_Wc_{BL} & s_W s_{BL} \\
s_W &  c_W c_{BL} & -c_W s_{BL} \\
0 & s_{BL} & c_{BL}
\end{array}
\right)
\left(
\begin{array}{c}
A_\mu  \\
Z_\mu   \\
Z'^{\mu}
\end{array}
\right).
\end{eqnarray}
Where $s_W (c_W) = \sin\theta_W (\cos\theta_W)$
and $s_{BL} (c_{BL}) 
= \sin\theta_{BL} (\cos\theta_{BL})$ are mixing 
angles of gauge bosons. In the above formulas, 
we have the following parameters:
%%%%%%%%%%%%%%%%%%%%%%%%%%%%%%%
\begin{align}
 t_{2(BL)} = 
 \frac{2\tilde{g}_t\sqrt{g^{2}+g'^{2}}}
 {\tilde{g}_t^{2} 
 + 16\left(\frac{v_\chi}{2v_\Phi}
 \right)^{2}g_{1}^{''2}
 -g^{2}-g'^{2}} \text{with} \;
 g_1^{''}
 =\frac{g_1'}{\sqrt{1-\kappa^2}}
 =\frac{g_X q_X}{\sqrt{1-\kappa^2}}.
\end{align}
Masses of physical gauge bosons 
$A$, $Z$ and $Z^{'}$ are expressed as 

\begin{eqnarray}
 M_A=0,\,\,M_{Z,Z^{'}}^{2}=\frac{1}{8}
 \left(Cv_\Phi^{2}\mp\sqrt{-D+v_\Phi^{4}C^{2}}\right),
\label{AZZp}
\end{eqnarray}
where 
\begin{align}
 C=g^{2}+g'^{2}+\tilde{g}_t^{2}
 +
 16\left(\frac{v_\chi}{2v_\Phi}
 \right)^{2}g_{1}^{''2},
 \hspace{0.5cm}
 D=16 v_\Phi^{2}
 v_\chi^{2}(g^{2}
 +g'^{2})g_{1}^{''2}.
\end{align}
%%%%%%%%%%%%%%%%%%%%%%%%%%%%%%%%%%%%%%%%%%%%%%%%%%
The covariant derivative with the kinetic mixing 
can be expressed in terms of the orthogonal 
fields $\tilde{B}$ and $\tilde{X}$ as 
%%%%%%%%%%%%%%%%%%%%%%%%%%%%%%%%%%%%%%%%
\begin{align}
D_{\mu}&=\partial_{\mu}-ig_{s}T^{a}G^{a}_{\mu}
-igT^{a}W^{a}_{\mu}-ig'Y \tilde{B}_{\mu}
-i\left(g_{X}Y_{X}\frac{1}{\sqrt{1-\kappa^2}}
-g' Y \frac{\kappa}{\sqrt{1-\kappa^2}}\right)\tilde{X}_{\mu}
\nonumber \\
&=\partial_{\mu}-ig_{s}T^{a}G^{a}_{\mu}
-igT^{a}W^{a}_{\mu}-ig'Y \tilde{B}_{\mu}
-i\left(\tilde{g}_{X} Y_{X}
+ \tilde{g}_t Y \right)\tilde{X}_{\mu}.
\end{align}
Where $\tilde{g}_{X} = g_{X} 
\frac{1}{\sqrt{1-\kappa^2}}$.
%%%%%%%%%%%%%%%%%%%%%%%%%%%%%%%%
In the case of none of the kinematic mixing gauge, taking 
$\kappa \rightarrow 0$, one has 
the covariant derivative as 
\begin{align}
D_{\mu}=\partial_{\mu}-ig_{s}T^{a}G^{a}_{\mu}
-igT^{a}W^{a}_{\mu}-ig'Y B_{\mu}-ig_{X}Y_{X}X_{\mu}.
\label{Dnon}
\end{align}

Neutrino masses in this model are generated 
by a seesaw mechanism that will be shown 
in the Yukawa Lagrangian, namely
%%%%%%%%%%%%%%%%%%%%%%%%%%%%%%%%%%
\begin{eqnarray}
-\mathcal{L}_Y&=&
\mathcal{L}_{SM} 
+ Y_\nu{\bar{L}}\tilde{\Phi}\nu_R
+ \frac{Y_M}{2}\bar{\nu_R^c}\nu_R\chi
+ H.c \nonumber\\
&{\supset}&
\frac{Y_{\nu}v_\Phi}{\sqrt{2}}\bar{\nu_L}\nu_R
+\frac{Y_{\nu}c_{\theta}}{\sqrt{2}}\bar{\nu_L}h\nu_R
+\frac{y_Mv_\Phi}{2\sqrt{2}}\bar{\nu_{R}^c}\nu_{R}
-\frac{y_Ms_{\theta}}{2\sqrt{2}}\bar{\nu_{R}^c}h\nu_{R}
+
H.c.
\end{eqnarray}
%%%%%%%%%%%%%%%%%%%%%%%%%%%%%%%%%%
where $\tilde{\Phi}=i\sigma_2\Phi^*$, 
the Dirac mass is $m_D=\frac{Y_\nu{v_\Phi}}{\sqrt{2}}$ 
and Majorana mass is $M_R=\frac{y_Mv_\chi}{\sqrt{2}}$. 
The Yukawa Lagrangian shows that $\chi$ field 
generates Majorana mass and $\Phi$ generates 
Dirac masses for RHNs.

The fermion Lagrangian is given by
%%%%%%%%%%%%%%%%%%%%%%%%%%%%%%%%%%
\begin{eqnarray}
\mathcal{L}_f&=&\mathcal{L}_{f}^{SM}
+i\bar{\nu}_R\gamma_{\mu}D^{\mu}\nu_{R} \\
&=&i\bar{q}_L\gamma_{\mu}D^{\mu}q_{L}
+i\bar{u}_R\gamma_{\mu}D^{\mu}u_{R}
+i\bar{d}_R\gamma_{\mu}D^{\mu}d_{R}
+i\bar{l}_L\gamma_{\mu}D^{\mu}l_{L}
+i\bar{e}_R \gamma_{\mu}D^{\mu}e_{R}
+i\bar{\nu}_R\gamma_{\mu}D^{\mu}\nu_{R}.
\nonumber
\end{eqnarray}

All couplings involving
to the processes under consideration
are shown in the following
Tables~\ref{couplingBL1}, \ref{couplingBL2}, 
\ref{couplingBL3}, \ref{couplingBL4}. 
In the Table~\ref{couplingBL1},
the third column is presented for the 
couplings in the second column 
with changing bare parameters to 
physical parameters accordingly.
The last column shows for the results
of the corresponding couplings in 
the limit
of none of the kinematic mixing. 
%%%%%%%%%%%%%%%%%%%%%%%%%%%%%%%%%%
\begin{table}[H]
\begin{center}
\begin{tabular}{|l|l|l|l| }
\hline\hline
Vertices 
& Mixing  
& Mixing (physical parameters) 
& Non-mixing  \\ 
\hline\hline
$hW^-W^+$ 
& 
$\frac{g^2v_{\Phi}}{2} 
c_{\theta} g_{\mu\nu}$
&$\frac{e M_W}{s_W} 
c_\theta g_{\mu\nu} $ & $-$ \\
\hline
$hW^+G^-$ & 
$
\frac{g}{2}c_{\theta} 
(p_{\mu}^h-p_{\mu}^{G^-})$
&
$\frac{e}{2s_W}c_{\theta} 
(p_{\mu}^h-p_{\mu}^{G^-})$ & $-$\\
\hline
$hW^-G^+$ & 
$\frac{g}{2} c_{\theta} 
(p_{\mu}^{G^+}-p_{\mu}^{h})$
&
$\frac{e}{2s_W}c_{\theta} 
(p_{\mu}^{G^+}-p_{\mu}^{h})$ & $-$\\
\hline
$AW^{\pm}G^{\mp}$ & 
$\frac{gg'c_Wv_\Phi}{2}\; g_{\mu\nu}$
&
$eM_W \; g_{\mu\nu} $ & $-$ \\
\hline
$AG^+G^-$ & $\frac{gs_W+g'c_W}{2}
(p_{\mu}^+-p_{\mu}^-) 
$
&$e (p_{\mu}^+-p_{\mu}^-)$ & $-$\\
\hline
$ZG^+G^-$ 
& $
[
\frac{gc_W-g's_W}{2} c_{BL}
+ \frac{\tilde{g}_t}{2} s_{BL}
]
(p_{\mu}^+ - p_{\mu}^-)$ 
&
$
[
\frac{e c_{2W}}{ s_{2W} } c_{BL}
+ \frac{\tilde{g}_t}{2}  s_{BL}
]
(p_{\mu}^+-p_{\mu}^-)$
&  
$
\frac{e\;c_{2W}}{s_{2W} }
(p_{\mu}^+-p_{\mu}^-)$ \\
\hline
$Z'G^+G^-$    
& 
$
[
\frac{-gc_W+g's_W}{2}s_{BL}
+ \frac{\tilde{g}_t}{2} c_{BL}
]
(p_{\mu}^+ - p_{\mu}^-)$ 
&
$
[
\frac{-e c_{2W}}{s_{2W} }s_{BL}
+ \frac{\tilde{g}_t}{2} c_{BL}
]
(p_{\mu}^+ - p_{\mu}^-)$ 
& $0$ \\
\hline
$ZW^{\pm}G^{\mp}$    
& $\frac{gv_\Phi}{2}
[
-g's_W c_{BL}
+ \tilde{g}_t  s_{BL}
]
\; g_{\mu\nu}$
&
$M_W
[
-e\frac{s_W}{c_W} c_{BL}
+ \tilde{g}_t  s_{BL}
]
\; g_{\mu\nu}$
& 
$-\frac{e M_W 
s_{W}}{c_{W}} \;
g_{\mu\nu}$
\\
\hline
$Z'W^{\pm}G^{\mp}$    
& 
$\frac{gv_\Phi}{2}
[
g's_Ws_{BL} 
+ \tilde{g}_t  c_{BL}
]
\; g_{\mu\nu}$
&
$M_W
[
e\frac{s_W}{c_W} s_{BL}
+ \tilde{g}_t  c_{BL}
]
\; g_{\mu\nu}$
& $0$
\\
\hline
$hG^+G^-$ & $\lambda_\Phi{v_\Phi}
\Big(c_\theta+\frac{v_\Phi}{v_{\chi}}s_\theta)
+2m_\Phi^2\frac{s_\theta}{v_\chi}$
& $\frac{M_{h}^2 }{v_{\Phi}}
c_\theta$ & $-$ 
\\ 
\hline \hline
\end{tabular}
\end{center}
\caption{\label{couplingBL1} 
The couplings 
relate to the processes
under consideration. 
In some cases, we have used the 
relations: $M_W=\frac{gv_{\Phi}}{2}, 
e=gs_W=g'c_W$ and $\tilde{g}_t
= -g'\frac{\kappa}{\sqrt{1-\kappa^2}}$.
}
\end{table}
%%%%%%%%%%%%%%%%%%%%%%%%%%%%%%%%%%
In the Table~\ref{couplingBL2}, 
the couplings of SM-like Higgs to 
$Z$ and $Z'$
relate to the computed processes 
are shown. The last column shows 
for the results
of the corresponding couplings in 
the case of non-kinematic mixing. 
%%%%%%%%%%%%%%%%%%%%%%%%%%%%%%%%%%
\begin{table}[H]
\begin{center}
\begin{tabular}{|l|l|l|}
\hline\hline
Vertices & Mixing & Non-mixing 
($\kappa=0, s_{BL}=0$)  \\ 
\hline\hline
$hZZ'$ 
& $
[ 
\tilde{g}^2_X Y^2_Xv_{\chi}s_{\theta}
s_{2(BL)} $
% & \\
% & 
$
-
\frac{1}{2}
(
\frac{e}{s_Wc_W}\; 
c_{BL}
- \tilde{g}_t s_{BL} 
)
\times $
& $0$  \\
& 
$\hspace{3.2cm}
\times 
(
\frac{e}{s_Wc_W}\;
s_{BL}
+ \tilde{g}_t c_{BL}
)
v_\Phi{c_\theta}
]
g_{\mu\nu}$ 
&   \\
\hline
%%%%%%%%%%%%%%%%%%%%%%
 $hZZ$  
 & $
 [ 
 2 \tilde{g}_X^2Y^2_Xv_{\chi}
 s_{\theta}\; s_{BL}^2
 $
%  & \\
%  & 
$ + 
\frac{v_{\Phi}c_{\theta}}{2}
(
 \frac{e}{s_Wc_W}\;
 c_{BL}
 -\tilde{g}_t s_{BL} 
)^2
]
g_{\mu\nu}$ 
 %%%%%%%%%%%%%%%%%%%
 & $\left(
 \frac{eM_W}{c^2_Ws_W}c_{\theta}
 \right)
 g_{\mu\nu}$ 
 \\
\hline
%%%%%%%%%%%%%%%%%%%%
$hZ'Z'$   
& $
[ 
2
\tilde{g}_X^2Y^2_Xv_{\chi}s_{\theta}\;
c_{BL}^2$
% & \\
% &
$
+\frac{v_{\Phi}c_{\theta}}{2}
(
\frac{e}{s_Wc_W}\;
s_{BL}
+ \tilde{g}_t c_{BL} 
)^2
]
g_{\mu\nu}$ 
& $
\left(
2 g_X^2Y^2_Xv_{\chi}s_{\theta}
\right)
g_{\mu\nu}$ 
\\
\hline\hline
\end{tabular}
\end{center}
\caption{\label{couplingBL2} 
The couplings of SM-like Higgs to $Z$ and $Z'$
relate to the processes under consideration. 
In some cases, 
we have used the following relations 
$M_W=\frac{gv_{\Phi}}{2}, e=gs_W=g'c_W$.
Here $s_{2(BL)} = 2 s_{BL}c_{BL}$.}
\end{table}
%%%%%%%%%%%%%%%%%%%%%%%%%%%%%%%%%%
In the Table~\ref{couplingBL3}, 
three gauge boson 
vertices relating to the processes
are presented.
The corresponding couplings in 
none of the kinematic mixing are 
shown 
in the last column results.
%%%%%%%%%%%%%%%%%%%%%%%%%%%%%%%%%%
\begin{table}[H]
\begin{center}
\begin{tabular}{ |l | l| l|}
\hline\hline
 Vertices & Mixing  & Non-mixing\\
 \hline \hline
 $A_{\rho}W^{-}_{\nu}W^{+}_{\mu}$ 
 & $ e
 [ (p_1-p_3)^{\nu} g^{\rho\mu}
 -(p_1-p_2)^{\rho}g^{\mu\nu}
 -(p_2-p_3)^{\mu}g^{\rho\nu}
 ]
 $ 
 & $-$\\
 \hline
 $Z_{\rho}W^{-}_{\nu}W^{+}_{\mu}$ 
 & $e\frac{c_W}{s_W} c_{BL} 
 [ 
 (p_1-p_3)^{\nu} g^{\rho\mu}
 -(p_1-p_2)^{\rho}g^{\mu\nu}
 -(p_2-p_3)^{\mu}g^{\rho\nu}
 ]
 $ 
 & $-|_{c_{BL}\rightarrow 1}$\\
  \hline
  ${Z'}_{\rho}W^{-}_{\nu}W^{+}_{\mu}$  
  & 
  $-e\frac{c_W}{s_W} s_{BL} 
 [ 
 (p_1-p_3)^{\nu} g^{\rho\mu}
 -(p_1-p_2)^{\rho}g^{\mu\nu}
 -(p_2-p_3)^{\mu}g^{\rho\nu}
 ]
 $ 
  & $0$
  \\
  \hline\hline
 \end{tabular}
\end{center}
\caption{\label{couplingBL3} 
The couplings three gauge 
boson vertices
relating to the processes
under consideration.}
\end{table}
%%%%%%%%%%%%%%%%%
In the Table~\ref{couplingBL4}, all 
couplings of $Zff$ and $Z' ff$ are shown. 
Hypercharge $Y_X^f$ is taken the 
corresponding values for $f$ 
showing in Table~\ref{MatterBL}. 
Here
$P_{L,R} = \frac{1-\gamma_5}{2}$ and 
$\tilde{g}_{X}=\frac{g_X}{\sqrt{1-\kappa^2}}$.
Again, the corresponding couplings in 
the case 
of none of the kinematic mixing are presented 
in the last column results.
%%%%%%%%%%%%%%%%%%%%%%
\begin{table}[H]
\begin{center}
\begin{tabular}{|l|l|l|  }
\hline\hline
 Vertices &   Mixing  & Non-mixing\\
 \hline\hline 
  $Z_\mu\bar{f}f$
  & 
  $-
  \frac{e}{s_Wc_W}
  \gamma_{\mu}
  [
  (I_{3}^{f} - s_W^2 Q_f)P_L
  -  s_W^2 Q_f P_R
  ]c_{BL} $
& \\
&
$
-
\gamma_{\mu}
[
( 
\tilde{g}_{X} Y_X^f
+\tilde{g}_t Y_{f_L}
) P_L
+
(
\tilde{g}_{X}
Y_X^f
+\tilde{g}_t
Y_{f_R}
) P_R
]s_{BL} 
$ 
& $-|_{\kappa \rightarrow0,
c_{BL}\rightarrow 1}$      
\\ \hline
%%%%%%%%%%%%%%%%%
 $Z'_\mu\bar{f}f$
  & 
  $\frac{e}{s_Wc_W}
\gamma_{\mu}
[
(I_{3}^{f} - s_W^2 Q_f)P_L
-s_W^2 Q_f P_R
]s_{BL} $
& \\
&
$
-
\gamma_{\mu}
[
(
\tilde{g}_X Y_X^f
+ \tilde{g}_t
Y_{f_L}
) P_L
+
(
\tilde{g}_X Y_X^f
+\tilde{g}_t Y_{f_R}
) P_R
]c_{BL} 
$ 
& $-g_X\; Y_X^f\gamma_{\mu}$ 
\\  \hline \hline
\end{tabular}
\end{center}
\caption{\label{couplingBL4}
The couplings related to the processes
under consideration. In some cases, 
we have used the 
relations: $M_W=\frac{gv_{\Phi}}{2},\quad e=gs_W=g'c_W$.
Hypercharge $Y_X^f$ is taken the 
corresponding values
for $f$ showing in Table~\ref{MatterBL}. 
Here
$P_{L,R} = \frac{1-\gamma_5}{2}$ and 
$\tilde{g}_{X}=\frac{g_X}{\sqrt{1-\kappa^2}}$.
}
\end{table}
%%%%%%%%%%%%%%%%%%%%%%%%%%%%%%%%%%%%%%%%%%%%%%%%
\subsection{Chiral $B-L$ model} %% 
%%%%%%%%%%%%%%%%%%%%%%%%%%%%%%%%%%
We are going to discuss another version of
$U(1)_{B-L}$ extension of the SM which is 
chiral $B-L$ model. In this version of 
$B-L$ model, together
with three right-handed neutrinos, 
we have an extra doublet
$\varphi$ and an scalar singlet $\sigma$. 
The matter field contents, scalar fields
and their quantum numbers are presented
in the Table~\ref{MatterCBL}. 
The generation index for matter particles
are suppressed.
%%%%%%%%%%%%%%%%%%%%%%%%%%%%%%%%%%%%%
\begin{table}[H]
\centering
 \begin{tabular}{|c|c|c|}
 \hline \hline
 ~~~Fields~~~ & $SU(3)_{C} 
 \otimes SU(2)_{L} 
 \otimes U(1)_{Y} 
 \otimes U(1)_{B-L}$     \\ 
  \hline \hline
 $L_{L}$  & ($1,2,-1/2,-1$) \\ 
 \hline 
 $Q_{L}$  & ($3,2,1/6,1/3$) \\ 
 \hline 
 $e_{R}$  & ($1,1,-1,-1$) \\ 
 \hline 
 $u_{R}$  & ($3,1,2/3,1/3$) \\ 
 \hline 
 $d_{R}$  & ($3,1,-1/3,1/3$) \\ 
 \hline 
 \hline
 $\nu_{R}^1$  & ($1,1,0,5$) \\ 
 \hline 
 $\nu_{R}^{2,3}$  & ($1,1,0,-4$) \\ 
 \hline
  \hline 
 $\Phi$  & ($1,2,1/2,0$) \\ 
 \hline 
 $\varphi$  & ($1,2,1/2,-3$) \\  
  \hline 
 $\sigma$  & ($1,1,0,3$) \\  
 \hline 
 $\chi_{d}$  & ($1,1,0,1/2$) \\  
 \hline 
 \end{tabular}  
\caption{\label{MatterCBL} Matter fields 
and their 
quantum numbers of the chiral $B-L$ model.
We omit the index of generations of 
matter particles in this Table. 
In this work, 
we note right handed neutrinos as
$N_i$ for $i=1,2,3$ for latter uses.}
\end{table}
%%%%%%%%%%%%%%%%%%%%%%%%%%%%%%%%%%
In the other hand, this model is also added
one more a scalar Dark matter $\chi_{d}$ which 
plays a role of dark matter. In general, the 
scalar sector is then taken the form of
%%%%%%%%%%%%%%%%%%%%%%%%%%%%%%%%%%
\begin{eqnarray}
\mathcal{L}_H 
&=& 
\mathcal{L}_K 
- \mathcal{V}(\Phi,\varphi,\sigma,\chi_d)   \\
&=&
(D_\mu{\Phi})^\dagger{D_{\mu}\Phi}
+(D_\mu{\varphi})^\dagger{D_{\mu}\varphi}
+(D_\mu{\sigma})^\dagger{D_{\mu}\sigma}
+(D_\mu{\chi_d})^\dagger{D_{\mu}\chi_d} 
-\mathcal{V}(\Phi,\varphi,\sigma,\chi_d).
\nonumber
\end{eqnarray}
%%%%%%%%%%%%%%%%%%%%%%%%%%%%%%%%%%
Where the covariant derivative is defined in 
(\ref{Dnon}). The scalar potential is 
expressed as follows:
%%%%%%%%%%%%%%%%%%%%%%%%%%%%%%%%%%
\begin{eqnarray}
\mathcal{V}(\Phi,\varphi,\sigma,\chi_d)
&=&
m_\sigma^2(\sigma^*\sigma)
+\frac{1}{2}\lambda_\sigma(\sigma^*\sigma)^2
+ m_\Phi^2(\Phi^\dagger\Phi) 
+ \frac{1}{2}\lambda_\Phi(\Phi^\dagger\Phi)^2
+ m_\varphi^2(\varphi^\dagger\varphi)
+\frac{1}{2}\lambda_\varphi(\varphi^\dagger\varphi)^2 
\nonumber\\
&& 
+ m_{\chi_d}^2(\chi_d^*\chi_d)
+ \frac{1}{2}\lambda_{\chi_d}(\chi_d^*\chi_d)^2
-\mu(\Phi^\dagger\varphi)\sigma
-\mu(\varphi^\dagger\Phi)\sigma^*
+\lambda_{\Phi\sigma}(\Phi^\dagger\Phi)(\sigma\sigma^*) 
\nonumber\\
&&
+ \lambda_{\varphi\sigma}(\varphi^\dagger\varphi)(\sigma\sigma^*)
+ \lambda_{\Phi\varphi_1}(\Phi^\dagger\Phi)(\varphi^\dagger\varphi)
+ \lambda_{\Phi\varphi_2}(\Phi^\dagger\varphi)(\varphi^\dagger\Phi)
+ \lambda_{\Phi\chi_d}
(\Phi^\dagger\Phi)(\chi_d^*\chi_d) 
\nonumber\\
&& 
+\lambda_{\varphi\chi_d}(\varphi^\dagger\varphi)
(\chi_d^*\chi_d)+\lambda_{\sigma\chi_d}
(\sigma^*\sigma)(\chi_d^*\chi_d).
\end{eqnarray}
%%%%%%%%%%%%%%%%%%%%%%%%%%%%%%%%%%
The fields $\Phi,\varphi$ and 
$\sigma$ can be written for
the SSB as follows:
%%%%%%%%%%%%%%%%%%%%%%%%%%%%%%%%%%
\begin{eqnarray}
\Phi=\frac{1}{\sqrt{2}}
\left(\begin{array}{c}
\sqrt{2}G^{\pm}_1  \\
v_\Phi + R_1 + i I_1 
\end{array}\right), 
\quad 
\varphi = \frac{1}{\sqrt{2}}
\left(\begin{array}{c}
\sqrt{2}G^{\pm}_2  \\
         v_\varphi+R_2+iI_2 
\end{array}\right),
\quad \sigma 
=
\frac{1}{\sqrt{2}}(v_\sigma+R_3+iI_3).
\end{eqnarray}
%%%%%%%%%%%%%%%%%%%%%%%%%%%%%%%%%%
In this role, the fields $\Phi$ and $\varphi$ will 
break electroweak symmetry. While the scalar fields 
$\sigma$ and $\varphi$ will break 
$U(1)_{B-L}$ symmetry. 
The minimization for the scalar potential
leads to the following system of equations:
%%%%%%%%%%%%%%%%%%%%%%%%%%%%%%%%%%
\begin{eqnarray}
2m_\Phi^2+\lambda_\Phi{v_\Phi^2}
+{\lambda_{\Phi\sigma}}{v_\sigma^2}
+(\lambda_{\Phi\varphi_1}
+\lambda_{\Phi\varphi_2}){v_\varphi^2}
-\frac{\sqrt{2}{\mu}v_\varphi{v_\sigma}}{v_\Phi}
&=&0,
\\
%%%%%%%%%%%%%%%%%%%%%%%%%%%%%%%%%%%
2m_\sigma^2+\lambda_\sigma{v_\sigma^2}
+{\lambda_{\Phi\sigma}}{v_\Phi^2}
+\lambda_{\varphi\sigma}{v_\varphi^2}
-\frac{\sqrt{2}{\mu}v_\Phi{v_\varphi}}{v_\sigma}
&=& 0,
\\
%%%%%%%%%%%%%%%%%%%%%%%%%%%%%%%%%%%
2m_\varphi^2+\lambda_\varphi{v_\varphi^2} 
+ {\lambda_{\varphi\sigma}}{v_\sigma^2} 
+ (\lambda_{\Phi\varphi_1}
+\lambda_{\Phi\varphi_2}){v_\Phi^2}
- \frac{\sqrt{2}{\mu}v_\Phi{v_\sigma}}{v_\varphi}
&=& 0.
\end{eqnarray}
%%%%%%%%%%%%%%%%%%%%%%%%%%%%%%%%%%
The mass matrix for charged Higgs is then collected
in term of the basics $(G^{\pm}_1, G^{\pm}_2)$ as 
\begin{equation}
\mathcal{M}_{\pm}^2 = \frac{1}{2}
\begin{pmatrix}
\frac{\sqrt{2}\mu v_{\sigma}v_{\varphi}}{v_{\Phi}}-v_{\varphi}^{2}\lambda_{\Phi \varphi_{2}} ~&~ v_{\Phi}v_{\varphi}\lambda_{\Phi \varphi_{2}} - \sqrt{2}\mu v_{\sigma}\\\\
 v_{\Phi}v_{\varphi}\lambda_{\Phi \varphi_{2}} - \sqrt{2}\mu v_{\sigma} ~&~ \frac{\sqrt{2}\mu v_{\sigma}v_{\Phi}}{v_{\varphi}}-v_{\Phi}^{2}\lambda_{\Phi \varphi_{2}} 
\end{pmatrix}.
\end{equation}
After symmetry breaking, this model 
contains Goldstone bosons $G^{\pm}$
and charged Higgs $H^{\pm}$ 
which are mixed by $G^{\pm}_1, G^{\pm}_2$. 
The Goldstone bosons $G^{\pm}$
will give the masses to the $W^{\pm}$ bosons. 
The mixing matrix is given by
%%%%%%%%%%%%%%%%%%%%%%%%%%%%%%%%%%
\begin{eqnarray}
\left(\begin{array}{c}
G^{\pm}  \\
H^{\pm} 
\end{array}\right)
=\left(\begin{array}{cc}
c_\alpha & s_\alpha  \\
-s_\alpha & c_\alpha
\end{array}\right)
\left(\begin{array}{c}
G^{\pm}_1  \\
G^{\pm}_2 
\end{array}\right).
\end{eqnarray}
The mass matrix in the basis 
$(I_{1},I_{2},I_{3})$ can be written as
\begin{equation}
\mathcal{M}_{I}^2=\frac{1}{\sqrt{2}}
\begin{pmatrix}
\frac{\mu v_{\varphi} v_{\sigma}}{v_{\Phi}}~
&~ -\mu v_{\sigma} ~&~ -\mu v_{\varphi}\\
-\mu v_{\sigma} ~ 
&~ \frac{\mu v_{\Phi} v_{\sigma}}{v_{\varphi}} 
~&~ \mu v_{\Phi}\\
-\mu v_{\varphi}  ~&~ \mu v_{\Phi} 
~&~ \frac{\mu v_{\Phi} v_{\varphi}}{v_{\sigma}} 
\end{pmatrix}.
\end{equation}
For the neutral components of Higgs fields, 
we have the following relation:
%%%%%%%%%%%%%%%%%%%%%%%%%%%%%%%%%%
\begin{eqnarray}
\left(\begin{array}{c}
G^{0}_1  \\
G^{0}_2 \\
A_0
\end{array}\right)
=\left(\begin{array}{ccc}
c_\alpha & s_\alpha & 0 \\
-s_\alpha{c_\beta}  & 
c_\alpha{c_{\beta}} & 
-s_{\beta} \\
-s_\alpha{s_\beta}  & 
c_\alpha{s_\beta}   & 
c_\beta
\end{array}\right)
\left(\begin{array}{c}
I_1  \\
I_2  \\
I_3
\end{array}\right),
    \quad \text{with} 
    \quad t_\alpha 
=\frac{v_\varphi}{v_\Phi}, 
\quad 
t_\beta
=\frac{v_\sigma{v}}
{v_\Phi{v_\varphi}}.
\end{eqnarray}
After the EWSB, $G^{0}_1, G^{0}_2$ 
will give the masses for
$Z$ and $Z'$, respectively. The 
remaining physical 
filed $A_0$ becomes CP-odd Higgs boson. 
The masses of the charged and CP-odd 
Higgses are determined as follows:
%%%%%%%%%%%%%%%%%%%%%%%%%%%%%%%%%%
\begin{eqnarray}
M^2_{H^{\pm}}&=&\frac{v^2}{2v_{\Phi}
v_{\varphi}}(\sqrt{2}\mu{v_\sigma}
-v_\Phi{v_{\varphi}}\lambda_{\Phi\varphi_2}),
\\
M^2_{A_0}
&=&
\frac{\mu}{\sqrt{2}v_{\Phi}v_{\varphi}
v_{\sigma}}(v_{\Phi}^2v_{\varphi}^2 
+ v_\sigma^2v^2).
\label{MA0}
\end{eqnarray}
%%%%%%%%%%%%%%%%%%%%%%%%%%%%%%%%%%
The mass matrix for neutral scalars 
is collected
in the basis $(R_{1},R_{2},R_{3})$. 
It can be 
expressed in the form of
\begin{equation}
\mathcal{M}_{S}^2=
\frac{1}{2}
\begin{pmatrix}
 2v_{\Phi}^{2}\lambda_{\Phi}
 +\frac{\sqrt{2}\mu}{v_{\Phi}}v_{\varphi}v_{\sigma}  
 &~~  2v_{\Phi}v_{\varphi}\lambda_{12} 
 - \sqrt{2}\mu v_{\sigma}  
 &~~   2v_{\Phi}v_{\sigma}\lambda_{\Phi \sigma} 
 -\sqrt{2}v_{\varphi}\mu \\\\
  2v_{\Phi}v_{\varphi}\lambda_{12} 
  -\sqrt{2}\mu v_{\sigma}  
  &~~  2v_{\varphi}^{2}\lambda_{\varphi}
  + \frac{\sqrt{2} \mu}{v_{\varphi}}v_{\Phi}v_{\sigma}    &~~  2 v_{\varphi}v_{\sigma}\lambda_{\varphi \sigma}- \sqrt{2}v_{\Phi}\mu\\\\
 2 v_{\Phi}v_{\sigma}\lambda_{\Phi\sigma}-\sqrt{2}v_{\varphi}\mu  &~~   2v_{\varphi}v_{\sigma}\lambda_{\varphi \sigma}  - \sqrt{2}v_{\Phi}\mu &~~  2v_{\sigma}^{2}\lambda_{\sigma} +\frac{\sqrt{2}\mu}{v_{\sigma}}v_{\Phi}v_{\varphi}, \\
\end{pmatrix},
\end{equation}
%%%%%%%%%%%%%%%%%%%%%%%%%%%%%%%%%%%%%%%%%%%
where, $\lambda_{12} 
= \lambda_{\Phi \varphi_{1}}
+ \lambda_{\Phi \varphi_{2}}.$
%%%%%%%%%%%%%%%%%%%%%%%%%%%%%%%%%%%%%
The matrix $\mathcal{M}^2_{S}$ can be 
diagonalized by an orthogonal matrix as 
\begin{eqnarray}
\mathcal{O}_{S}^T M_{S}^2 \mathcal{O}_S 
= \text{diag}(M_{H_1}^2,M_{H_2}^2,M_{H_3}^2).
\end{eqnarray}
 Where the relation between two basics
 are given by
%%%%%%%%%%%%%%%%%%%%%%%%%%%%%%%%%%%%%%%%%%%
\begin{eqnarray}
\label{rot1}
\left( \begin{array}{c} 
H_1\\ H_2\\ H_3\\ 
\end{array} \right) 
= 
\mathcal{O}_{S} 
\left(\begin{array}{c} R_1\\ R_2\\ R_3\\ 
\end{array} \right). 
\end{eqnarray}
%%%%%%%%%%%%%%%%%%%%%%%%%%%%%%%%%%%%
For CP-even scalar Higges, 
the mass eigenstates to be ordered 
by their masses 
$M_{H_1}\leq M_{H_2}\leq M_{H_3}$. 
In this notation, we note that 
$h=H_{1}$ is identified as the SM 
Higgs of $125$ GeV. 
The rotation matrix is defined as 
$\mathcal{O}_{S} = \mathcal{R}_{23} 
\mathcal{R}_{13} \mathcal{R}_{12}$ 
where each matrix is expressed as follows:
%%%%%%%%%%%%%%%%%%%%%%%%%%%%%%%%%%%%%
\begin{equation}
\mathcal{R}_{12} = \left(
\begin{array}{ccc}
c_{12} & -s_{12} & 0\\
s_{12} & c_{12} & 0\\
0 & 0 & 1
\end{array} \right), 
\quad \mathcal{R}_{13} = \left(
\begin{array}{ccc}
c_{13} & 0 & -s_{13}\\
0 & 1 & 0\\
s_{13} & 0 & c_{13}
\end{array} \right), 
\quad \mathcal{R}_{23} = \left(
\begin{array}{ccc}
1 & 0 & 0\\
0 & c_{23} &  -s_{23}\\
0 & s_{23} & c_{23}
\end{array} \right)
\end{equation}
%%%%%%%%%%%%%%%%%%%%%%%%%%%%%%%%%%%%%%
where $c_{ij} = \cos \theta_{ij}, s_{ij} 
= \sin \theta_{ij}$ with 
$-\frac{\pi}{2}\leq\theta_{ij}\leq \frac{\pi}{2}$.

%%%%%%%%%%%%%%%%%%%%%%%%%%%%%%%%%%
The rotation matrix for neutral
gauge bosons
are given 
%%%%%%%%%%%%%%%%%%%%%%%%%%%%%%%%%%
\begin{eqnarray}
\left(\begin{array}{c}
A^\mu  \\
Z^\mu  \\
Z'^\mu
\end{array}
\right)
=\left(
\begin{array}{ccc}
c_W & s_W & 0     \\
-c'_{BL} s_W & c'_{BL}c_W  & -s'_{BL}    \\
-s'_{BL}s_W & s'_{BL}c_W & c'_{BL}    \\
\end{array}
\right) 
\left(
\begin{array}{c}
B^\mu    \\
W_3^\mu  \\
X^\mu
\end{array}
\right),
\quad \text{with}
\quad t'_{2(BL)}=\frac{C'}{B'}
\end{eqnarray}
%%%%%%%%%%%%%%%%%%%%%%%%%%%%%%%%%%
which the masses of photon 
$A, Z, Z'$ bosons are obtained 
%%%%%%%%%%%%%%%%%%%%%%%%%%%%%%%%%%
\begin{eqnarray}
M_A=0,\quad\quad 
M_Z^2=\frac{v^2}{8}(A'-\sqrt{{B'}^2+{C'}^2}), 
\quad\quad 
M_{Z'}^2=\frac{v^2}{8}(A'+\sqrt{{B'}^2+{C'}^2}).
\end{eqnarray}
%%%%%%%%%%%%%%%%%%%%%%%%%%%%%%%%%%
Where 
\begin{eqnarray}
A'=36\frac{v^2_\varphi+v_{\sigma}^2}{v^2}g_X^2+(g^2+{g'}^2), 
\quad B'=36\frac{v^2_\varphi
+v_{\sigma}^2}{v^2}g_X^2
-(g^2+{g'}^2), \quad 
C'= 12 g_X\frac{v_\varphi^2}{v^2}
\sqrt{g^2+{g'}^2}.
\nonumber\\
\end{eqnarray}
%%%%%%%%%%%%%%%%%%%%%%%%%%%%%%%%%%
It is noted that we haven't consider the
kinematic mixing of two gauges $U(1)_Y$ 
and $U(1)_{B-L}$. But in this report, 
we assume the mixing case of two neutral 
gauge bosons $Z$, $Z'$.

The Yukawa Lagrangian of the model 
can be written 
%%%%%%%%%%%%%%%%%%%%%%%%%%%%%%%%%%
\begin{eqnarray}
-\mathcal{L}_Y&=&
Y_e\bar{L}\Phi{e_R}+Y_u\bar{Q}\tilde{\Phi}{u_R}
+Y_e\bar{Q}\Phi{d_R} 
+Y_\nu\bar{L}\tilde{\varphi}{\nu_R} 
+ H.c \nonumber\\
&{\supset}&
\frac{Y_\nu{v_{\varphi}}}{\sqrt{2}}\bar{\nu_L}\nu_R
+ H.c
\end{eqnarray}
%%%%%%%%%%%%%%%%%%%%%%%%%%%%%%%%%%
where the $\varphi$ field generates Dirac 
neutrino mass for RHNs. This model doesn't 
have Majorana mass term for $\nu_R$.

Finally, the mass of dark matter from 
the Higgs potential is given as
%%%%%%%%%%%%%%%%%%%%%%%%%%%%%%%%%%
\begin{eqnarray}
M_{DM}^2 &=& 
\frac{2m_{\chi_d}^2 
+ v_{\Phi}^2\lambda_{\chi_d} 
+ v_{\sigma}^2\lambda_{\sigma\chi_d} 
+ v_{\varphi}^2\lambda_{\varphi\chi_d}}{2}
\end{eqnarray}
%%%%%%%%%%%%%%%%%%%%%%%%%%%%%%%%%%

In this version, we have two CP-even Higges, 
one CP-odd Higgs, and two charged Higsses. 
Furthermore, we have three more right handed
neutrinos and a neutral gauge boson $Z'$. 
All these new particles are exchanged in the loop
Feynman diagrams of the processes under 
consideration. 

All couplings relating to the 
the computed processes are 
shown in the following 
Tables~\ref{couplingchiralBL1},
~\ref{couplingchiralBL2},
~\ref{couplingchiralBL3},
~\ref{couplingchiralBL4},
~\ref{couplingchiralBL5}.
In several cases, we have used
$M_W=\frac{1}{2}g\sqrt{v_{\Phi}^2+v_{\varphi}^2},
\quad s_\alpha=\frac{v_\varphi}{v},
\quad c_\alpha=\frac{v_{\Phi}}{v},
\quad e=gs_W=g'c_W$.
In Table~\ref{couplingchiralBL1}, 
the third column results are corresponding
the couplings in the second column with 
replacing bare parameters in the Lagrangian
to the physical parameters. The last column
shows for the results of the non-mixing 
two gauge bosons. 
%%%%%%%%%%%%%%%%%%%%%%%%%%%%%%%%%%
\begin{table}[H]
\begin{center}
\begin{tabular}{|l|l|l|l|}
 \hline\hline
 \text{Vertices} 
 & Mixing 
 & Mixing (physical parameters) 
 & Non-mixing \\ 
 \hline\hline
$hW^{\pm}W^{\mp}$        
& $\frac{g^2}{2}(v_{\Phi}c_{13}c_{12}
-v_{\varphi}c_{13}s_{12}) \; g_{\mu\nu}$  
& $ \frac{e M_W}{s_W} c_{\alpha + \widehat{12}}
c_{13}$ & $-$
\\ \hline
$Z'W^{\pm}G^{\mp}$   
& 
$-\frac{g g's_W (c_{\alpha}v_\Phi
+s_{\alpha}v_\varphi) }{2}s_{BL}'$
&
$-\frac{e M_W s_{W}}{c_W}s_{BL}'$
& $0$
\\  \hline
$ZW^{\pm}G^{\mp}$    
& 
$-\frac{g g' s_W (c_{\alpha}v_\Phi 
+ s_{\alpha}v_\varphi)
c_{BL}'}{2}$
&
$-\frac{e M_W s_{W}}{c_W}c_{BL}'$
& $-|_{c_{BL}' =1}$
\\
\hline
$hW^+G^-$        
& $\frac{gc_{13}(c_{12}c_{\alpha}
-s_{12}s_{\alpha})}{2}
(p_{\mu}^{h}-p_{\mu}^{G^-})$ 
& $\frac{e}{2s_W}
c_{\alpha+\widehat{12}}
c_{13}
(p_{\mu}^{h}-p_{\mu}^{G^-})$
& $-$
\\
\hline
$hW^-G^+$        
& $-\frac{gc_{13}(c_{12}c_{\alpha}
- s_{12}s_{\alpha})}{2} 
(p_{\mu}^{h}-p_{\mu}^{G^+})$  
&
$-
\frac{e}{2s_W}
c_{\alpha+\widehat{12}}
c_{13}
(p_{\mu}^{h}-p_{\mu}^{G^+})
$
& 
$-$
\\
\hline
$AG^{\pm}G^{\mp}$        
& $\frac{(gs_W+g'c_W)}{2}
(p_{\mu}^{+}-p_{\mu}^-)^G 
$
&$e(p_{\mu}^{+}-p_{\mu}^-)^G$ 
& $-$
\\
\hline
$ZG^{\pm}G^{\mp}$        
& $\frac{(gc_W-g's_W )c_{BL}'}{2}
(p_{\mu}^{+}-p_{\mu}^-)^G$
&$e \frac{c_{2W}}{s_{2W}}
c_{BL}' 
(p_{\mu}^{+}-p_{\mu}^-)^G$  
& $-|_{c_{BL}' =1}$
\\
\hline
$Z'G^{\pm}G^{\mp}$  
& $\frac{(gc_W-g's_W) 
s_{BL}'}{2}(p_{\mu}^{+}-p_{\mu}^-)^G
$ &
$e\frac{c_{2W}}{s_{2W}}s_{BL}' 
(p_{\mu}^{+}-p_{\mu}^-)^G$
& $0$
\\
\hline
 $AH^{\pm}H^{\mp}$        
& $\frac{(gs_W+g'c_W)}{2} 
(p_{\mu}^{+}-p_{\mu}^-)^H 
$
&$e (p_{\mu}^{+}-p_{\mu}^-)^H$  
& $-$
\\ \hline
$ZH^{\pm}H^{\mp}$        
& $\frac{(gc_W-g's_W) c_{BL}'}{2}
\; (p_{\mu}^{+}-p_{\mu}^-)^H
$
&$e \frac{c_{2W}}{s_{2W}}
c_{BL}' 
(p_{\mu}^{+}-p_{\mu}^-)^H$ 
&$-$
\\ \hline
$Z'H^{\pm}H^{\mp}$       
& $\frac{(gc_W-g's_W) s_{BL}'}{2}
\; (p_{\mu}^{+}-p_{\mu}^-)^H$
&$e \frac{c_{2W}}{s_{2W} }s_{BL}' 
\;(p_{\mu}^{+}-p_{\mu}^-)^H$ 
& $0$
\\ \hline \hline
\end{tabular}
\end{center}
\caption{\label{couplingchiralBL1} 
The couplings relate to the processes 
under consideration. Some cases we have used
$M_W=\frac{1}{2}g\sqrt{v_{\Phi}^2+v_{\varphi}^2},
\quad s_\alpha=\frac{v_\varphi}{v},
\quad c_\alpha=\frac{v_{\Phi}}{v},
\quad e=gs_W=g'c_W$. The vector
$B-L$ can be obtained by taking the limits of
$c_{12}\rightarrow1, c_{23}\rightarrow 1$ and
$s_{BL}' \rightarrow - s_{BL}$.
}
\end{table}
% %%%%%%%%%%%%%%%%%%%%%%%%%%%%%%%%%%
\begin{table}[H]
\begin{center}
\begin{tabular}{|l|l|l|}
 \hline\hline
 \text{Vertices} & Mixing & Non-mixing\\ 
 \hline\hline
$hZZ$ & 
$
[ 2\; g_X^2  Y^2_X (v_{\sigma} s_{13}
- v_{\varphi}c_{13}s_{12} 
)s_{BL}'^2
+\frac{e M_W}{s_W c^2_W}
c_{\alpha+\widehat{12}}
c_{13}
c_{BL}'^2
] 
g_{\mu\nu}$ 
& $-|_{c_{BL}'=1}$
\\ \hline
%%%%%%%%%%%%%%%%%%%%%
$hZ'Z'$ 
&
$
[
2 g_X^2Y^2_X (v_{\sigma} s_{13}
-v_{\varphi} c_{13}s_{12})
c_{BL}'^2
+\frac{e M_W}{s_W c^2_W}
c_{\alpha+\widehat{12}}
c_{13}
s_{BL}'^2
] 
g_{\mu\nu}$  
& $-|_{c_{BL}'=1}$
\\ \hline 
%%%%%%%%%%%%%%%%%%%%%
$hZZ'$ & 
$
[ 
\frac{e M_W}{2 s_W c^2_W}
c_{\alpha+\widehat{12}}
c_{13}
+g_X^2Y^2_X
(v_{\varphi} 
c_{13}s_{12}-v_{\sigma} s_{13}) 
]s_{2(BL)'} 
\; g_{\mu\nu} $  
&$0$
\\ \hline\hline 
\end{tabular}
\end{center}
\caption{\label{couplingchiralBL2}
The couplings $hZZ, h Z'Z'$ in chiral
$B-L$ model. We have already taken
into account $Y^2_X=Y^2_X(\sigma) = 
Y^2_X(\varphi)=(\pm 3)^2$. 
The vector
$B-L$ can be obtained by taking the limits of
$c_{12}\rightarrow1, c_{23}\rightarrow 1$ and
$s_{BL}' \rightarrow - s_{BL}$.  }
\end{table}
% %%%%%%%%%%%%%%%%%%%%%%%%%%%%%%%%%%%%%%%%%%%%%%%
\begin{table}[H]
\begin{center}
\begin{tabular}{|l|l|l| }
\hline\hline
 Vertices & Mixing & Non-mixing   \\
 \hline \hline
 $A_{\rho}W^{-}_{\nu}W^{+}_{\mu}$ 
 & $e\; 
 [
 (p_1-p_3)^{\nu} g^{\rho\mu}
 -(p_1-p_2)^{\rho}g^{\mu\nu}
 -(p_2-p_3)^{\mu}g^{\rho\nu}
 ]
 $ 
 & $-$
 \\   \hline
 $Z_{\rho}W^{-}_{\nu}W^{+}_{\mu}$
 & $\frac{ec_W}{s_W} c_{BL}'\; 
 [
 (p_1-p_3)^{\nu} g^{\rho\mu}
 -(p_1-p_2)^{\rho}g^{\mu\nu}
 -(p_2-p_3)^{\mu}g^{\rho\nu}
 ]
 $ 
 & $-|_{c_{BL}'=1}$
 \\  \hline
  ${Z'}_{\rho}W^{-}_{\nu}W^{+}_{\mu}$  
  & $\frac{ec_W}{s_W}s_{BL}' 
  [
  (p_1-p_3)^{\nu} g^{\rho\mu}
 -(p_1-p_2)^{\rho}g^{\mu\nu}
 -(p_2-p_3)^{\mu}g^{\rho\nu}
 ]
  $ 
  & $0$
  \\ \hline \hline
  \end{tabular}
\end{center}
\caption{\label{couplingchiralBL3} 
The couplings of three gauge bosons
in chiral $B-L$ model which are related 
to the processes 
under consideration. The vector $B-L$ model
can be derived by applying 
$s_{BL}' \rightarrow 
-s_{BL}$.}
\end{table}
%%%%%%%%%%%%%%%%%%%%%%%%%%%%
\begin{table}[H]
\begin{center}
\begin{tabular}{|l|l|l|  }
\hline\hline
 Vertices &   Mixing  & Non-mixing\\
 \hline\hline 
$Z_\mu\bar{f}f$
& 
$
-
\frac{e}{s_Wc_W}
\gamma_{\mu}
[
  (I_{3}^{f} - s_W^2 Q_f)P_L
-  s_W^2 Q_f P_R
]c'_{BL} $
% % & \\
% &
$
+
\gamma_{\mu}
(g_X Y_X^f) s'_{BL} 
$ 
& $-|_{
c'_{BL}\rightarrow 1}$      
\\ \hline
%%%%%%%%%%%%%%%%%
 $Z'_\mu\bar{f}f$
  & 
  $-\frac{e}{s_Wc_W}
\gamma_{\mu}
[
  (I_{3}^{f} - s_W^2 Q_f)P_L
-  s_W^2 Q_f P_R
]s'_{BL} $
$
-
\gamma_{\mu}
(g_X Y_X^f) c'_{BL} 
$  
& $-g_XY_X^f\gamma_{\mu}$ 
\\  \hline \hline
\end{tabular}
\end{center}
\caption{
\label{couplingchiralBL4} 
The couplings of $Z$ and $Z'$ 
bosons to fermion 
pair in chiral $B-L$ model which are 
related to the processes 
under consideration. The vector 
$B-L$ model
can be derived by applying 
$s_{BL}' \rightarrow 
-s_{BL}$.
Hypercharge $Y_X^f$ is taken the corresponding 
values for $f$ showing in Table~\ref{MatterBL}. 
Here
$P_{L,R} = \frac{1-\gamma_5}{2}$ and 
$\tilde{g}_{X}=\frac{g_X}{\sqrt{1-\kappa^2}}$.
}
\end{table}

%%%%%%%%%%%%%%%%%%%%%%%%%%%%%%%%%%%%%%%%%%%%
\begin{table}[H]
\begin{center}
\begin{tabular}{|l|l|l|}
 \hline\hline
 \text{Vertices } & Mixing & Non-mixing  \\ 
 \hline\hline
$A\bar{u}^{\pm}u^{\mp}$ & $\mp{ie \;p^{\mu}}$  
& $ -$
\\ \hline
$Z'\bar{u}^{\pm}u^{\mp}$ & 
$\mp{i}\frac{e}{s_W}c_W s_{BL}' \;p^{\mu} $ 
& $0$
\\ \hline
$Z\bar{u}^{\pm}u^{\mp}$ 
& $\mp{i}\frac{e}{s_W}c_W c_{BL}' \; p^{\mu}$ 
& $\mp{i}\frac{e}{s_W}c_W \; p^{\mu}$
\\
\hline \hline
\end{tabular}
\end{center}
\caption{
\label{couplingchiralBL5} 
The couplings of neutral gauge bosons to 
Ghost particles
in chiral $B-L$ model which are related to 
the processes under consideration. The vector 
$B-L$ model can be derived by applying 
$s_{BL}' \rightarrow -s_{BL}$.
$A$ is noted for photon field 
and $p$ is four momentum of $u^+$. }
\end{table}
%%%%%%%%%%%%%%%%%%%%%%%%%%%%%%%%%%%%%%

The couplings of $hH^{\pm}H^{\mp}$
and $hG^{\pm}G^{\mp}$ 
are given by
\begin{eqnarray}
g_{hH^{\pm}H^{\mp}} 
&=&
\lambda_{\Phi}v_{\Phi}c_{13}c_{12}s_{\alpha}^2
-\lambda_{\varphi}v_{\varphi}c_{13}s_{12}c_{\alpha}^2
-\frac{\mu}{\sqrt{2}}s_{13}s_{\alpha}c_{\alpha}
-\frac{\mu}{\sqrt{2}}s_{13}s_{\alpha}c_{\alpha} 
\nonumber\\
&&
-\lambda_{\Phi\sigma}v_{\sigma}s_{\alpha}^2s_{13}
-\lambda_{\varphi\sigma}v_{\sigma}c_{\alpha}^2s_{13}
+\lambda_{\Phi\varphi_1} 
[-c_{13}s_{12}s_{\alpha}^2v_{\varphi}
+c_{13}c_{12}c_{\alpha}^2v_{\Phi}] 
\nonumber\\
&&
-\lambda_{\Phi\varphi_2}
(-v_{\Phi}c_{13}s_{12}
+
v_{\varphi}c_{13}c_{12})
s_{\alpha}c_{\alpha}, \\
%%%%%%%%%%%%%%%
g_{hG^{\pm}G^{\mp}} &=&
\lambda_{\Phi}v_{\Phi}c_{13}c_{12}c_{\alpha}^2 
-\lambda_{\varphi}v_{\varphi}c_{13}s_{12}s_{\alpha}^2
+\frac{\mu}{\sqrt{2}}s_{13}s_{\alpha}c_{\alpha}
+\frac{\mu}{\sqrt{2}}s_{13}s_{\alpha}c_{\alpha} 
\nonumber\\
&&
-\lambda_{\Phi\sigma}v_{\sigma}c_{\alpha}^2s_{13} 
-\lambda_{\varphi\sigma}v_{\sigma}s_{\alpha}^2s_{13} 
+ \lambda_{\Phi\varphi_1}\qty[-c_{13}s_{12}c_{\alpha}^2v_{\varphi} 
+ c_{13}c_{12}s_{\alpha}^2v_{\Phi}] \nonumber\\
&& 
+\lambda_{\Phi\varphi_2}\qty(-v_{\Phi}c_{13}s_{12}
+v_{\varphi}c_{13}c_{12})s_{\alpha}c_{\alpha}.
\end{eqnarray}

We stress that the corresponding
couplings in the vector
$B-L$ can be derived by taking the limits of
$c_{12}\rightarrow1, c_{23}\rightarrow 1$, 
$c_{13} \rightarrow c_{\theta}$, 
$v_{\Phi} \rightarrow v$ and
$s_{BL}' \rightarrow - s_{BL}$. 
In these limits, for example, 
the third column results in 
Table~\ref{couplingchiralBL1}
are back to the third column 
couplings
in Table~\ref{couplingBL1}.
%%%%%%%%%%%%%%%%%%%%%%%%%%%%%%%%%%%%%%%%%%%%%
\section{One-loop form factors for         %%
$h\rightarrow \ell \bar{\ell}\gamma$ 
and $e^-e^+    %%
\rightarrow h\gamma$ in the 
$U(1)_{B-L}$ models}%%
%%%%%%%%%%%%%%%%%%%%%%%%%%%%%%%%%%%%%%%%%%%%%
One-loop form factors for         
$h\rightarrow \ell \bar{\ell}\gamma$ and $e^-e^+   
\rightarrow h\gamma$ in $U(1)_{B-L}$ models
are presented in detail in this section.
First, we arrive the calculations for one-loop
form factors for $h\rightarrow 
\ell \bar{\ell}\gamma$. 
We then show that cross-sections for $e^-e^+   
\rightarrow h\gamma$ can be derived
by using one-loop form factors 
in the decay process 
$h\rightarrow \ell \bar{\ell}\gamma$.
%%%%%%%%%%%%%%%%%%%%%%%%%%%%%%%%%%%%%%%
\subsection{Form factors for         %%
the decay process                    %%
$h\rightarrow \ell \bar{\ell}\gamma$}%%
%%%%%%%%%%%%%%%%%%%%%%%%%%%%%%%%%%%%%%%
In this section, we are going to present
the calculations for one-loop form factors 
for the decay channels 
$h\rightarrow \ell \bar{\ell}\gamma$ 
in the $U(1)_{B-L}$ extension for the SM. 
In this computation, all new couplings 
appear in the $U(1)_{B-L}$ models 
are denoted as $g_{\text{Vertices}}$. 
For the cases of the couplings of the 
SM-like Higgs to vector boson pair 
and fermion-pair, we parameterize
as 
\begin{eqnarray}
\label{factorSM}
g_{hVV}
=\kappa_{hVV}\cdot 
g^{\text{SM}}_{hVV},  
\quad 
g_{hff}
=\kappa_{hff}\cdot 
g^{\text{SM}}_{hff}. 
\end{eqnarray}
All couplings from 
the $B-L$ models involving to
the processes 
are listed in all the above 
Tables in the section $2$. 
The factored couplings 
$\kappa_{hff}$ and 
$\kappa_{hVV}$ can be collected
easily from the above Tables. 

\begin{figure}[H]
\centering
\includegraphics[width=8cm, height=5cm]
{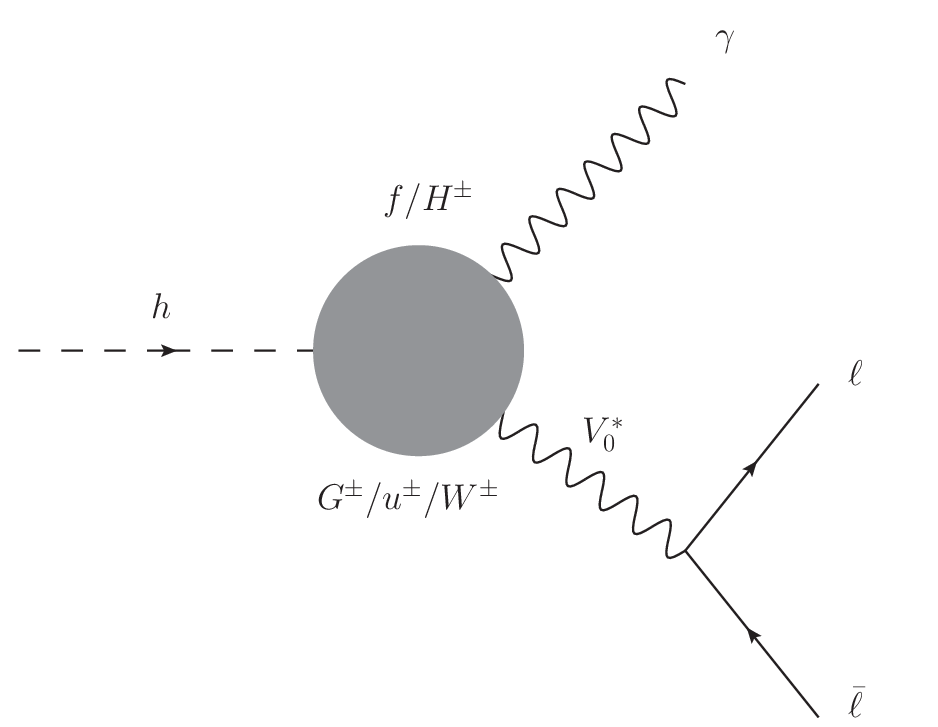}
\caption{\label{G1a} Group $1 (a)-$One-loop 
Feynman diagrams $V_{0}^{*}$-poles
contributing to the processes. Considering 
all fermions, 
$W$ bosons, charged Higgs, Goldstone bosons 
and Ghost particles
exchanging in the loop. $V_{0}^{*}$ can be 
$\gamma^*, Z^*$ and $Z'^{*}$ in this calculation.}
\end{figure}

\begin{figure}[H]
\centering
\includegraphics[width=16cm, height=10cm]
{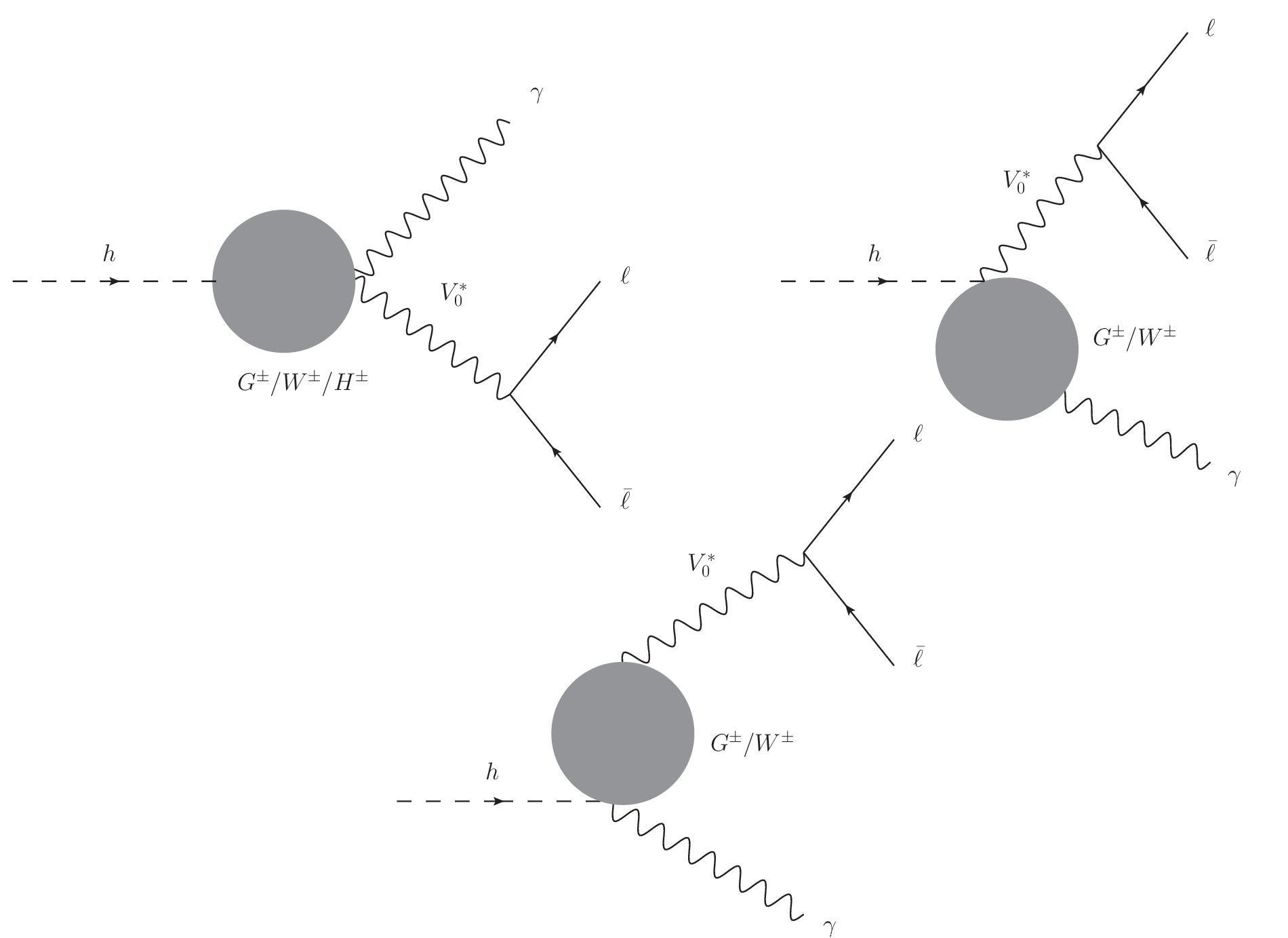}
\caption{\label{G1b} Group $1 (b)-$One-loop 
Feynman diagrams $V_{0}^{*}$-poles
contributing to the processes. Considering 
all fermions, 
$W$ bosons, charged Higgs, Goldstone 
bosons and Ghost particles
exchanging in the loop. $V_{0}^{*}$ can be 
$\gamma^*, Z^*$ and $Z'^{*}$ in this 
calculation.}
\end{figure}

\begin{figure}[H]
\centering
\includegraphics[width=16cm, height=6cm]
{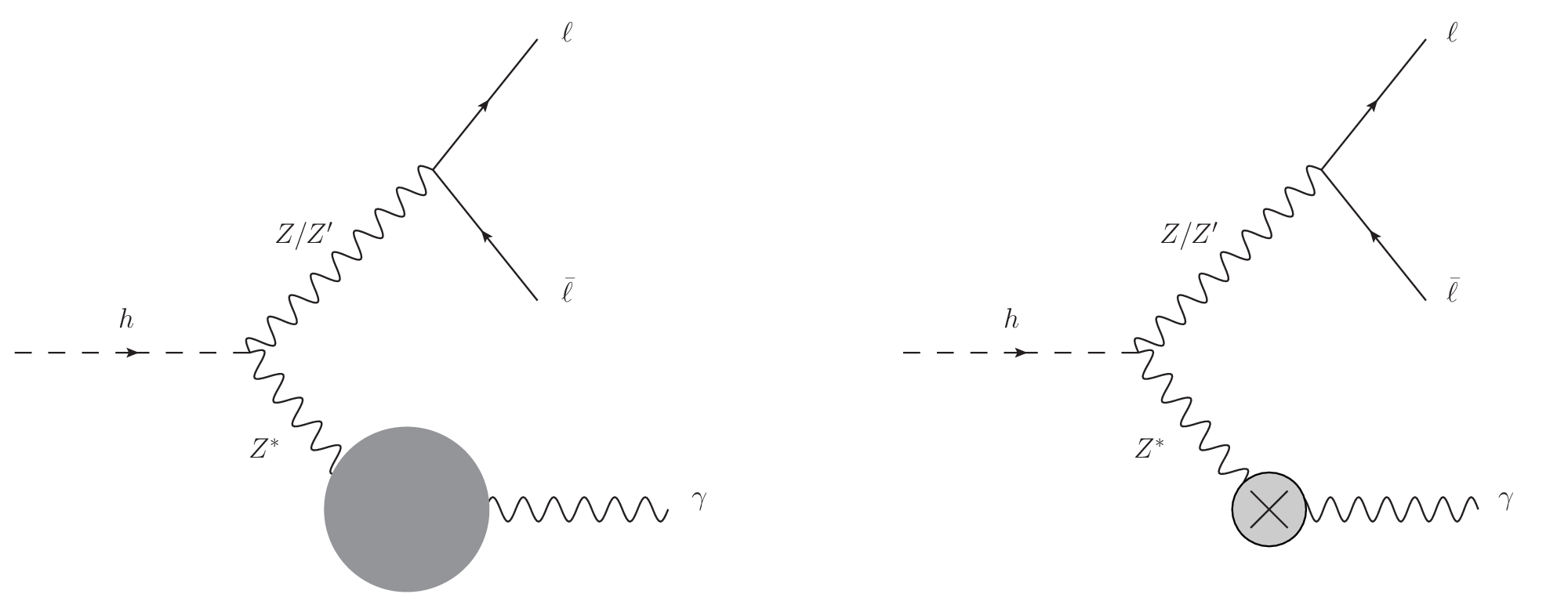}
\caption{\label{G1bb} Group $1 (c)-$One-loop 
Feynman diagrams $V_{0}^{*}$-poles
contributing to the processes. Considering 
all fermions, 
$W$ bosons, charged Higgs, Goldstone bosons 
and Ghost particles
exchanging in the loop. $V_{0}^{*}$ can be 
$\gamma^*, Z^*$ and $Z'^{*}$ in this calculation.}
\end{figure}

Within the HF gauge, all one-loop Feynman 
diagrams contributing to the decay process 
are plotted in the following paragraphs. 
These one-loop diagrams can 
be classified into several groups. 
In group $1~(a)$ (as shown in 
Fig.~\ref{G1a}), we include all one-loop 
Feynman diagrams having $V_{0}^{*}$-poles 
contributing to the decay process.
It means that we have $V_{0}^{*}
\rightarrow \ell \bar{\ell}$ 
involving in these diagrams. In the 
current work, $V_{0}^{*}$ 
can be $\gamma^*, Z^*$ and $Z'^{*}$.
%%%%%%%%%%%%%%%%%%% 
In the groups $1~(b), 1~(c)$ (as presented
in Figs.~\ref{G1b},~\ref{G1bb}), 
other one-loop Feynman diagrams with 
$V_{0}^{*}$-poles are also plotted. 
In the HF gauge, we also take into 
account all fermions, charged Higgs, 
vector bosons, Goldstone bosons and 
Ghost particles
propagating in the loop. 
As proved in Ref.~\cite{Phan:2021xwc}, 
we only collect one-loop form factors which are
proportional to $q_i^\mu q_3^{\nu}$ (for $i=1,2$)
appearing in Eq.~\ref{V0*pole}. All diagrams in 
Fig.~\ref{G1b} and Fig.~\ref{G1bb} can be
hence ignored in the present work.

\begin{figure}[H]
\centering
\includegraphics[width=16cm, height=10cm]
{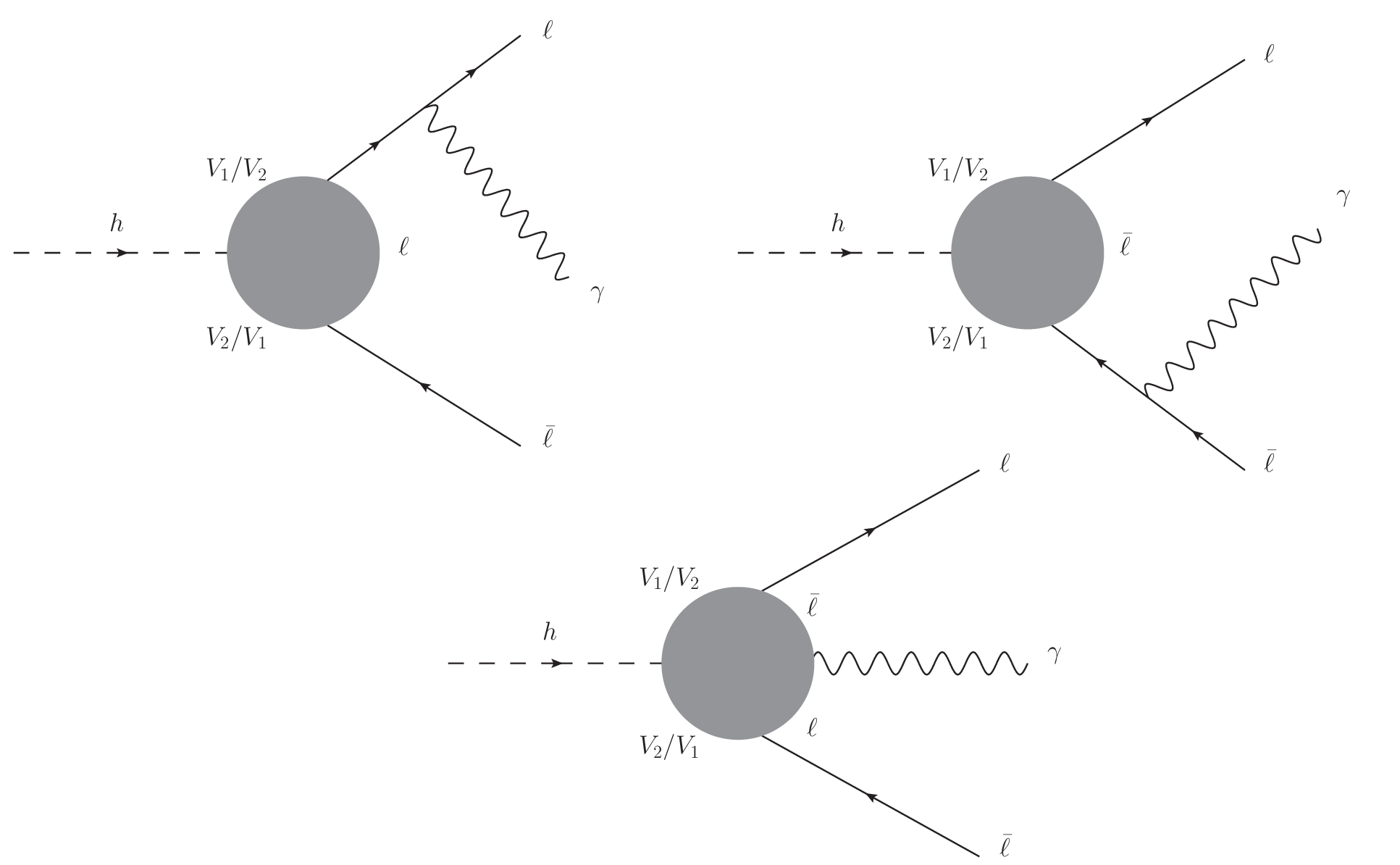}
\caption{\label{G2V1V2} One-loop Feynman 
non $V_{0}$-pole diagrams
with two neutral gauge bosons
$V_1, V_2$ exchanging in the loop. 
In the calculation, $V_1, V_2$ can 
be $Z^*$ and $Z'^{*}$ in the loop.}
\end{figure}

\begin{figure}[H]
\centering
\includegraphics[width=16cm, height=10cm]
{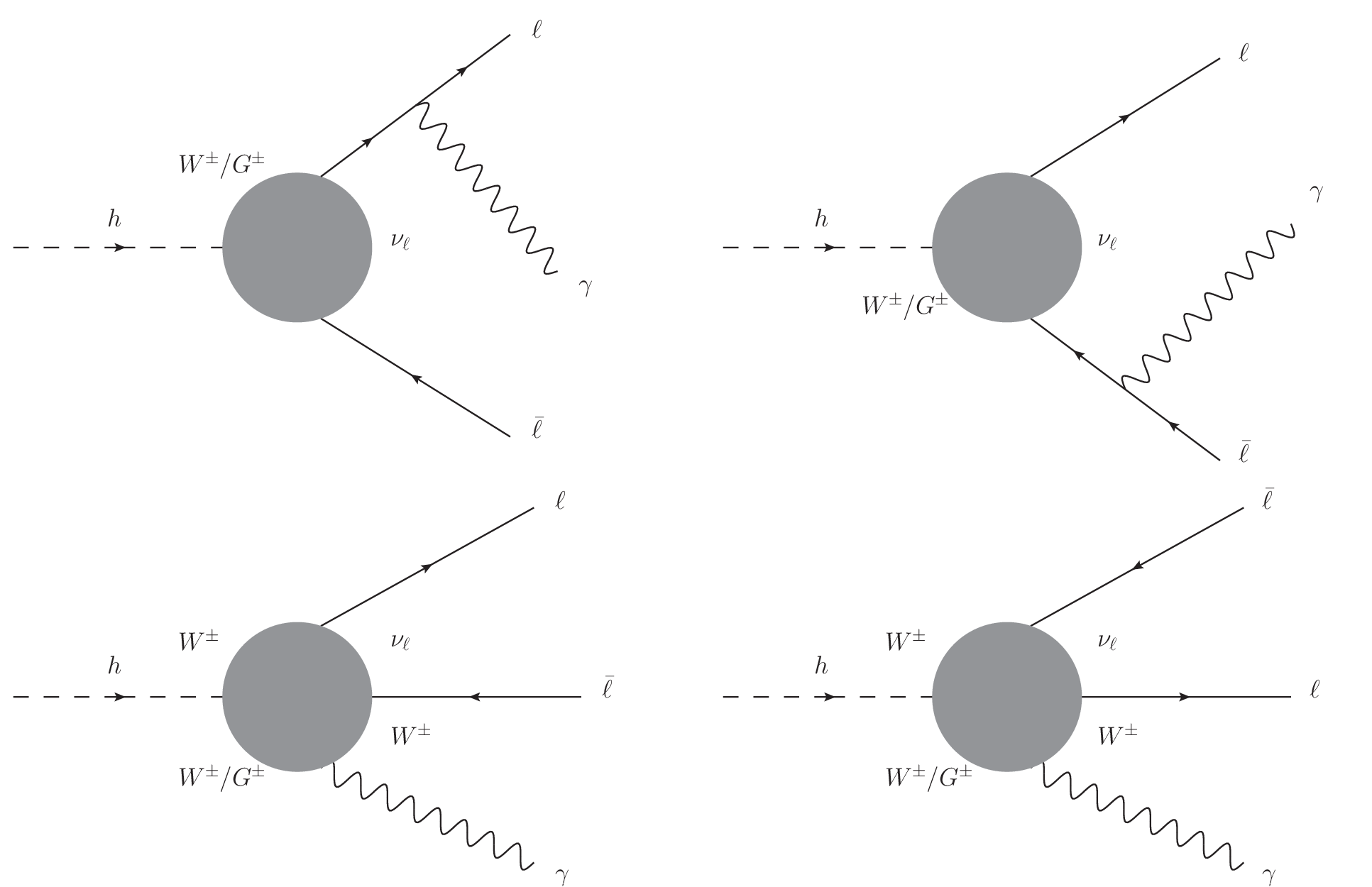}
\caption{\label{G2WG} One-loop Feynman 
non $V_{0}$-pole diagrams 
with $W$ boson exchanging in the loop.}
\end{figure}

\begin{figure}[H]
\centering
\includegraphics[width=16cm, height=10cm]
{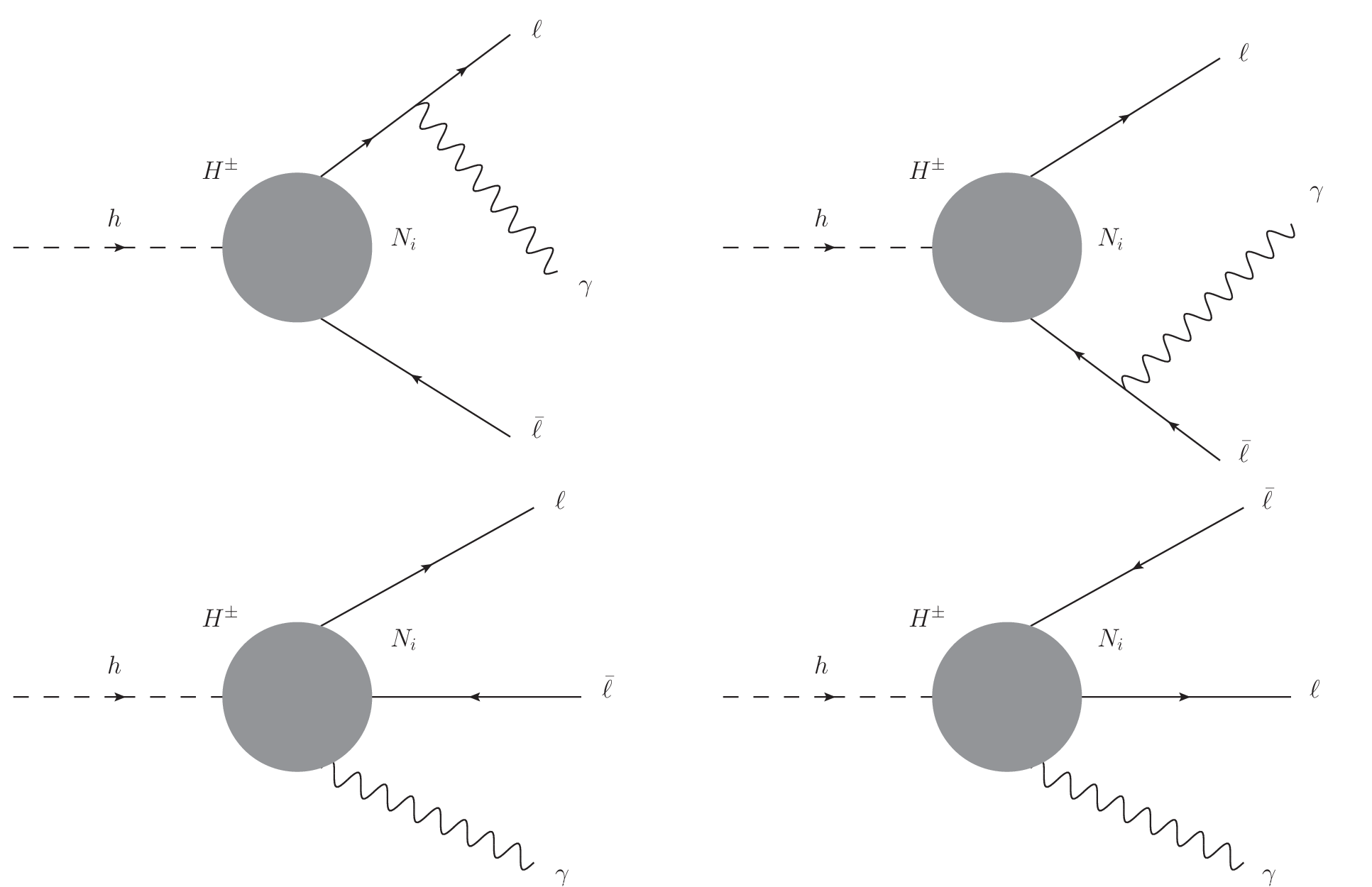}
\caption{\label{G5HPM} One-loop Feynman 
non $V_{0}$-pole diagrams
with charged Higgs exchanging in the loop.}
\end{figure}

There are also three kinds of one-loop Feynman 
diagrams without $V_{0}$-poles (called as 
non $V_{0}$-pole diagrams hereafter) 
contributing to the processes 
under consideration. The first classification 
is to non $V_{0}$-pole diagrams with two 
neutral gauge bosons $V_1, V_2$ exchanging 
in the loop. In the calculation, 
$V_1, V_2$ can be $Z^*$ and $Z'^{*}$ in the loop 
(as depicted in Fig.~\ref{G2V1V2}). 
The second group is to non $V_{0}$-pole
with $W$ bosons propagating in the loop
(as presented in Fig.~\ref{G2WG}).
The last categorization is to
non $V_{0}$-pole diagrams
with charged Higgs propagating in the loop
(as plotted in Fig.~\ref{G5HPM}).
%%%%%%%%%%%%%%%%%%%%%%%%%%%%%%%%%%%%%%%%%%
%%%%%%%%%%%%%%%%%%%
%%%%%%%%%%%%%%%%%%%%%%%%%%

%%%%%%%%%%%%%%%%%%%%%%%%%%%%%%%%
%%%%%%%%%%%%%%%%%%%

%%%%%%%%%%%%%%%%%%%%%%%%%%%%%%%
%%%%%%%%%%%%
%%%%%%%%%%%%%%%%%%%%%%%%%%
%%%%%%%%%%%%%%%%%%%%%%%%%%%%%%%%%%%%%%%

In general, one-loop amplitude for 
the decay process
of SM-like Higgs $h(p) \rightarrow \ell(q_1) \, 
\bar{\ell}(q_2)\, \gamma_\mu(q_3)$ can be expressed 
in terms of the Lorentz structure as follows:
%%%%%%%%%%%%%%%%%%%%
\begin{eqnarray}
\label{V0*pole}
\mathcal{A}^{h\rightarrow 
\ell \bar{\ell}\gamma}_{1-\text{loop}}
&=&
\sum \limits _{i = \{1, 2\}}
\sum \limits _{P = \{L, R\}}
\Big\{
\bar{v} (q_2)
\Big(
\gamma_\nu \mathcal{O}_P
\Big)
\Big[
F^P_{i, L} 
(q_i^\mu q_3^\nu)
-
F^P_{i, T} 
(
q_i \cdot q_3
)
g^{\mu \nu}
\Big]
u (q_1)
\Big\}
\varepsilon^*_\mu (q_3)
\end{eqnarray}
where $\mathcal{O}_{P=L,R} 
= \frac{1 \mp \gamma_5}{2}$. 
Due to on-shell photon at final state, 
the amplitude have to follow ward identiy, 
we can derive the relation
$F^P_{i, T} = F^P_{i, L}$. As a result, 
decay rates for the processes can be computed
via one of the above form factors. 
This is advantage to select
one-loop form factors 
$F^P_{i, L}$ for the later analysis since 
these form factors don't have the 
ultraviolet divergent ($UV$-divergent). 
Subsequently, all diagrams in 
Fig.~\ref{G1b} and Fig.~\ref{G1bb} can be
hence ignored in this paper, as our 
previous stateterment. We denote that 
$F^P_{i} =F^P_{i, L}$, 
hereafter, 
in the remaining of the paper. 
One-loop form factors $F_i^P$ 
are separated into two parts 
as follows:
\begin{eqnarray}
F_i^P
&=&
\sum \limits_{V_0 \;\equiv\; A , Z, Z'}
F_{i, \text{$V_0^*$-pole}}^P
+
F_{i, \text{Non-pole}}^P,
\quad \textrm{for}\quad i=1,2.
\end{eqnarray}
The related kinematic variables
for this process calculations are 
included as:
\begin{eqnarray}
 p^2=M_h^2, \quad q_1^2=q_2^2 =m_{\ell}^2, \quad 
 q_3^2 =0, \quad q_{12} = (q_1+q_2)^2, \quad
 q_{13} = (q_1+q_3)^2.
\end{eqnarray} 
The kinematic variables also follow the relation
\begin{eqnarray}
 q_{12} + q_{13}+ q_{23} = M_h^2 +2 m_{\ell}^2.
\end{eqnarray}

We first implement $B-L$ models into 
the package {\tt FeynArt}~\cite{Hahn:2000kx}. 
One-loop amplitude can be then 
generated automatically by this package. 
The amplitude is then decomposed 
one-loop tensor integrals. The tensor 
integrals are then reduced to scalar 
one-loop PV-functions by using the 
packages {\tt FormCalc}~\cite{Mertig:1990an}. 
One-loop form factors $F_{i, \text{$V_0^*$-pole}}^P$ 
and $F_{i, \text{Non-pole}}^P$ are collected and 
presented as functions
of the kinematic variables $M_h^2, q_{12}, q_{13}, 
m_{\ell}^2$ and the squared internal masses. 
Analytical expressions for the form factors
are written in terms of  scalar PV-functions
in the notations of 
{\tt LoopTools}~\cite{Hahn:1998yk}.
As a result, one-loop decay rates can be 
evaluated numerically by using 
{\tt LoopTools}. 
In the next paragraphs, 
we show analytical results for the 
form factors
in group by group of Feynman diagrams. 
\begin{itemize}
%%%%%%%%%%%%%%%%%%%%%%%%%%%%%%%%%%%%%%%
\item 
\underline{$V_0^*$-pole contributions:}
%%%%%%%%%%%%%%%%%%%%%%%%%%%%%%%%%%%%%%%

We first consider all $V_0^*$-pole 
diagrams as shown in Fig.~\ref{G1a}
including decay of 
$V_0^* \rightarrow \ell \bar{\ell}$
in these diagrams. Form factors 
$F_{i, \text{$V_0^*$-pole}}^P$ 
can be expressed in the form of 
\begin{eqnarray}
F_{i, \text{$V_0^*$-pole}}^P
&=&
- \frac{\alpha}{
(8\pi) M_W s_W }
\frac{1}
{(q_{12} - M_{V_0}^2) 
+ i M_{V_0} \Gamma_{V_0}}
\;
(g^{P}_{V_0^* \ell \bar{\ell}})
\times
\\
&&
\times
\Bigg\{
\kappa_{hW^\pm W^\mp}
\times
F_{i, \text{$V_0^*$-pole}} ^{P,W^\pm}
-
\sum \limits_{f}
\kappa_{h\bar{f}{f}}
\left(
N^C_f Q_f \, m_f^2
\right)
F_{i, \text{$V_0^*$-pole}}^{P,f}
+
F_{i, \text{$V_0^*$-pole}}^{P,H^\pm}
\Bigg\}
\nonumber
\end{eqnarray}
for $P = \{L, R\}$.

From fermions $f$-loop contributions, 
the form factor is given as follows:
%%%%%%%%%%%%%%%%%%%%%%%%%%%%%%%%%%
\begin{eqnarray}
\label{V0f}
F_{i, \text{$V_0^*$-pole}}^{P,f}
&=&
4
\Big(\sum\limits_{j=L,R}
g^{j}_{V_0^* f \bar{f}} 
% +
% g^{ R}_{V_0^* f \bar{f}}
\Big)
\Big[
C_{0} 
+
4
\big(
C_{2} 
+
C_{12} 
+
C_{22}
\big)
\Big] (0,q_{12},M_h^2,m_f^2,m_f^2,m_f^2).
% \nonumber\\
\end{eqnarray}
Where $g^{L, R}_{V_0^* f \bar{f}}$ are
the couplings of $V_0^*$ with 
fermion-pair in the loop. 
%%%%%%%%%%%%%%%%%%%%%%%%%%%%%%%%%%
We next concern the $W$-boson loop 
contributions. Different from our
previous work
in~\cite{Hue:2023tdz} 
which we have 
included new parameters to unify 
one-loop form factors for both 
photon and $Z$-poles in one 
analytic expression. 
In this work, in order to avoid
definning new parameters, we present
one-loop form factors for $\gamma^*$-pole
and $Z^*$- or $Z'^*$-pole separately. 
In detail, 
the form factors 
are shown for $V_0^* \equiv \gamma^*$-pole
as follows:
\begin{eqnarray}
\label{V0A}
\dfrac{F_{i, \text{$A^*$-pole}}^{P,W^\pm}}
{g_{A^*WW} }
&=& 8 
\Big[ 4 M_W^2 C_0 
+
(M_h^2+6 M_W^2)
(C_{2} 
+
C_{12} 
+
C_{22} )
\Big](0,q_{12},M_h^2,M_W^2,M_W^2,M_W^2).
\nonumber\\
\end{eqnarray}
Here $A^*$ is noted for $\gamma^*$-pole. 
%%%%%%%%%%%%%%%%%%%%%%
One form factors for 
$V_0^* \equiv Z^*$-pole
and $V_0^* \equiv Z'^{*}$-pole
are taken the form of
\begin{eqnarray}
\label{V0ZZp}
\dfrac{
F_{i, \text{$V_0^*$-pole}} ^{P,~W^\pm}
}
{ g_{V_0^*WW} }
&=& 
4
\Big\{ 
2 M_W^2(3-t_W^2) C_0 
+
[
(M_h^2+ 10 M_W^2)
-
t_W^2 (M_h^2+ 2 M_W^2)
] 
\times
\\
&& 
\hspace{3.7cm}
\times 
(C_{2} 
+
C_{12} 
+
C_{22} )
\Big\}
(0,q_{12},M_h^2,M_W^2,M_W^2,M_W^2).
\nonumber
\end{eqnarray}
Where we have used $t_W =s_W/c_W$.
In this formulas, $g_{V_0^*WW}$
shows for the couplings of $Z$ or $Z'$
with $W$-pair which are given in 
Tables~\ref{couplingBL3}
and \ref{couplingchiralBL3} accordingly.

In addition, one-loop form factor
with the charged Higgs bosons $H^\pm$ 
in the loop reads
\begin{eqnarray}
\label{V0S}
F_{i, \text{$V_0^*$-pole}}^{P,~H^\pm}
&=&
\frac{4 M_W s_W}{\pi \alpha}
(
g_{h \, H^\pm H^\mp} 
)
\,
(
g_{A \, H^\pm H^\mp} 
)
\,
(
g_{V_0^* \, H^\pm H^\mp}
)
\times
\\
&&\hspace{0cm} 
\times
\Big[
C_{12} 
+
C_{2} 
+
C_{22}
\Big] 
(0,q_{12},M_h^2,M_{H^\pm}^2,
M_{H^\pm}^2,M_{H^\pm}^2).
\nonumber
\end{eqnarray}
Here, it is noted 
that $g_{V_0^* \, H^\pm H^\mp}$ stands 
for the couplings of $V_0^* \, H^\pm H^\mp$
with $V_0^* \equiv \gamma^*, Z^*$ and $Z'^*$.

We emphasize the following 
inportant points. First, for the case 
of $V_0^*$-pole being $\gamma^*$-
and $Z^*$-poles, reducing PV functions 
in Eqs.~\ref{V0f}, \ref{V0A}, \ref{V0ZZp}, 
\ref{V0S} to scalars one-loop functions, 
we then reprodure the 
corresponding results 
in~\cite{Hue:2023tdz}.
Surely,  
we have to replace
all the couplings 
factored out in all the above formulas
from this model
to the HESM respectively.
Secondly, the contributions from 
$Z'^*$-pole are new results from this work. 
Thirdly, we derive alternative presentations
for one-loop form factors for 
$h\rightarrow \ell \bar{\ell}\gamma$ in comparison
with our previous work in~\cite{VanOn:2021myp}
which the results have shown in the unitary
gauge. 
Last but not least, taking the limits 
of $V^*_0$ to on-shell $\gamma$ or 
on-shell $Z$, we have one-loop form factors
for loop-induced $h\rightarrow \gamma\gamma$
and $h\rightarrow Z\gamma$ in the 
$U(1)_{B-L}$ extension of the SM.
%%%%%%%%%%%%%%%%%%%%%%%%%%%%%%%%%%%%%%%%%%%
\item \underline{Non $V_0$-pole          %%
contributions:}                          %%
%%%%%%%%%%%%%%%%%%%%%%%%%%%%%%%%%%%%%%%%%%%

We turn to one-loop 
diagrams without $V_0$-pole contributions. 
Form factors 
$F_{i, \text{Non-pole}}^P$ 
can be separated into the form of
\begin{eqnarray}
F_{i, \text{Non-pole}}^P
&=&
\sum \limits_{V_1 , V_2 
\;\equiv\; \{ Z, Z' \} }
F_{i, \text{Non-pole} \, V_1 , V_2}^P
+
F_{i, \text{Non-pole} \, W^\pm}^P
+
F_{i, \text{Non-pole} \, H^\pm}^P.
\end{eqnarray}

We arrive at one-loop diagrams with two
neutral gauge bosons $V_1$ and $V_2$ 
contributions, the form factors 
$F_{i, \text{Non-pole} \, V_1 , V_2}^P$ 
are also expressed in terms of 
PV-functions as follows:
\begin{eqnarray}
\label{boxZZp}
F_{1, \text{Non-pole} 
\; V_1 , V_2}^L
&=&
\frac{1}{n!}
\times
\frac{
(
g_{h V_1 V_2}
)
(
g^{ L}_{A \ell \bar{\ell}}
)
}{4\pi^2}
\,
\Big(
\prod\limits_{i=1}^2
g^{L}_{V_i \ell \bar{\ell}}
\Big)
\times \\
&&\times
\Big\{
[
D_{2}
+
D_{12}
+
D_{23} 
] 
(0,q_{13},M_h^2,q_{23},
0,0,m^2_\ell,
m^2_\ell,M_{V_1}^2,M_{V_2}^2)
\nonumber \\
&&\hspace{0.25cm}
+
[
D_{3}
+
D_{13} 
+
D_{23} 
] 
(0,0,M_h^2,0,q_{23},q_{13},
m^2_\ell,m^2_\ell,M_{V_1}^2,M_{V_2}^2)
\Big\},
\n \\
%%%%%%%%%%%%%%%%%%%%%%%%%%%%%%%%%%%%%%
F_{1, \text{Non-pole}\; V_1 , V_2}^R
&=&
F_{1, \text{Non-pole}\; V_1 , V_2}^L
\,
\{ 
(
g^{L}_{V_1 \ell \bar{\ell}}
,
g^{ L}_{V_2 \ell \bar{\ell}})
\rightarrow 
(
g^{ R}_{V_1 \ell \bar{\ell}} 
,
g^{ R}_{V_2 \ell \bar{\ell}})
\},
\nonumber \\
%%%%%%%%%%%%%%%%%%%%%%%%%%%%%%%%%%%%%%
F_{2, \text{Non-pole} \; V_1 , V_2}^{L/R}
&=&
F_{1, \text{Non-pole} \; V_1 , V_2}^{L/R}
\,
\{
q_{13} \leftrightarrow q_{23} 
\}.
\nonumber
%%%%%%%%%%%%%%%%%%%%%%%%%%%%%%%%%%%%%%
\end{eqnarray}
where factor $(n!)^{-1}$ denoting for two 
identical internal particles in loop 
$(\, n = 2 \,)$ for the case of 
$V_1 \equiv V_2 \equiv Z$ or $Z'$. 
Next, one-loop contributions of charged 
gauge bosons $W^\pm$ internal lines are 
concerned. The form factors 
$F_{i, \text{Non-pole} \, W^\pm}^P$ 
read as:
\begin{eqnarray}
\label{boxWW}
F_{1, \text{Non-pole} 
\, W^\pm}^L
&=&
-
\frac{\alpha M_W}{\pi s_W}
\,
\kappa_{hW^\pm W^\mp}
\,
(
g_{W^\pm \ell 
\bar{\nu}_{\ell}}
)^2
\times  \\
&&\times
\Big\{
[
D_{1} 
+
D_{13} 
] (0,q_{12},0,q_{23},
0,M_h^2,m_{\ell'}^2,M_W^2,M_W^2,M_W^2)
\nonumber\\
&&\hspace{0.25cm}
+
[
D_{2} 
-
D_{23} 
-
D_{33} 
] (0,q_{12},0,q_{13},
0,M_h^2,m_{\ell'}^2,M_W^2,M_W^2,M_W^2)
\Big\},
\nonumber \\
%%%%%%%%%%%%%%%%%%%%%%%%%%%%
F_{2, \text{Non-pole} \, W^\pm}^L
&=&
F_{1, \text{Non-pole} \, W^\pm}^L
\,
\big\{ q_{13} \leftrightarrow q_{23} \big\}, 
\nonumber \\
%%%%%%%%%%%%%%%%%%%%%%%%%%%%
F_{i, \text{Non-pole} \, W^\pm}^R
&=&
0. 
\nonumber
%%%%%%%%%%%%%%%%%%%%%%%%%%%%
\end{eqnarray}
We note that $m_{\ell'}^2= m_{\nu_{\ell}}^2=0$
in the decay channels $h\rightarrow
\ell \bar{\ell}\gamma$
and $m_{\ell'}^2= m_{\ell}^2\neq 0$
in the decay channels $h\rightarrow \nu_{\ell} 
\bar{\nu}_{\ell}\gamma$ in this case.
Finally, we consider one-loop 
diagrams with non-pole $V_0$ which
the charged 
Higgs bosons $H^\pm$, right
handed neutrino $N_i$ 
exchanging in the loop.
\begin{eqnarray}
\label{boxSN}
F_{1, \text{Non-pole} 
\, H^\pm}^L
&=&
- \frac{
M_{N_i}^2 
}{8 \pi^2}
\,
( 
g_{h \, H^\pm H^\mp}
)
\,
(
g_{A H^\pm H^\mp}
)
\,
(
g_{H^\pm \ell 
\bar{N}_i }
)^2
\times
\\
&&
\hspace{0cm}
\times 
\Big\{
[
D_{2} 
+
D_{12} 
+
D_{22} 
] 
(q_{12},0,q_{13},0,M_h^2,
0,M_{H^\pm}^2,M_{H^\pm}^2,
M_{H^\pm}^2, M_{N_i}^2)
\nonumber\\
&&
\hspace{0.25cm}
+
[
D_{2} 
+
D_{12} 
+
D_{22} 
+
D_{23} 
] 
(q_{12},0,q_{23},0,M_h^2,
0,M_{H^\pm}^2,M_{H^\pm}^2,
M_{H^\pm}^2, M_{N_i}^2)
\Big\},
\nonumber 
\\
%%%%%%%%%%%%%%%%%%%%%%%%%%%%
F_{2, \text{Non-pole} \, H^\pm} ^L
&=&
F_{1, \text{Non-pole} \, H^\pm} ^L
\,
\{ 
q_{13} \leftrightarrow q_{23} 
\}, 
\nonumber \\
%%%%%%%%%%%%%%%%%%%%%%%%%%%%
F_{i, \text{Non-pole} \, H^\pm} ^R
&=&
0.
\nonumber
%%%%%%%%%%%%%%%%%%%%%%%%%%%%
\end{eqnarray}
\end{itemize}

It is important to
note the following arguments. 
When we take only 
$Z, W$ in the non-$V_0$ pole 
diagrams and replace
overall couplings respectively to the
HESM, we then reprodure the corresponding
results in \cite{Hue:2023tdz}. Furthermore,
including all contributions 
from  $Z, Z', W$ and charged scalar Higgs
in the non $V_0$-pole diagrams, we derive 
alternative results from our previous 
reference~\cite{VanOn:2021myp}.

Having all one-loop form factors, 
we are going
to check the calculations by 
verifying the $UV$ finiteness
and $IR$ finiteness of the form factors.
We verify that all form factors presented in 
the calculations are $UV$-finite and $IR$-finite.
Since there is no 
virtual photon exchanging
in the loop, we don't have the $IR$-divergent 
in the processes. We have confirmed 
the $UV$ finiteness for the form factors 
analytically
in \cite{Phan:2021xwc,Hue:2023tdz}.

After confirming the correctness 
of the results, differential decay 
rate for 
$h \rightarrow \ell\bar{\ell}
\gamma$ with including one-loop 
contributions can be computed as 
follows
\begin{eqnarray}
\dfrac{d \Gamma_{h\rightarrow 
\ell \bar{\ell}\gamma}}{d q_{12} 
\, d q_{13}}
&=&
\dfrac{q_{12}}{512 \pi^3 M_h^2}
\Big[
q_{13}^2 
\Big(
\big| F^L_1 \big|^2
+
\big| F^R_1 \big|^2
\Big)
+
q_{23}^2 
\Big(
\big| F^L_2 \big|^2
+
\big| F^R_2 \big|^2
\Big)
\Big].
\end{eqnarray}
Integration over 
$(m_{\ell \bar{\ell}}^{\text{cut}})^2 
\leq q_{12} \leq M_h^2$ and $0 \leq q_{13} 
\leq M_h^2 - q_{12}$, one then obtains  
the total decay rates. At the LHC, signal 
strengths for $h\rightarrow
\ell \bar{\ell}\gamma$ can be derived as follows:
\begin{eqnarray}
 \mu_{B-L} = \dfrac{\sigma^{B-L}(pp\rightarrow
 h) }{\sigma^{SM}(pp\rightarrow
 h) }
 \dfrac{\Gamma_{h\rightarrow 
\ell \bar{\ell}\gamma}^{B-L} }
{\Gamma_{h\rightarrow 
\ell \bar{\ell}\gamma}^{SM} }
\dfrac{\Gamma_{h\rightarrow 
\textrm{all} }^{SM} }
{\Gamma_{h\rightarrow 
\textrm{all}}^{B-L} }.
\end{eqnarray}
Where $\Gamma_{h\rightarrow 
\textrm{all}}^{B-L}  = \Gamma_{h\rightarrow 
\textrm{all} }^{SM} + \Gamma^{B-L}[h\rightarrow 
 Z'Z']
 + \Gamma^{B-L}[h\rightarrow 
 \sum\limits_{i=1}^3 N_i \bar{N}_i]$. 
 In the scenario of heavy neutral 
 gauge boson $Z'$ and 
 heavy neutrinos $N_i$ ($M_{N_i}\sim 1$ TeV), 
 we then confirm that 
 $\Gamma_{h\rightarrow 
\textrm{all}}^{B-L} \simeq 
\Gamma_{h\rightarrow 
\textrm{all} }^{SM}$. 
Furthermore, 
we verify that 
\begin{eqnarray}
 \dfrac{\sigma^{B-L}(pp\rightarrow
 h) }{\sigma^{SM}(pp\rightarrow
 h) }  \simeq  \kappa_{hff}^2. 
\end{eqnarray}
Where $\kappa_{hff}$ is 
defined as in Eq.~(\ref{factorSM}).
Therefore, the signal strengths
can be evaluated as 
\begin{eqnarray}
\label{Sig}
 \mu_{B-L} \simeq
 \kappa_{hff}^2 \times 
 \dfrac{\Gamma_{h\rightarrow 
\ell \bar{\ell}\gamma}^{B-L} }
{\Gamma_{h\rightarrow 
\ell \bar{\ell}\gamma}^{SM} }.
\end{eqnarray}
In the scope of this paper, we are 
interested in this limit and take 
the signal strengths in 
Eq.~(\ref{Sig}) for phenomenological 
analysis at the LHC.
%%%%%%%%%%%%%%%%%%%%%%%%%%%%%%%%%%%%%%%
\subsection{Form factors for         %%
$e^-e^+                              %%
\rightarrow h\gamma$ }               %%
%%%%%%%%%%%%%%%%%%%%%%%%%%%%%%%%%%%%%%%
We turn our attention to evaluate 
cross-sections for 
$e^-e^+ \rightarrow h\gamma$ at future lepton
colliders. Applying the same method for 
$h\rightarrow \ell \bar{\ell}\gamma$, 
one-loop amplitude
for the production process can be written  
in terms of the following Lorentz structure
:
\begin{eqnarray}
\mathcal{A}^{e^- e^+ \rightarrow
h \gamma}_{1-\text{loop}}
&=&
\sum \limits _{i = \{1, 2\}}
\sum \limits _{P = \{L, R\}}
\Big\{
\bar{v} (q_2)
\Big(
\gamma_\nu \mathcal{O}_P
\Big)
K_i^P 
\Big[
q_i^\mu q_3^\nu
-
\Big(
q_i \cdot q_3
\Big)
g^{\mu \nu}
\Big]
u (q_1)
\Big\}
\varepsilon^*_\mu (q_3).
\end{eqnarray}
Where one-loop form factors $K_i^P$ 
are collected as coefficient of 
$q_i^\mu q_3^\nu$. By comparing 
with one-loop form factors $F_i^P$, 
we find that 
the form factors $K_i^P$ can be derived
from $F_i^P$ by applying 
the following relations: 
\begin{eqnarray}
K_{i, \text{$V_0^*$-pole}}^P
&=&
F_{i, \text{$V_0^*$-pole}}^P;
\\
%%%%%%%%%%%%%%%%%%%%%%%%%%%%%%%%%%%%%%%%%
K_{i, \text{Non-pole}}^P
&=&
F_{i, \text{Non-pole}}^P
\;
|_{q_{13} \leftrightarrow q_{23}}
\quad \textrm{for}\; P=L,R.
\end{eqnarray}
The above relations can be explained by 
applying cross-symmetry for converting
the decay process
$h\rightarrow e^- e^+ \gamma$ to the 
production scattering $e^- e^+ \rightarrow  
h \gamma$, see~\cite{Kanemura:2018esc} 
for more detail. 
Differential cross section for the 
production process including initial beam 
polarizations can be expressed in terms of 
one-loop form factors as follows:
%%%%%%%%%%%%%%%%%%
\begin{eqnarray}
\dfrac{d \sigma_{\text{pol}}}
{d\cos\theta_{e^-\gamma} }
&=&
\frac{q_{12} - M_h^2}{256 \pi \, q_{12}}
\Big\{
\Big(
1 + \lambda_+
\Big)
\Big(
1 - \lambda_-
\Big)
\Big[
q_{13}^2 
\big| F^L_1 \big|^2
+
q_{23}^2 
\big| F^L_2 \big|^2
\Big]
\n \\
&&\hspace{2cm}
+
\Big(
1 - \lambda_+
\Big)
\Big(
1 + \lambda_-
\Big)
\Big[
q_{13}^2 
\big| F^R_1 \big|^2
+
q_{23}^2 
\big| F^R_2 \big|^2
\Big]
\Big\}
\end{eqnarray}
where $\lambda_-$ and $\lambda_+$ are 
the polarized degrees of the 
initial $e^-$ and $e^+$ beams, respectively.
In the above expression, $q_{23}, q_{13}$
are given by
\begin{eqnarray}
q_{23}, q_{13}
&=&
- \dfrac{q_{12} - M_h^2}{2}
\Big(
1 \mp \cos\theta_{e^-\gamma}
\Big).
\end{eqnarray}
Here, $\theta_{e^-\gamma}$ is the angle 
between initial electron and external photon.

The signal strengths
for the Higgs production associated
with a photon for the models 
at future lepton colliders can be estimated
as follows
\begin{eqnarray}
 \mu_{B-L}(\sqrt{s}, \lambda_+, \lambda_-) 
 = \dfrac{ \sigma_{B-L} }
 {\sigma_{SM}} (\sqrt{s},\lambda_+, \lambda_-).
\end{eqnarray}
Where $\sqrt{s}$ is center-of-mass energy. 
In the next section, we are going to present
phenomenological results for the calculations. 
%%%%%%%%%%%%%%%%%%%%%%%%%%%%%%%%%%%%%
\section{Phenomenological results}%%%
%%%%%%%%%%%%%%%%%%%%%%%%%%%%%%%%%%%%%
In the phenomenological results, we use 
the following input parameters:
$M_Z = 91.1876$ GeV, 
$\Gamma_Z  = 2.4952$ GeV, 
$M_W = 80.379$ GeV, $\Gamma_W  = 2.085$ GeV, 
$M_h =125$ GeV, $\Gamma_H = 4.07\cdot 10^{-3}$ GeV. 
The lepton masses are selected
as $m_e =0.00052$ GeV,
$m_{\mu}=0.10566$ GeV and $m_{\tau} = 1.77686$ GeV.
For quark masses, one takes $m_u= 0.00216$ GeV
$m_d= 0.0048$ GeV, $m_c=1.27$ GeV, $m_s = 0.93$ GeV, 
$m_t= 173.0$ GeV, and $m_b= 4.18$ GeV. 
We work in the so-called $G_{\mu}$-scheme
in which the Fermi constant 
is taken $G_{\mu}=1.16638\cdot 10^{-5}$ 
GeV$^{-2}$ and the 
electroweak coupling can be calculated 
appropriately
as follows:
\begin{eqnarray}
\alpha = 
\sqrt{2}/\pi G_{\mu} M_W^2(1-M_W^2/M_Z^2)
=1/132.184.
\end{eqnarray}
As we show in later, the mass of $Z'$ boson is 
considered as a parameter. The total decay
width of $Z'$  can be calculated as 
in~\cite{Deppisch:2019kvs}. As indicated in 
the previous sections, the models 
$B-L$ contain an additional neutral gauge boson
$Z'$ with associated the coupling $g_X = g_{B-L}$
in the gauge sector. The limit on 
$M_{Z'}$ and $g_{B-L}$ reported by 
Tevatron~\cite{Carena:2004xs} is 
\begin{eqnarray}
 \frac{M_{Z'} }{2 g_{B-L}} \gtrsim 3 \;
 \text{TeV}.
\end{eqnarray}
Moreover, from ElectroWeak Precision Test 
(EWPT)~\cite{Cacciapaglia:2006pk,
Electroweak:2003ram}, 
one has the further constraint 
on $M_{Z'}-g_{B-L}$
parameter space as follows:
\begin{eqnarray}
 \frac{M_{Z'} }{2 g_{B-L}} \gtrsim 3.5 \;
 \text{TeV}.
\end{eqnarray}
In the LHC era, we can probe $Z'$ via 
the processes $pp\rightarrow Z'
\rightarrow l\bar{l}$. Recently, ATLAS
~\cite{ATLAS:2019erb}
and CMS~\cite{CMS:2019kaf} have 
reported a lower bound of $M_{Z'}$
as $M_{Z'} \gtrsim 4.8$ GeV.  
In scalar 
sector, $B-L$ models include
neutral Higgs $H$, mixing angle $\sin\theta$
of SM-like Higgs with the scalar singlet
$\chi$. As summarized results from LHC run II 
data~\cite{Bechtle:2015pma, Robens:2016xkb,
Chalons:2016jeu,Ilnicka:2018def}, 
we have the limits $|\sin\theta| <0.37$
for $150$ GeV $\leq M_{H}\leq 1000$ GeV. 
We also have two heavy neutrinos $N_i$ in
the neutrino sector. Direct searches for 
heavy neutrinos can be performed at lepton
colliders via channels 
$e^-e^+ \rightarrow N_i \nu$ where 
$N_i \rightarrow 
l W, \nu Z, \nu h$, etc. At LEP and 
LEP II\cite{L3:1992xaz,DELPHI:1996qcc},
the limits
for $M_{N_i}$ are $10$ GeV $\leq M_{N_i}\leq 80$ GeV.
At the LHC~\cite{CMS:2012wqj,ATLAS:2012yoa,CMS:2015qur,
CMS:2018iaf, CMS:2018jxx,CMS:2017ybg}, heavy neutrinos 
mixing to
light neutrinos can be probed via processes like
$pp \rightarrow W^{\pm, *} \rightarrow Nl^{\pm}
\rightarrow l^{\pm}l^{\pm}jj$. The data give 
the limits
on $M_{N_i}$ as 
$1$ GeV $\leq M_{N_i}\leq 1000$ GeV.
%%%%%%%%%%%%%%%%%%%%%%%%%%%%%%%%%%%%%%%%%%%%%%%%%%
\subsection{Numerical checks           %%%%%%%%%%%
--confirming gauge invariance }        %%%%%%%%%%%               
%%%%%%%%%%%%%%%%%%%%%%%%%%%%%%%%%%%%%%%%%%%%%%%%%%
We first confirm the results of decay rates
of $h\rightarrow \ell \bar{\ell}\gamma$ in vector $B-L$
in our previous work~\cite{VanOn:2021myp}
in which the computations have performed in 
the unitary gauge. We use the same input 
parameters as in~\cite{VanOn:2021myp}. 
The cross-check results are 
shown in Table~\ref{decayUniHF}. 
In the first column, we apply the cuts
$m_{e^-e^+}^{\text{cut}} = 
k^{\text{cut}} M_h$ and $E_\gamma^{\text{cut}}$
as same in \cite{VanOn:2021myp}. 
As a result, the integration region is now
as $(m_{e^-e^+}^{\text{cut}})^2 \leq q_{12}
\leq M_h^2(1-2 E_\gamma^{\text{cut}} )$
and $(m_{e^-e^+}^{\text{cut}})^2\leq q_{13}
\leq M_h^2 -q_{12}$.
The second column shows the results from this 
work in HF gauge. While the last column
presents for the results in ~\cite{VanOn:2021myp}.
We find that two results are good agreements. 
The checks confirm gauge
invariance--the results are unchanged in 
different gauges.
%%%%%%%%%%%%%%%%%%%%%%%%%%%%%%%%
\begin{table}[H]
\begin{center}
\begin{tabular}{|l|l|l|}  
\hline \hline 
$\big(k^{\text{cut}} , 
E_\gamma^{\text{cut}} \big)$
& This work
& Ref.~\cite{VanOn:2021myp}
\\ \hline \hline
$(0.05, 1)$
& 
$0.265513$
&
$0.265498$
\\ \hline
$(0.1, 5)$
& 
$0.239109$
&
$0.239101$
\\ \hline
$(0.2, 10)$
& 
$0.198871$
&
$0.198859$
\\ \hline
$(0.3, 15)$
& 
$0.160219$
&
$0.160212$
\\ \hline
$(0.4, 20)$
& 
$0.122837$
&
$0.122831$
\\ \hline   
\hline      
\end{tabular}
\caption{
\label{decayUniHF}
The Table shows decay rate
(presents in KeV)
$\Gamma_{h\rightarrow \ell \bar{\ell}\gamma}$ 
for $U(1)_{B-L}$ non-mixing in 
unitary and HF gauges.
In this, we denote that 
$m_{e^-e^+}^{\text{cut}} = 
k^{\text{cut}} M_h$ and we 
consider in the case of 
$c_\theta = 0.98, g' = 10^{-3}$ 
and $M_{Z'} = 10$ GeV. 
Energy cut for external 
photon is presented in GeV.}
\end{center}
\end{table}

We then cross-check cross sections 
for $e^-e^+ \rightarrow h \gamma$
in this work 
with previous computations within 
the SM. The results for this test 
are presented in Table~\ref{crosssectionUNi}. 
In the first column, we select polarized
degrees for initial beams of electron 
and positron. The second column shows for 
the results from this work. The last column
results are from Ref.~\cite{Aoki:2019sht}.
Both results are good agreements up to last 
digits. 
%%%%%%%%%%%%%%%%%%%%%%%%%%%%%%%%%%%%%%%%%%%%%%%%%%
\begin{table}[H]
\begin{center}
\begin{tabular}{|l|l|l|}  
\hline \hline 
$\big( \lambda_- , \lambda_+ \big)$
&\text{This work}
&\text{Ref.~\cite{Aoki:2019sht} }
\\ \hline \hline
$\big( - 100\% , + 100\% \big)$
& 
$0.34999$
& 
$0.35$
\\ \hline   
$\big( + 100\% , - 100\% \big)$
& 
$0.01591$
& 
$0.016$
\\ \hline   
$\big( - 80\% , + 30\% \big)$
& 
$0.20234$
& 
$0.20$
\\ \hline   
\hline      
\end{tabular}
\caption{\label{crosssectionUNi} The Table 
shows the numerical 
result  in femtobarn (fb) for cross-checking
of the cross section $\sigma_{\text{pol}}$
with various different beam polarizations
$\big( \lambda_-, \lambda_+ \big)$ at 
CMS energy $\sqrt{s} = 250$ GeV 
compared with the numerical values 
$\big( P_{e^-} , P_{e^+} \big)$ in 
Table $1$in Ref.~\cite{Aoki:2019sht}
within the framework of Standard Model.}
\end{center}
\end{table}
%%%%%%%%%%%%%%%%%%%%%%%%%%%%%%%%%%%
The effects of one-loop contributing
to the differential decay rates
of $h\rightarrow \bar{l}l\gamma$ in vector 
$B-L$ model have presented 
in~\cite{VanOn:2021myp}. In this work, 
we will focus on the signal strengths
$h\rightarrow \ell \bar{\ell}\gamma$ in 
both vector 
and chiral $B-L$ model at the LHC. 
Moreover,
we also examine the signal strengths
for the Higgs production associated
with a photon for the models 
at future lepton colliders. 
%%%%%%%%%%%%%%%%%%%%%%%%%%%%%%%%%%%%%%%%%%%%%%%%%%
\subsection{Decay process 
$h\rightarrow \ell \bar{\ell}
\gamma$ at the LHC}
%%%%%%%%%%%%%%%%%%%%%%%%%%%%%%%%%%%%%%%%%%%%%%
We arrive at the phenomenological results
for decay process 
$h\rightarrow \ell \bar{\ell}\gamma$ 
at the LHC in vector and chiral $B-L$ models. 
The signal
strengths given in (\ref{Sig}) are 
examined in the following subsections. 
%%%%%%%%%%%%%%%%%%%%%%%%%%%%%%%%%%
\subsubsection{Vector $B-L$ model}
%%%%%%%%%%%%%%%%%%%%%%%%%%%%%%%%%%
In vector $B-L$ model, 
new physical parameters are included 
mixing angle between two neutral Higgses 
($c_{\theta}$), the $U(1)_{B-L}$ coupling 
$g_X = g_{B-L}$, a new neutral gauge boson
$M_Z'$ and the kinematic mixing parameter 
$\kappa$ as well as mixing angle ($s_{BL}$) 
of two gauge bosons $Z$ and $Z'$.
For numerical results in this section, 
we only concern the non-mixing gauge bosons 
(non-mixing between two $U(1)$ gauges)
in which we apply the limits of 
$\kappa \rightarrow 0,
s_{BL}\rightarrow 0$.

\begin{figure}[H]
\centering
\includegraphics[width=14cm, height=7cm]
{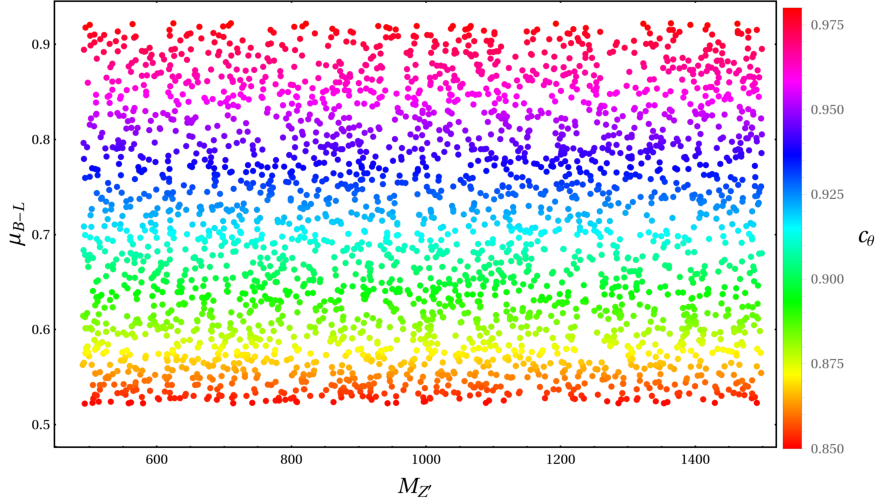}
\caption{\label{muLHC_BL} Signal 
strengths for
the decay process in vector $B-L$ are shown 
as functions of
$c_{\theta}$ and $M_Z'$ at the LHC.}
\end{figure}

In Fig.~\ref{muLHC_BL}, 
the signal strengths as functions of
$c_{\theta}$ and $M_Z'$ are generated.
In this plot, we set $g_{B-L} =0.07$. 
By considering the experimental 
constraint like 
$\frac{M_Z'}{2 g_{B-L}} \geq 3.5$ TeV, 
the mass of neutral gauge boson 
$M_Z' \geq 490$ GeV. For this reason, 
the signal strengths are shown in the 
ranges of $490$ GeV $\leq M_Z' \leq 1000$ 
GeV. The mixing angle of 
two neutral Higgses is taken 
$0.85\leq c_{\theta}\leq 0.98$. 
We find that the signal strengths 
are proportional to mixing angle and 
depend slightly
on $M_Z'$. It shows that 
the contributions of neutral
gauge boson $Z'$ are rather small.
The effects of neutral
gauge boson $Z'$ to the decay 
rates for the processes are 
hard to probe
at the LHC.
%%%%%%%%%%%%%%%%%%%%%%%%%%%%%%%
%%%%%%%%%%%%%%%%%%%%%%%%%%%%%%%%%%
\subsubsection{Chiral $B-L$ model}
%%%%%%%%%%%%%%%%%%%%%%%%%%%%%%%%%%
In this version of $B-L$, 
together with $g_{B-L}, M_Z'$ mixing angle 
($s_{BL}'$) of two gauge bosons
of $U(1)$ gauge groups we include
three more mixing angles of neutral Higges
as $c_{12}, c_{13}, c_{23}$, mixing angle of 
charged Higgs with charged Goldstone boson
$t_{\alpha}$, mixing angle between neutral 
Goldstone bosons and CP-odd Higgs 
$t_{\beta}$. All masses of new particles in 
this version are 
$M_{H_{2,3}}, M_{A_0}, M_{H^{\pm}}, M_{N_i}$.
For numerical results, considering non-mixing 
gauge boson case (or $s_{BL}'\rightarrow 0$), 
we also take the simplest case as 
$c_{13}=c_{23}=1$. The mixing angle 
between $H_1$ (treated as SM-like Higgs boson) 
and $H_2$ is as $c_{\widehat{12} }$. In this 
limit, the 
coupling of $hH^+H^-$ is then derived as
follows:
\begin{eqnarray}
\label{gHpmHmp}
 g_{hH^+H^-} = \frac{1}{v}\Big\{
 \Big[-\frac{\sqrt{2} \mu v_{\sigma}}{s_{\alpha}c_{\alpha}}
 + 2 M_{H^\pm}^2 - M_{h}^2\Big] c_{\alpha + \widehat{12}}
 %%%%%
 + \Big[
 \frac{\sqrt{2} \mu v_{\sigma}}{s_{\alpha}c_{\alpha}}
 \cot(2\alpha) \; s_{\alpha + \widehat{12}}
 + \frac{M_{h}^2}{s_{\alpha}c_{\alpha}}
  s_{\alpha-\widehat{12}}
 \Big]
 \Big\}.
\end{eqnarray}
In this model, we have 
$\kappa_{hVV} = c_{\alpha+\widehat{12}}$. 
For the phenomenological studies, 
we are interested in the 
the case (the SM limit) 
$c_{\alpha+\widehat{12}} \rightarrow 1$.

\begin{figure}[H]
\centering
\includegraphics[width=14cm, height=7cm]
{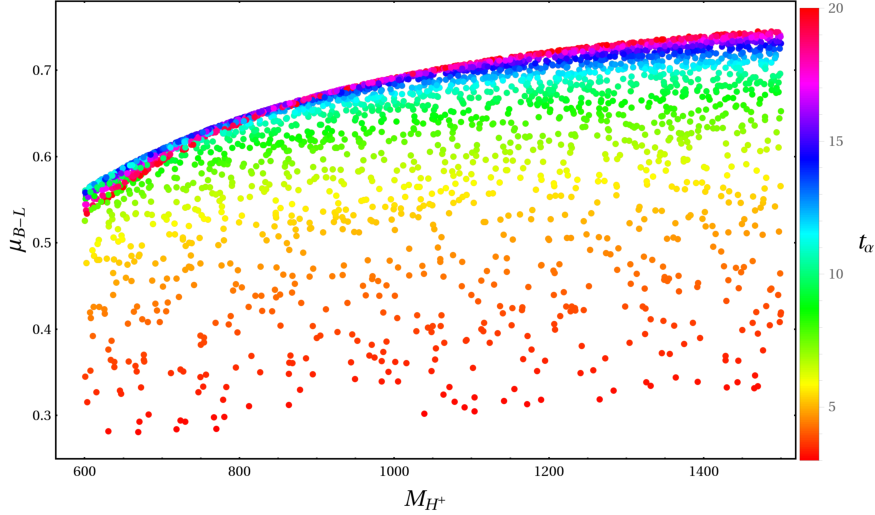}
\caption{\label{muLHC_cBL} Signal 
strengths for
the decay process in chiral 
$B-L$ are shown 
as functions of 
$t_{\alpha}$ and 
charged Higgs mass
$M_{H^\pm}$ at the LHC.
}
\end{figure}

In Fig.~\ref{muLHC_cBL}, 
the signal strengths 
are presented as functions of 
$t_{\alpha}$ and 
charged Higgs mass
$M_{H^\pm}$. The values of $\mu_{B-L}$
are generated in the ranges of
$3 \leq t_{\alpha} \leq 20$
 and $600$ GeV 
$\leq M_{H^{\pm}} \leq 1500$ GeV. 
In further, one selects $g_{B-L} = 0.005$, 
$M_Z' = 500$ GeV, masses of right 
handed neutrinos $M_{N_i} \sim 
\mathcal{O}(1)$ TeV.
Furthermore, $v_{\sigma} =3.5$ TeV 
in combination with the values of 
$v\sim 246$ GeV and $t_{\alpha}$, 
we then get $v_{\Phi}$
and $v_{\phi}$. Having the value of 
$v_{\sigma}$, 
taking $M_{A_0}=800$ GeV, the value of 
$\mu$ can be then obtained appropriately
(follows Eq.~\ref{MA0}). 
In principle $-\pi/2 \leq \widehat{12} \leq \pi/2$,
as mentioned in above we are interested in the 
the SM limit $c_{\alpha+\widehat{12}} \rightarrow 1$. 
For example, one takes 
$c_{\alpha + \widehat{12}} =0.95$
for the following plot. In this plot, we 
find that the signal strengths are 
increased with developing $M_{H^{\pm}}$. 
At the fix value of $M_{H^{\pm}}$, 
$\mu_{B-L}$
is also proportional to $t_{\alpha}$. 
The scatter plot shows that the contributions
from charged Higgs to the processes
are significant and the effects can be probed
at HL-LHC.
%%%%%%%%%%%%%%%%%%%%%%%%%%%%%%%%%%%%%%%%%
%%%%%%%%%%%%%%%%%%%%%%%%%%%%%%%%%%%

\begin{figure}[H]
\centering
\includegraphics[width=14cm, height=7cm]
{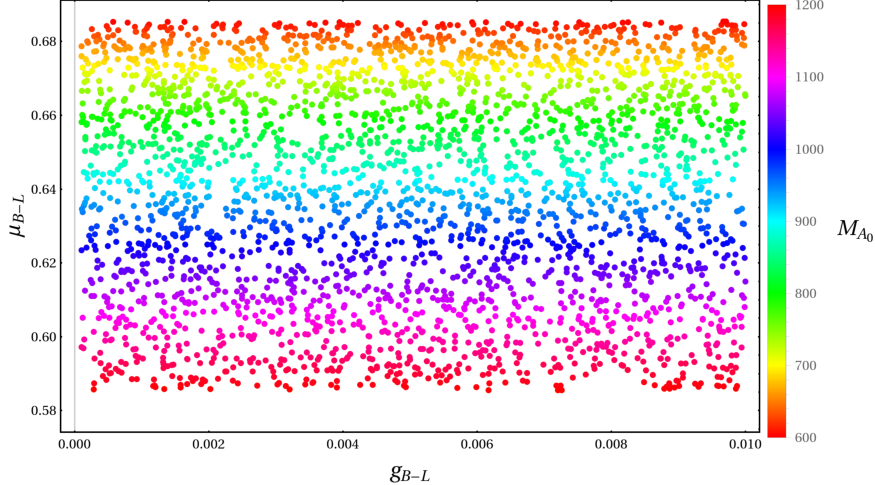}
\caption{\label{muLHC_cBLMA0} Signal 
strengths for
the decay process in chiral $B-L$ are shown 
as functions of 
$g_{B-L}$ and 
CP-odd Higgs mass
$M_{A_0}$ at the LHC.
The vertical grey line is 
shown in this Figure 
for distinguishing
from
the plotting 
boundary.
}
\end{figure}

We concern interestingly
for the contributions
of CP-odd Higgs to 
the signal strengths. 
In Fig.~\ref{muLHC_cBLMA0}, 
the signal strengths
are presented as functions of 
$g_{B-L}$ and CP-odd Higgs 
mass $M_{A_0}$ at the LHC. 
For the scatter plot, we take
$t_{\alpha}=10$, 
$M_{H^{\pm}} = 1000$ GeV
and $M_Z' = 500$ GeV. 
In further, the values of 
$\mu_{B-L}$
are generated in the ranges 
of $10^{-4}\leq g_{B-L} 
\leq 10^{-2}$
and $600$ GeV $\leq M_{A_0} 
\leq 1200$ GeV.
The masses of right handed 
neutrinos are selected as 
$M_{N_i} \sim 
\mathcal{O}(1)$ TeV and the mixing 
angle is taken as 
$c_{\alpha + \widehat{12}} =0.95$
for the following plot. 
We find 
that the signal strengths 
depend significantly on $M_{A_0}$
but slightly change with $g_{B-L}$.
It shows that the contributions
from $Z'$ is rather small in 
comparison with the ones from 
chiral Higgs scalars. The indirect 
impacts of $M_{A_0}$ on the signal
strengths can be explained as follows. 
By changing $M_{A_0}$, the values of 
$\mu$ in Eq.~\ref{MA0} are varied 
appropriately. This leads to the 
change of the coupling 
$hH^{\pm}H^{\mp}$ in 
Eq.~\ref{gHpmHmp}. 
This fact explains for 
the signal strengths changing 
crucially to $M_{A_0}$.
%%%%%%%%%%%%%%%%%%%
%%%%%%%%%%%%%%%%%%%%%%%%%%%%%%%%%%%%%%%
\subsection{Process $e^-e^+
\rightarrow h\gamma$
at future lepton colliders}          %%             
%%%%%%%%%%%%%%%%%%%%%%%%%%%%%%%%%%%%%%%
We turn to Higgs production associated 
with a photon
at future lepton colliders within vector 
and chiral $B-L$ versions.
%%%%%%%%%%%%%%%%%%%%%%%%%%%%%%%%%%%%%
\subsubsection{Vector $B-L$ model}%%%  
\begin{figure}[H]
\centering
$
\begin{array}{cc}
\includegraphics[width=8.5cm, height=5cm]
{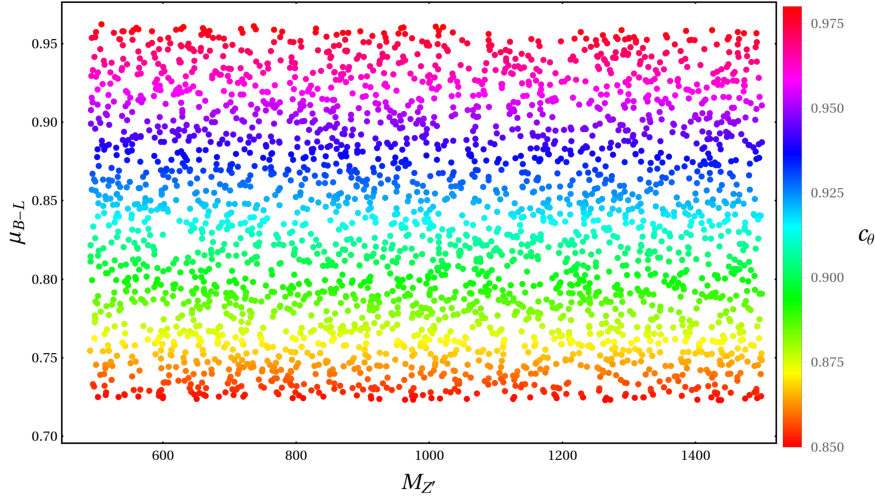}
& 
\includegraphics[width=8.5cm, height=5cm]
{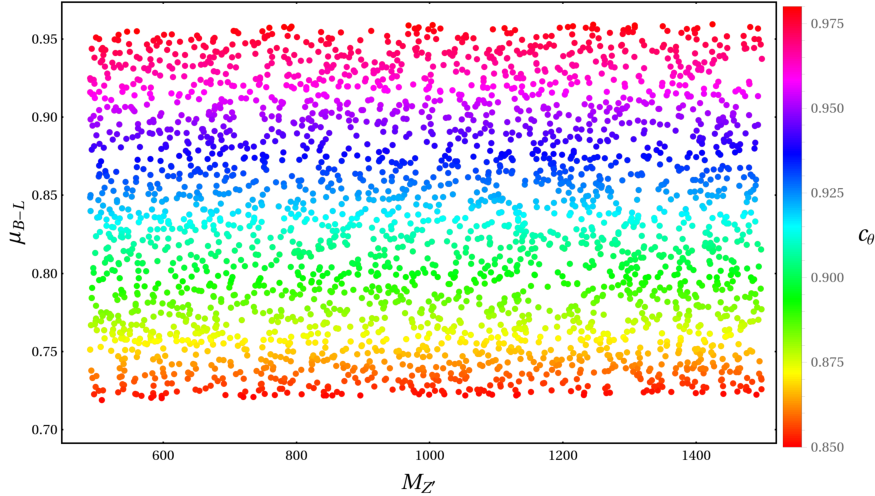}
%%%%%%%%%%%%%%%%%%%%
\end{array}$
\caption{\label{vectorBL250GEV} Signal 
strengths for
the process $e^-e^+\rightarrow h\gamma$ %%       
in vector $B-L$ model at center-of-mass 
energy $\sqrt{s}=250$ GeV. The left panel 
shows for 
$LR$ polarization case and right panel 
is for $RL$
polarization case.}
\end{figure}

\begin{figure}[H]
\centering
$
\begin{array}{cc}
\includegraphics[width=8.5cm, height=5cm]
{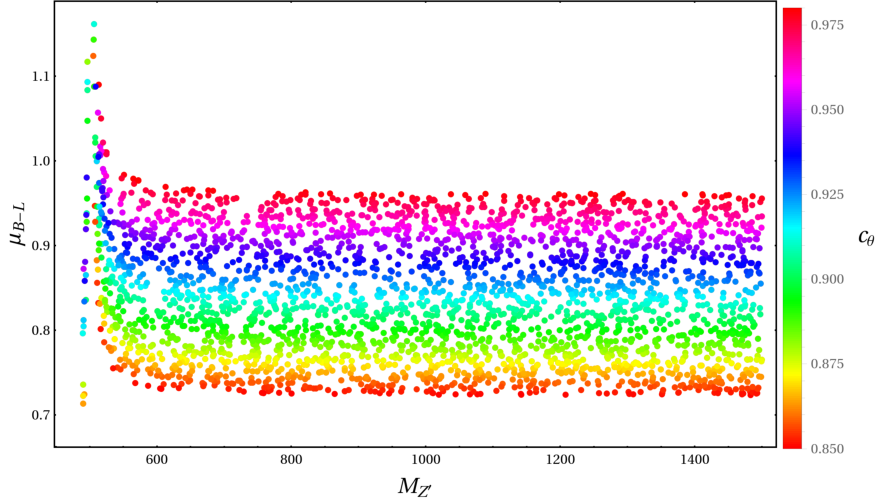}
& 
\includegraphics[width=8.5cm, height=5cm]
{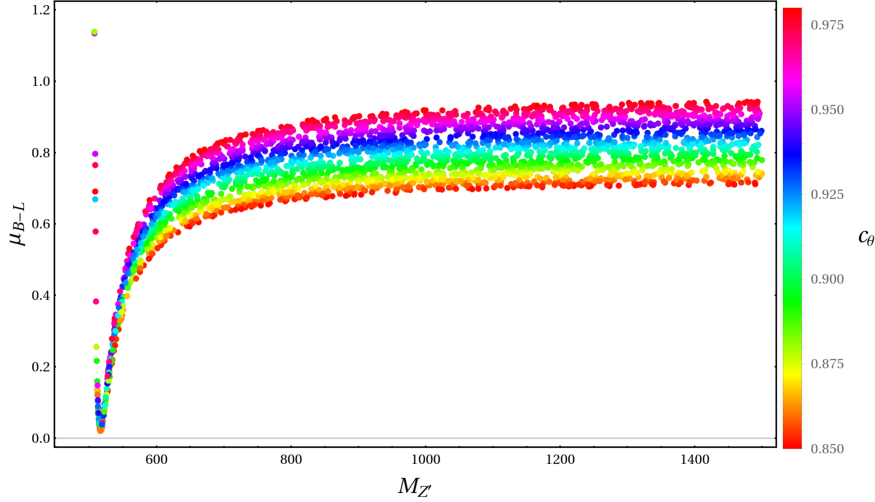}
%%%%%%%%%%%%%%%%%%%%
\end{array}$
\caption{\label{vectorBL500GEV} Signal 
strengths for
the process $e^-e^+\rightarrow h\gamma$ %%       
in vector $B-L$ model at center-of-mass energy 
$\sqrt{s}=500$ GeV. The left panel shows for 
$LR$ polarization case and right panel is for 
$RL$
polarization case.
The horizontal grey line 
appears for distinguishing
from the boundary of the right scatter plot.
}
\end{figure}

%%%%%%%%%%%%%%%%%%%%%%%%%%%%%%%%%%%%%   
As previous cases, we also
consider the non-mixing gauge bosons, 
$\kappa \rightarrow 0,
s_{BL}\rightarrow 0$, for the following 
numerical results. 
In Figs.~\ref{vectorBL250GEV}, 
\ref{vectorBL500GEV}, 
the signal strengths as functions of
$c_{\theta}$ and $M_Z'$ are presented
at $\sqrt{s}=250$ GeV and at  
$\sqrt{s}=500$ GeV, respectively. 
In these plots,
we take the same previous 
input parameters as 
$g_{B-L} =0.07$, 
$490$ GeV $\leq M_Z' 
\leq 1000$ GeV and 
$0.85 \leq c_{\theta}\leq 0.98$.
In the left panel, we show for 
the signal strengths in the 
$LR$ polarization case 
($e^-_L e^+_R$) and for 
$RL$ polarization case
($e^-_R e^+_L$)
in the 
right panel, respectively.

At $\sqrt{s} =250$ GeV, we find 
that the signal strengths are 
proportional to the mixing angle 
$c_{\theta}$ and slightly change with 
$M_ Z'$. The plots indicate that 
the contributions from $Z'$ are rather 
small. As a result, 
the effects of $Z'$ to the decay processes 
are hard to probe at future LC.
%%%%%%%%%%%%%%%%%%%%%%%%%%%%%%%%%%%%%%%%%
At $\sqrt{s} =500$ GeV, 
around the $Z'$-peak, 
$\sqrt{s}=M_Z'\sim 500$ GeV,
we also find the peak of 
the signal strengths. Beyond the $M_Z'$-peak
region, the same behavior of signal strengths
as previous case is observed the signal 
strengths are 
proportional to the mixing angle 
$c_{\theta}$ and slightly change
with $M_ Z'$. It is indicated that 
the contributions from $Z'$ are rather 
small beyond the $Z'$-peak. 
Consequence, 
the impacts of $Z'$ on the decay rates
for the processes
are hard to measured at future LC.
One may probe the contributions
of $Z'$ to $\mu_{B-L}$ around
the $Z'$-peak at future LC.
%%%%%%%%%%%%%%%%%%%%%%%%%%%%%%%%%%%%%%%%%
%%%%%%%%%%%%%%%%%%%%%%%%%%%%%%%%%%%%
\subsubsection{Chiral $B-L$ model}%%               
\begin{figure}[H]
\centering
$
\begin{array}{cc}
\includegraphics[width=8.5cm, height=5cm]
{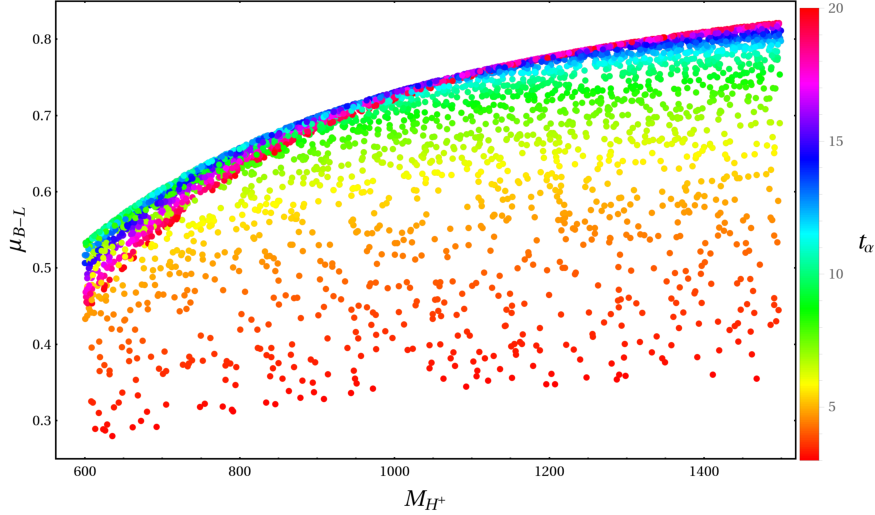}
& 
\includegraphics[width=8.5cm, height=5cm]
{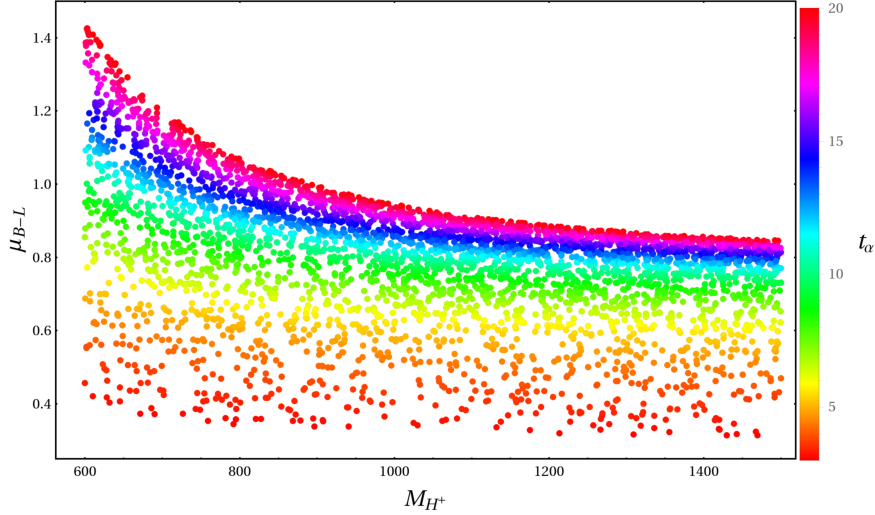}
%%%%%%%%%%%%%%%%%%%%
\end{array}$
\caption{\label{chiralBLMH250GEV} 
Signal strengths for
the process $e^-e^+\rightarrow h\gamma$ %%       
in chiral $B-L$ model at $\sqrt{s} = 250$ GeV.
The left panel is for $LR$ polarization case, 
the right
panel shows for the $RL$ polarization case.}
\end{figure}

\begin{figure}[H]
\centering
$
\begin{array}{cc}
\includegraphics[width=8.5cm, height=5cm]
{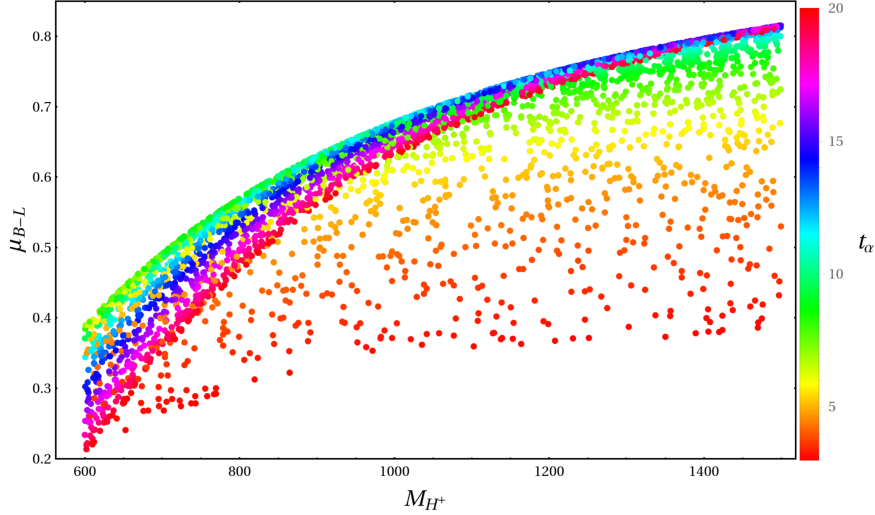}
& 
\includegraphics[width=8.5cm, height=5cm]
{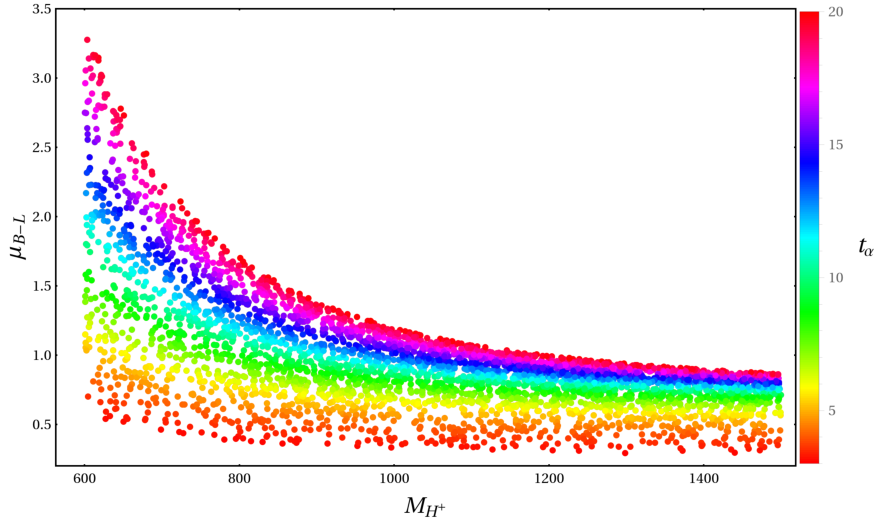}
%%%%%%%%%%%%%%%%%%%%
\end{array}$
\caption{\label{chiralBLMH500GEV} Signal 
strengths for
the process $e^-e^+\rightarrow h\gamma$ %%       
in chiral $B-L$ model at $\sqrt{s} = 500$ GeV.
The left panel is for $LR$ polarization case, 
the right
panel shows for the $RL$ polarization case.}
\end{figure}

%%%%%%%%%%%%%%%%%%%%%%%%%%%%%%%%%%%%
In Figs.~\ref{chiralBLMH250GEV}
(and Figs.~\ref{chiralBLMH500GEV}), 
the signal strengths 
are presented as functions of 
$t_{\alpha}$ and 
charged Higgs mass
$M_{H^\pm}$ at $\sqrt{s} =250$ GeV 
($\sqrt{s} =500$ GeV), respectively.
In these plots, the values of $\mu_{B-L}$
are generated in the ranges of
$3 \leq t_{\alpha} \leq 20$
 and $600$ GeV 
$\leq M_{H^{\pm}} \leq 1500$ GeV. 
In further, one selects $g_{B-L} = 0.005$, 
$M_Z' = 500$ GeV, masses of right 
handed neutrinos $M_{N_i} = 10$ GeV, 
$M_{A_0}=800$ GeV. 
As mentioned in above, 
we are also interested in the 
the SM limit $c_{\alpha +\widehat{12}}
\rightarrow 1$. 
For example, one takes 
$c_{\alpha + \widehat{12}} =0.95$
for the following plots.

In the left panel (right panel), 
signal strengths are 
for $LR$ ($RL$) polarization cases, 
respectively. By implying the 
initial beam
polarizations, there haven't 
the contributions of one-loop 
non $V_0$-pole diagrams with 
exchanging $W$ boson and 
charged Higgs in the loop for the $RL$ case
(see Figs.~\ref{G2WG},~\ref{G5HPM}).
This explains the different behavior 
of $\mu_{B-L}$ in the LR and RL cases. 
This also 
causes for the values of $\mu_{B-L}$ in $RL$ case
are bigger than the $LR$ case. Moreover,
it is interesting to find that 
the signal strengths are increased with 
the charged Higgs mass in 
$LR$ case.
While $\mu_{B-L}$ are decreased with 
the charged Higgs mass in $RL$ case. 
The signal strengths 
$\mu_{B-L}$ are also sensitive 
with $t_\alpha$ at 
the fix value of the charged Higgs mass.
The results show that the effects of 
charged Higgs bosons in the loop 
are significant contributions in
comparison with the corresponding 
ones from $Z'$ in $U(1)_{B-L}$. 
From the scatter plots, one can 
probe the charged Higgs
contributions via the production 
process at future LC. 
%%%%%%%%%%%%%%%%%%%%%%%%%%%%%%%%%%%%%%%%%%%%%
%%%%%%%%%%%%%%%%%%%%%%%%%%%%%%%%%%%%%%%%%%%%

%%%%%%%%%%%%%%%%%%%%%%%%%%%%%%%%%%%%%%%%%%%%
\begin{figure}[H]
\centering
$
\begin{array}{cc}
\includegraphics[width=8.5cm, height=5cm]
{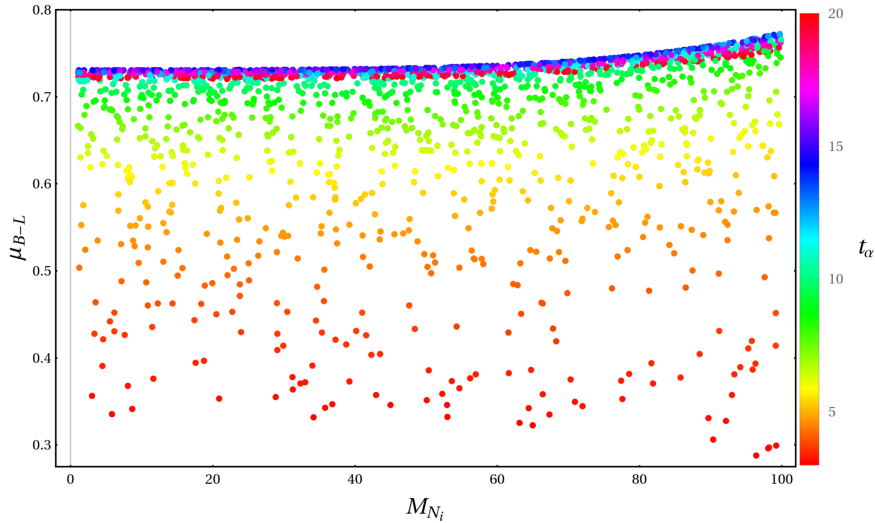}
& 
\includegraphics[width=8.5cm, height=5cm]
{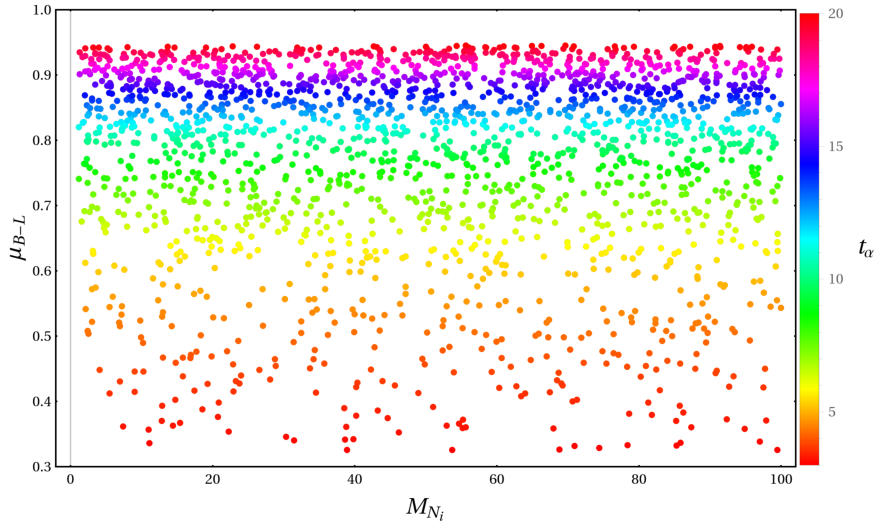}
%%%%%%%%%%%%%%%%%%%%
\end{array}$
\caption{\label{chiralBLMN250GEV} Signal 
strengths for
the process $e^-e^+\rightarrow h\gamma$ %%       
in chiral $B-L$ model at $\sqrt{s} = 500$ GeV.
The left panel is for $LR$
polarization case, the right
panel shows for the $RL$
polarization case. 
There are vertical grey lines in the Figures 
which are for distinguishing 
from
the plotting 
boundary.}
\end{figure}

\begin{figure}[H]
\centering
$
\begin{array}{cc}
\includegraphics[width=8.5cm, height=5cm]
{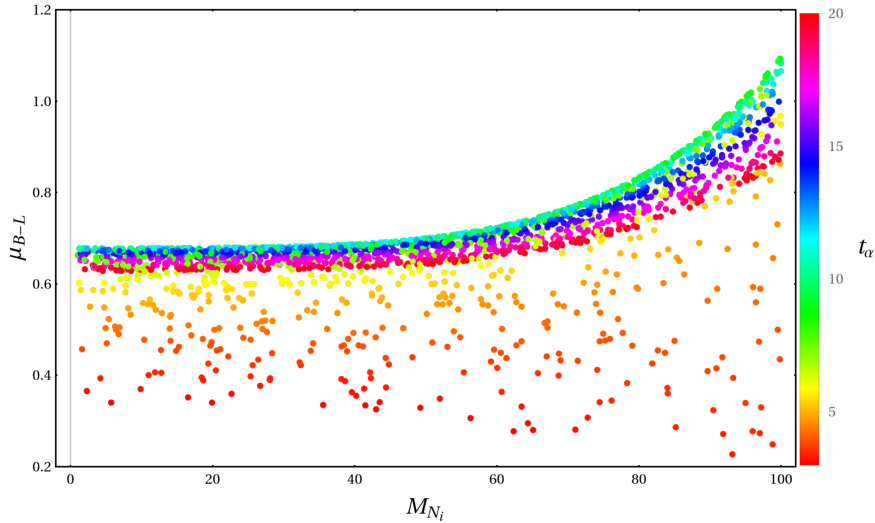}
& 
\includegraphics[width=8.5cm, height=5cm]
{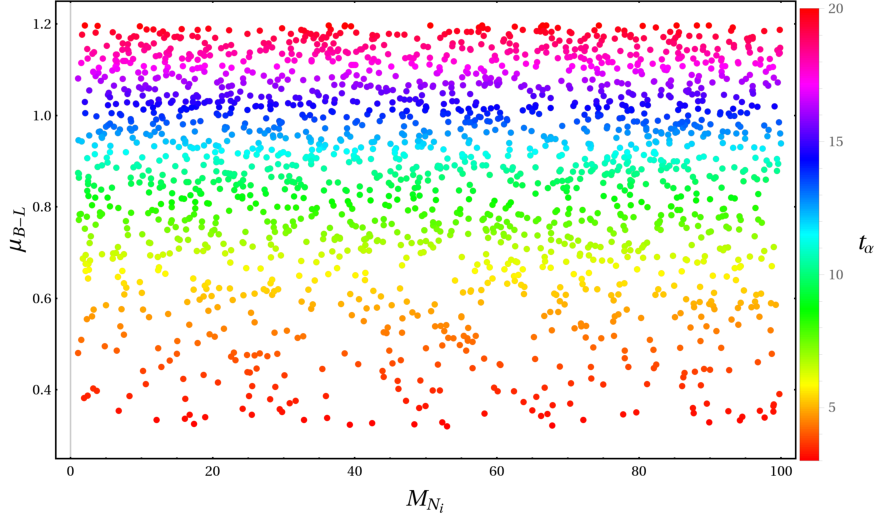}
%%%%%%%%%%%%%%%%%%%%
\end{array}$
\caption{\label{chiralBLMN500GEV} 
Signal strengths for the process 
$e^-e^+\rightarrow h\gamma$ in chiral 
$B-L$ model at $\sqrt{s} = 500$ GeV.
The left panel is for $LR$ polarization 
case, the right panel shows for the $RL$
polarization case. The vertical and horizontal
grey lines appear in the Figures for 
distinguishing from the plotting 
boundary.} 
\end{figure}

%%%%%%%%%%%%%%%%%%%%%%%%%%%%%%%%%%%%%%%%%%%%
In Figs.~\ref{chiralBLMN250GEV}
(and Figs. \ref{chiralBLMN500GEV}), 
the signal strengths are shown as functions of
$t_{\alpha}$ and right handed neutrino mass 
$M_{N_i}$. In these plots, we present
$\mu_{B-L}$ in the ranges of 
$3\leq t_{\alpha}\leq 20$ and 
$1$ GeV $\leq M_{N_i} \leq 150$ GeV. 
We take 
$g_{B-L} = 0.005$, 
$M_Z' = 500$ GeV, charged Higgs 
mass $M_{H^{\pm}} = 1000$ GeV.
As previous case,
having $v_{\sigma} =3.5$ TeV and 
$M_{A_0}=800$ GeV, the value of 
$\mu$ can be then derived
appropriately. 
We are also great of interest 
in the SM limit 
$c_{\alpha+\widehat{12}} \rightarrow 1$ and 
take, for example, 
$c_{\alpha+\widehat{12}} =0.95$ in these plots. 
In the left panel, we have signal strengths
for $LR$ polarization case. While the right 
panel shows for the signal strengths for
$RL$ polarization case. We find that 
the signal strengths 
are unchanged
in the case of $RL$ polarization because of
there haven't the contributions of 
charged Higgs box diagrams. As a result, 
$\mu_{B-L}$ must be unchanged
with varying $M_{N_i}$. In the left panel, 
the signal 
strengths depend slightly on $M_{N_i}$
at $\sqrt{s} =250$ GeV. One finds more
impacts of $M_{N_i}$ 
on signal strengths 
at $\sqrt{s} = 500$ GeV.
%%%%%%%%%%%%%%%%%%%%%%%%%%%%%%%%%%%%%%%%%%%%%%
%%%%%%%%%%%%%%%%%%%%%%%%%%%%%%%%%%%%%%%%%%%%

%%%%%%%%%%%%%%%%%%%%%%%%%%%%%%%%%%%%%%%%%
\begin{figure}[H]
\centering
$
\begin{array}{cc}
\includegraphics[width=8.5cm, height=5cm]
{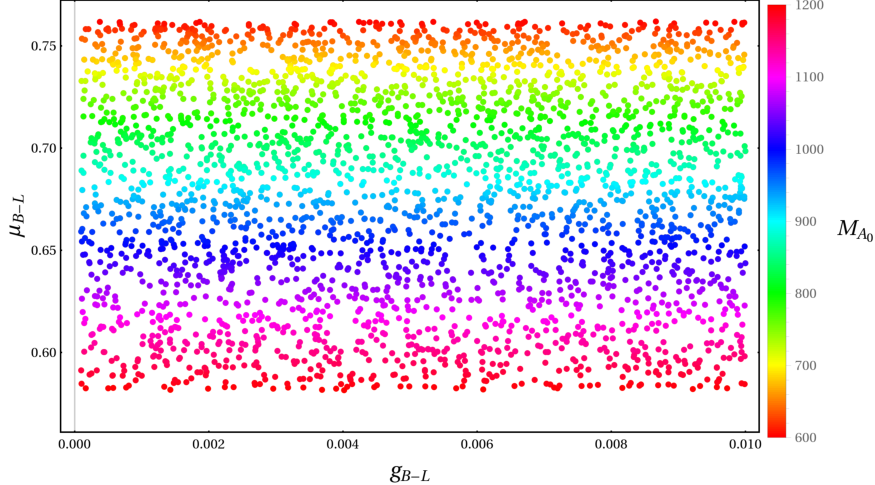}
& 
\includegraphics[width=8.5cm, height=5cm]
{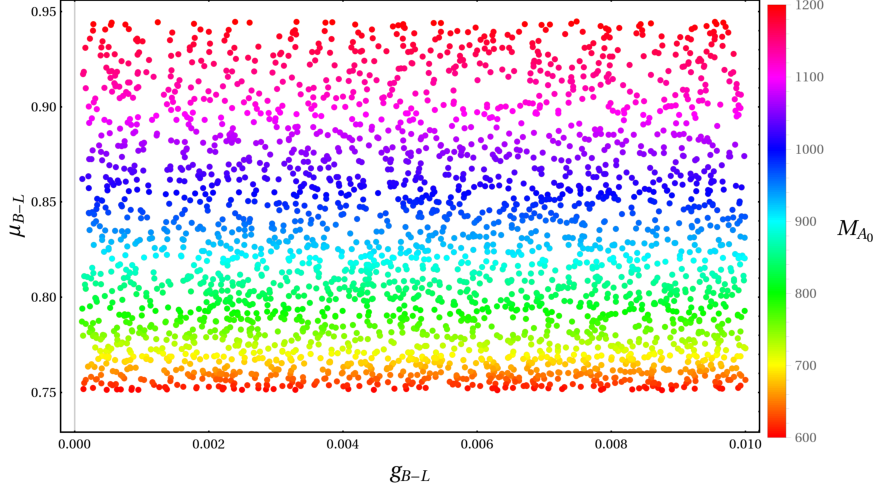}
%%%%%%%%%%%%%%%%%%%%
\end{array}$
\caption{\label{muLC_cBLMA0250} 
Signal strengths for the production 
process in chiral $B-L$ are shown 
as functions of $g_{B-L}$ and 
CP-odd Higgs mass $M_{A_0}$ at 
$\sqrt{s}=250$ GeV at the future LC.
The left panel is for $LR$ polarization 
case, the right panel shows for the $RL$
polarization case. There are vertical 
grey lines in the Figures which are 
for distinguishing from
the plotting boundary.
}
\end{figure}

\begin{figure}[H]
\centering
$
\begin{array}{cc}
\includegraphics[width=8.5cm, height=5cm]
{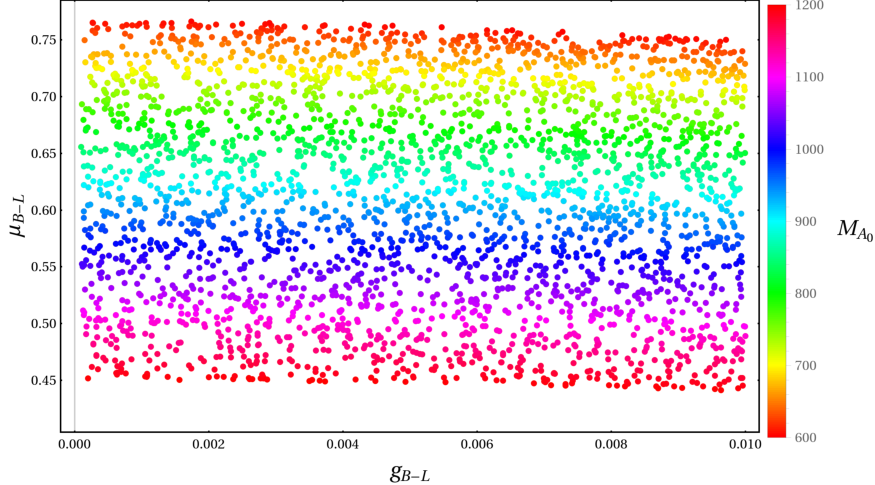}
& 
\includegraphics[width=8.5cm, height=5cm]
{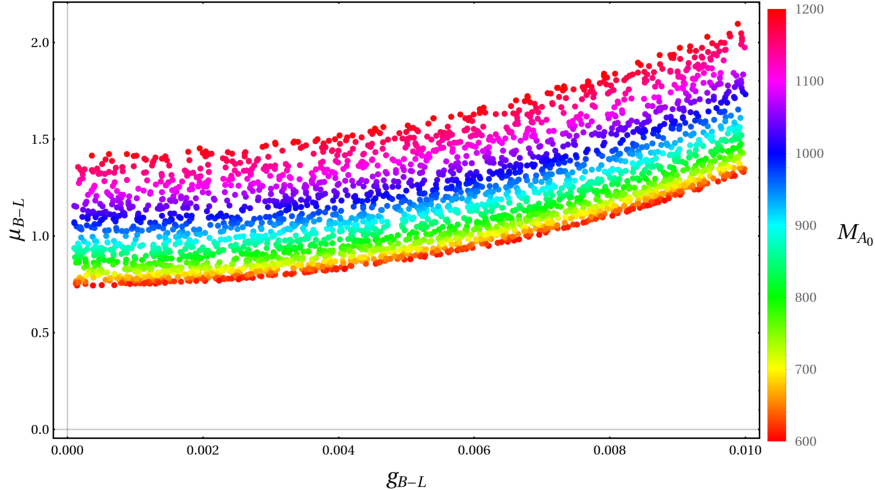}
%%%%%%%%%%%%%%%%%%%%
\end{array}$
\caption{\label{muLC_cBLMA0500} 
Signal strengths for
the production process in chiral 
$B-L$ are shown as functions of 
$g_{B-L}$ and $M_{A_0}$ at 
$\sqrt{s}=500$ GeV at 
the future LC. 
The left panel is for $LR$ polarization 
case, the right panel shows for the $RL$
polarization case. The vertical 
and horizontal grey lines appear 
in the Figures for distinguishing from
the 
boundary of the 
scatter plots.} 
\end{figure}

%%%%%%%%%%%%%%%%%%%%%%%%%%%%%%%%%%%%%%%
As previous discussions, 
we also consider interestingly for
the contributions of $M_{A_0}$
to the signal strengths at 
the future LC. In 
Figs.~\ref{muLC_cBLMA0250},
~\ref{muLC_cBLMA0500}, 
the signal strengths 
are shown as functions of 
$g_{B-L}$ and 
$M_{A_0}$ at $\sqrt{s}=250$ GeV 
and $\sqrt{s}=500$ GeV, respectively. 
In these Figures, we fix 
the following parameters as  
$t_{\alpha}=10$, 
$M_{H^{\pm}} = 1000$ GeV
and $M_Z' = 500$ GeV.  
Furthermore, the values 
of $\mu_{B-L}$ are 
generated in the regions of
$10^{-4}\leq g_{B-L} \leq
10^{-2}$ and 
$600$ GeV $\leq M_{A_0} \leq 1200$ GeV.
The masses of right handed neutrinos are
selected as
$M_{N_i}=10$ GeV and the mixing 
angle is taken as 
$c_{\alpha + \widehat{12}} =0.95$
for the following plots. 
At $\sqrt{s}=250$ GeV,
the signal strengths 
depend significantly on $M_{A_0}$
but slightly change with $g_{B-L}$.
It shows that the contributions
from $Z'$ is rather small in 
comparison with the 
contributions from chiral Higgs 
scalars. In the $LR$ polarization case 
at $\sqrt{s}=500$ GeV, 
we observe the same behavior of 
$\mu_{B-L}$ as the previous cases 
at $\sqrt{s}=250$ GeV. At high energy 
regions, the attributions from $Z'$ in 
the loop is more impact on the decay 
rates. Since we only consider the non-mixing
of $Z$ and $Z'$. Subsequently, there isn't
the $Z'$-pole with the $W$-loop in this case. 
Futhermore, 
the contributions of box diagrams 
with $Z'$ exchanging in the loop 
are much smaller in comparison with 
$W$ boson loop. As a result, this explains
for $\mu_{B-L}$ depend slightly on $g_{B-L}$
in this case. It is interesting to 
find that the signal strengths change 
significantly to both $M_{A_0}$ 
and $g_{B-L}$ in the $RL$ polarized
case at $\sqrt{s}=500$ GeV. 
With the absent of the 
$W$-boson, charged Higgs 
in the loop diagrams, 
the box diagrams with the $Z'$ 
internal lines
contribute to $\mu_{B-L}$ is more 
substantial effects. It is explained 
that the signal 
strengths depend crucially on $g_{B-L}$
at fixing $M_{A_0}$.
%%%%%%%%%%%%%%%%%%%
%%%%%%%%%%%%%%%%%%%
%%%%%%%%%%%%%%%%%%%%%%%%%%%%%
\section{Conclusions}   %%%%%
%%%%%%%%%%%%%%%%%%%%%%%%%%%%% 
In this work, we have presented 
one-loop form factors
for $h\rightarrow \ell 
\bar{\ell}\gamma$
in $U(1)_{B-L}$ 
extension of the standard model.
The computations have performed 
in 't Hooft-Feynman gauge. 
We have shown that 
production cross-sections for 
$e^-e^+\rightarrow h\gamma$ 
at future lepton colliders 
can be derived  
by using 
one-loop form factors
in the decay 
$h\rightarrow \ell \bar{\ell}\gamma$. 
The form factors are expressed 
in terms of one-loop scalar 
PV-functions in the standard notation 
of {\tt LoopTools}. As a result, 
one-loop decay rates, cross-sections
as well as the signal strengths can 
be evaluated numerically by using 
{\tt LoopTools}. In phenomenological 
results, the signal strengths for
$h\rightarrow \ell \bar{\ell}\gamma$ at 
Large Hadron Collider and 
for $e^-e^+\rightarrow h\gamma$
at future Linear Colliders
have investigated in physical 
parameter space for both 
vector and chiral $B-L$ models. 
We find that the contributions 
from neutral gauge boson $Z'$ 
to the signal strengths are 
rather small. 
Consequence, the effects from $Z'$
in the models are 
hard to probe at future colliders.
While the impacts 
of charged Higgs in the 
chiral $B-L$ model on 
the signal strengths
are significant and 
can be measured with the help 
of the initial polarization 
beams at 
future Lepton Colliders. 
%%%%%%%%%%%%%%%%%%%%%%%%%%%%%
\\

\noindent
{\bf Acknowledgment:}~
This research is funded by Vietnam National 
Foundation for Science and Technology Development 
(NAFOSTED) under the grant number $103.01$-$2023.16$.
Dzung Tri Tran and Khiem Hong Phan express their 
gratitude to all the valuable support from Duy Tan 
University, for the 30th anniversary of establishment 
(Nov. 11, 1994 - Nov. 11, 2024) towards "Integral, 
Sustainable and Stable Development.

%%%%%%%%%%%%%%%%%%%%%%%%%%%%%%%%%%%%%%%%%%%%%%%%%%%%
\section*{Appendix A: PV-functions 
and checks for the calculations}%%
%%%%%%%%%%%%%%%%%%%%%%%%%%%%%%%%%%
We follow the approach in~\cite{Denner:2005nn} 
for tensor reductions for one-loop integrals. 
Based on the method,  tensor one-loop 
integrals with $N$-external  lines can 
be expressed in terms of scalar one-loop 
with one-, two-, three- and  four-point 
functions.
The definition for tensor one-loop 
integrals with rank $R$ are given by
(upto four point functions):
\begin{eqnarray}
\{A; B; C; D\}^{\mu_1\mu_2\cdots \mu_R}
= (\mu^2)^{2-d/2}
\int \frac{d^dk}{(2\pi)^d} 
\dfrac{k^{\mu_1}k^{\mu_2}\cdots 
k^{\mu_R}}{\{P_1; P_1 P_2;P_1P_2P_3; 
P_1P_2P_3P_4\}}.
\end{eqnarray}
In this formula, 
$P_j^{-1}$ ($j=1,\cdots, 4$) 
are Feynman propagators
\begin{eqnarray}
 P_j = (k+ q_j)^2 -m_j^2 +i\epsilon.
\end{eqnarray}
The momenta 
$q_j = \sum\limits_{i=1}^j p_i$ are 
taken into account where 
external momenta $p_j$ and 
internal masses $m_j$ are involved.
The term $i\epsilon$ a Feynman 
prescription. Space-time dimension 
$d$ is taken the form of 
$d=4-2\varepsilon$ with 
$\varepsilon\rightarrow0$
at final results. 
The parameter $\mu^2$ in a 
overall factor of 
tensor integrals playing role of a 
renormalization scale. 
Reduction formulas for tensor 
one-loop one-, 
two-, three- and four-point 
integrals up to rank $R=3$ 
are shown~\cite{Denner:2005nn}:
\begin{eqnarray}
A^{\mu}        &=& 0, \\
A^{\mu\nu}     &=& g^{\mu\nu} A_{00}, \\
A^{\mu\nu\rho} &=& 0,\\
B^{\mu}        &=& q^{\mu} B_1,\\
B^{\mu\nu}     &=& g^{\mu\nu} B_{00} 
+ q^{\mu}q^{\nu} B_{11}, \\
B^{\mu\nu\rho} &=& \{g, q\}^{\mu\nu\rho} B_{001} 
+ q^{\mu}q^{\nu}q^{\rho} B_{111}, \\
C^{\mu}        &=& q_1^{\mu} C_1 + q_2^{\mu} C_2 
 = \sum\limits_{i=1}^2q_i^{\mu} C_i, 
\\
C^{\mu\nu}    &=& g^{\mu\nu} C_{00} 
 + \sum\limits_{i,j=1}^2q_i^{\mu}q_j^{\nu} C_{ij},
\\
C^{\mu\nu\rho} &=&
\sum_{i=1}^2 \{g,q_i\}^{\mu\nu\rho} C_{00i}+
\sum_{i,j,k=1}^2 q^{\mu}_i q^{\nu}_j q^{\rho}_k C_{ijk},\\
D^{\mu}        &=& q_1^{\mu} D_1 
+ q_2^{\mu} D_2 + q_3^{\mu}D_3 
 = \sum\limits_{i=1}^3q_i^{\mu} D_i, \\
D^{\mu\nu}    &=& g^{\mu\nu} D_{00} 
 + \sum\limits_{i,j=1}^3q_i^{\mu}q_j^{\nu} D_{ij}, 
\\
D^{\mu\nu\rho} &=&
	\sum_{i=1}^3 \{g,q_i\}^{\mu\nu\rho} D_{00i}+
	\sum_{i,j,k=1}^3 q^{\mu}_i q^{\nu}_j q^{\rho}_k D_{ijk}.
\end{eqnarray}
Tensor $\{g, q_i\}^{\mu\nu\rho}$ is defined
as follows: $\{g, q_i\}^{\mu\nu\rho} 
= g^{\mu\nu} q^{\rho}_i 
+ g^{\nu\rho} q^{\mu}_i + g^{\mu\rho} q^{\nu}_i$
and
$A_{00}, B_1, \cdots, D_{333}$ are so-called as 
Passarino-Veltman functions 
(PV-functions)~\cite{Denner:2005nn}. These functions
have been implemented into 
{\tt LoopTools}~\cite{Hahn:1998yk} for 
numerical computations. 
\section*{Appendix B: The couplings}%
%%%%%%%%%%%%%%%%%%%%%%%%%%%%%%%%%%%%%
In this Appendix, we present all 
related couplings to the processes
under consideration in this paper.
We begin with the vector $B-L$ model 
and show the couplings in chiral 
$B-L$ model in the following 
subsections.
%%%%%%%%%%%%%%%%%%%%%%%%%%%%%%%%%%
\subsection*{Vector $B-L$ Model}%%
%%%%%%%%%%%%%%%%%%%%%%%%%%%%%%%%%%
We first derive the coupling of 
$Z$ to fermion pair, $Z\bar{f}f$. 
For illustrate, we take 
$Z\bar{q}q$ with quark $q$ is taken
in the first generation of matter 
contents. In detail, the couplings 
are derived as follows:
%%%%%%%%%%%%%%%%%%%%%%%%%%%%%%%
\begin{eqnarray}
&&\mathcal{L}_{Z\bar{q}q} 
\supset
i \bar{q}_L  \slashed{D} q_L 
+ i\bar{u}_R \slashed{D} u_R 
+ i\bar{d}_R \slashed{D} d_R
\\
&&=
-\frac{1}{2}
\left(
\begin{array}{cc}
\bar{u}_L & \bar{d}_L
\end{array} 
\right)
\gamma_\mu Z^{\mu}
\left[
\begin{array}{cc}
gc_Wc_{BL}-(2Q-1)g's_Wc_{BL} & 0 \\
0 & -gc_Wc_{BL}-(2Q+1)g's_Wc_{BL}
\end{array}
\right]
\left(
\begin{array}{c}
        u_L \\
        d_L
\end{array}
\right) 
\nonumber\\
&&\hspace{0.5cm}
-g'Y_{u_R} (-s_Wc_{BL}) 
\bar{u}_R\gamma_\mu Z^{\mu}u_R
-g'Y_{d_R} (-s_Wc_{BL}) 
\bar{d}_R\gamma_\mu Z^{\mu}d_R 
\\
&&\hspace{0.5cm}
-s_{BL}
\Big(
\frac{g_XY_X}{\sqrt{1-\kappa^2}}
-g'Y_{u_L} \frac{\kappa}{\sqrt{1-\kappa^2}} 
\Big)
\bar{u}_L \gamma_{\mu}Z^{\mu} u_L
-s_{BL}
\Big(
\frac{g_XY_X}{\sqrt{1-\kappa^2}}
-g'Y_{d_L} \frac{\kappa}{\sqrt{1-\kappa^2}} 
\Big)
\bar{d}_L\gamma_{\mu}Z^{\mu} d_L 
\nonumber\\
&&\hspace{0.5cm}
-s_{BL}
\Big(
\frac{g_XY_X}{\sqrt{1-\kappa^2}}
-g'Y_{u_R} \frac{\kappa}{\sqrt{1-\kappa^2}} 
\Big)
\bar{u}_R\gamma_{\mu}Z^{\mu} u_R  
-s_{BL}
\Big(
\frac{g_XY_X}{\sqrt{1-\kappa^2}}
-g'Y_{d_R} \frac{\kappa}{\sqrt{1-\kappa^2}} 
\Big)
\bar{d}_R\gamma_{\mu}Z^{\mu} d_R 
\nonumber
\\
&&=
-\frac{1}{12}\frac{e}{s_Wc_W}
\bar{u}
\gamma_{\mu}\; Z^{\mu}
\Big[
  (1-\gamma_5)(3-6Q_{u_L}s_W^2)
- (1+\gamma_5)(6s_W^2Y_{u_R})
\Big]c_{BL}\;u   \nonumber\\
&&\hspace{0.5cm} 
-\frac{1}{12}\frac{e}{s_Wc_W}
\bar{d}
\gamma_{\mu} Z^{\mu}
\Big[-(1-\gamma_5)(3+6Q_{d_L}s_W^2)
- (1+\gamma_5)(6s_W^2Y_{d_R})\Big]c_{BL}\; d
\nonumber\\
&& \hspace{0.5cm}
-\Big[
\frac{g_XY_X}{\sqrt{1-\kappa^2}}
-\frac{g'}{2}[Y_{u_L} +Y_{u_R}
+(Y_{u_R} - Y_{u_L})\gamma_5
]\frac{\kappa}{\sqrt{1-\kappa^2}}
\Big]s_{BL} 
\bar{u}\gamma_{\mu}Z^{\mu} u
\nonumber\\
&&\hspace{0.5cm}
-\Big[\frac{g_XY_X}{\sqrt{1-\kappa^2}} 
- \frac{g'}{2}[Y_{d_L} + Y_{d_R} 
+ (Y_{d_R} - Y_{d_L})\gamma_5]
\frac{\kappa}{\sqrt{1-\kappa^2}}
\Big]s_{BL} 
\bar{d}\gamma_{\mu}Z^{\mu} d.
\end{eqnarray}
We next calculate the couplings of $Z' \bar{q}q$
as follows:
%%%%%%%%%%%%%%%%%%%%%%%%%%%%%%%%%%%%%%%%%%%%%%%
\begin{eqnarray}
&&\mathcal{L}_{Z'\bar{q}q} 
\supset 
i\bar{q}_L  \slashed{D} q_L 
+i\bar{u}_R \slashed{D} u_R
+i\bar{d}_R \slashed{D} d_R 
\\
&&= -\frac{1}{2}
\left(\begin{array}{cc}
\bar{u}_L & \bar{d}_L
\end{array}\right)\gamma_\mu Z'^{\mu} 
\left[
\begin{array}{cc}
-gc_Ws_{BL}+(2Q-1)g's_Ws_{BL} & 0 \\
0 & gc_Ws_{BL}+(2Q+1)g's_Ws_{BL}
\end{array}
\right]
\left(
\begin{array}{c}
u_L \\
d_L
\end{array}\right)
\nonumber\\
&&\hspace{0.5cm} 
-g' Y_{u_R} s_W s_{BL}  
\; \bar{u}_R\gamma_{\mu}Z'^{\mu} u_R 
-g' Y_{d_R} s_W s_{BL}  
\; \bar{d}_R\gamma_{\mu}Z'^{\mu} d_R 
\\
&&\hspace{0.5cm} 
-\Big(\frac{g_XY_X}{\sqrt{1-\kappa^2}}
-g'Y_{u_L} \frac{\kappa}{\sqrt{1-\kappa^2}}\Big) 
c_{BL} \; \bar{u}_L \gamma_\mu Z'^{\mu}u_L 
-\Big(\frac{g_XY_X}{\sqrt{1-\kappa^2}}
-g'Y_{d_L}\frac{\kappa}{\sqrt{1-\kappa^2}}\Big) 
c_{BL}
\; \bar{d}_L\gamma_\mu Z'^{\mu} d_L 
\nonumber\\
&&\hspace{0.5cm} 
-\Big(\frac{g_XY_X}{\sqrt{1-\kappa^2}}
-g'Y_{u_R}\frac{\kappa}{\sqrt{1-\kappa^2}}\Big) 
c_{BL}
\; \bar{u}_R\gamma_\mu Z'^{\mu} u_R
-\Big(\frac{g_XY_X}{\sqrt{1-\kappa^2}}
-g'Y_{d_R}\frac{\kappa}{\sqrt{1-\kappa^2}}\Big) 
c_{BL}
\; \bar{d}_R \gamma_\mu Z'^{\mu} d_R 
\nonumber\\
&&
=
-\frac{1}{12}\frac{e}{s_Wc_W}
\bar{u}\gamma_{\mu}\; Z'^{\mu}
\Big[(1-\gamma_5)(-3+6Q_{u_L}s_W^2)
+(1+\gamma_5)6s_W^2Y_{u_R} \Big] 
s_{BL}u
\nonumber\\
&&\hspace{0.5cm} 
-\frac{1}{12}\frac{e}{s_Wc_W}
\bar{u}\gamma_{\mu}\; Z'^{\mu}
\Big[(1-\gamma_5)(3+6Q_{d_L}s_W^2)
+(1+\gamma_5)s_W^26Y_{d_R} \Big]
s_{BL}d
\nonumber\\
&&\hspace{0.5cm} 
-\Big[
\frac{g_XY_X}{\sqrt{1-\kappa^2}}
-
\frac{g'}{2}[
Y_{u_L}+Y_{u_R}
+ (Y_{u_R}-Y_{u_L})\gamma_5
]
\frac{\kappa}{\sqrt{1-\kappa^2}}
\Big] 
c_{BL}
\bar{u}\gamma_{\mu}Z'^{\mu}u
\nonumber\\
&&\hspace{0.5cm} 
-\Big[
\frac{g_XY_X}{\sqrt{1-\kappa^2}}
-
\frac{g'}{2}[Y_{d_L}+Y_{d_R}
+(Y_{d_R}-Y_{d_L})\gamma_5
]\frac{\kappa}{\sqrt{1-\kappa^2}}
\Big] 
c_{BL}
\bar{d}\gamma_{\mu}Z'^{\mu}d .
\end{eqnarray}
%%%%%%%%%%%%%%%%%%%%%%%%%%%%%%%%%%%%%%
General couplings of $Z\bar{f}f$
and $Z'\bar{f}f$ with arbitrary fermions
$f$ are given in Table.~\ref{couplingBL4}.

The components of Higgs kinetic for 
vector $B-L$ model can be written as 
follows:
%%%%%%%%%%%%%%%%%%%%%%%%%%%%%%%%%%
\begin{eqnarray}
\mathcal{L}_{K} 
\supset 
(D_{\mu}\Phi)^\dagger(D_{\mu}\Phi)
&\supset&
\frac{g^2v_{\Phi}c_{\theta}}{2}
hW^{\mp,\mu}W^{\pm}_{\mu}
+
\frac{gg'c_wv_{\Phi}}{2}
A^{\mu}W^{\mp}_{\mu}G^{\pm} 
\nonumber\\
&&
+\frac{igc_{\theta}}{2}
(\partial^{\mu}hW^+_{\mu}G^- 
-hW^+_{\mu}\partial^{\mu}G^-
-\partial^{\mu}hW^-_{\mu}G^+
+hW^-_{\mu}\partial^{\mu}G^+) 
\nonumber\\
&&
+
\frac{i(gs_W+g'c_W)}{2}
(A^{\mu}G^-\partial_{\mu}G^+
-A^{\mu}G^+\partial_{\mu}G^-) 
\nonumber\\
&&
+
i
\Big[
\frac{(gc_W-g's_W)}{2}c_{BL}
-s_{BL}
\frac{g'\; \kappa}
{2 \sqrt{1-\kappa^2}} 
\Big]
(Z^{\mu}G^-\partial_{\mu}G^+
-Z^{\mu}G^+\partial_{\mu}G^-) 
\nonumber\\
&&
+i
\Big[ 
\frac{(g's_W-gc_W)}{2}s_{BL}
- c_{BL}
\frac{g'\; \kappa}
{2\sqrt{1-\kappa^2}} 
\Big]
(Z'_{\mu}G^-\partial^{\mu}G^+
-Z'_{\mu}G^+\partial^{\mu}G^-) 
\nonumber\\
&&
-
\frac{gv_\Phi}{2}
\Big(
g's_Wc_{BL}
+ s_{BL}
\frac{g'\; \kappa}
{\sqrt{1-\kappa^2}}
\Big)
Z^{\mu}W^{\pm}_{\mu}G^{\mp} 
\nonumber\\
&&
+\frac{gv_\Phi}{2} 
\Big(
g's_Ws_{BL}
-
c_{BL}
\frac{g'\; \kappa}
{\sqrt{1-\kappa^2}} \Big) 
Z'^{\mu}W^{\pm}_{\mu}G^{\mp} 
\nonumber\\
&& 
+
\frac{v_{\Phi}c_{\theta}}{2}
\Big[
( gc_W+g's_W) c_{BL}
- s_{BL}g' 
\frac{\kappa}{\sqrt{1-\kappa^2} 
}
\Big]^2
hZ^{\mu}Z_{\mu} 
\nonumber\\
&&
+
\frac{v_\Phi{c_\theta}}{2}
\Big[(gc_W+g's_W)s_{BL}
- c_{BL}g' 
\frac{\kappa}{\sqrt{1-\kappa^2}}
\Big)^2
h{Z'}^{\mu}{Z'}_{\mu}
\nonumber\\
&&
-\frac{v_\Phi{c_\theta}}{2}
\Big[ (gc_W+g's_W) c_{BL}
+ s_{BL}g' 
\frac{\kappa}
{\sqrt{1-\kappa^2}}\Big]
\times
\nonumber\\
&& 
\hspace{0.8cm}
\times 
\Big[ 
(gc_W+g's_W) s_{BL}
- c_{BL}g'
\frac{\kappa}
{\sqrt{1-\kappa^2}}
\Big]
hZ^{\mu}{Z'}_{\mu}.
\end{eqnarray}
%%%%%%%%%%%%%%%%%%%%%%%%%%%%%%%%%%
In further, we also have 
\begin{eqnarray}
\mathcal{L}_{K} 
&\supset&   
(D_{\mu}\chi)^\dagger(D_{\mu}\chi)
\supset
g_X^2Y^2_Xv_{\chi}s_{\theta}\;
h X^{\mu}X_{\mu} 
\\
&=&
g_X^2Y^2_Xv_{\chi}s_{\theta}
\frac{s_{(2BL)}}{{1-\kappa^2}}
hZ^{\mu}{Z'}_{\mu} 
+
g_X^2Y^2_Xv_{\chi}s_{\theta}
\frac{s_{BL}^2}{{1-\kappa^2}}
hZ^{\mu}Z_{\mu} 
+ g_X^2Y^2_Xv_{\chi}s_{\theta}
\frac{c_{BL}^2}{{1-\kappa^2}}
h{Z'}^{\mu}{Z'}_{\mu}.
\nonumber
\end{eqnarray}
%%%%%%%%%%%%%%%%%%%%%%%%%%%%%%%%%%
where we have already used the relation
%%%%%%%%%%%%%%%%%%%%%%%%%%%%%%%%%%
\begin{align}
X^{\mu} = \frac{1}{\sqrt{1-\kappa^2}}\tilde{X}^\mu
=\frac{1}{\sqrt{1-\kappa^2}}
(Z^{\mu}s_{BL}+{Z'}^{\mu}c_{BL}).
\end{align}
%%%%%%%%%%%%%%%%%%%%%%%%%%%%%%%%%%
The Higgs potential is expanded and we collect
the coupling of $2v_{\chi}s_{\theta}$
%%%%%%%%%%%%%%%%%%%%%%%%%%%%%%%%%%
\begin{align}
\mathcal{V}(\Phi,\chi){\supset}
\Big[
\lambda_\Phi{v_\Phi}
\Big(c_\theta+\frac{v_\Phi}{v_{\chi}}s_\theta)
+2m_\Phi^2\frac{s_\theta}{v_\chi}
\Big]\; 
hG^+G^-
\end{align}
We have used the following relations:
%%%%%%%%%%%%%%%%%%%%%%%%%%%%%%%%%%%%%%%%%%
\begin{align}
\frac{\lambda_\Phi}{2} 
&=\frac{M_{H_2}^2}{4v_\Phi^2}(1-c_{2\theta})
+\frac{M_{H_1}^2}{4v_{\Phi}^2}(1+c_{2\theta}),
\\
\frac{\lambda_\chi}{2}
&=\frac{M_{H_1}^2}{4v_{\chi}^2}(1-c_{2\theta})
+\frac{M_{H_2}^2}{4v_{\chi}^2}(1+c_{2\theta}), 
\\
\lambda_{\Phi\chi}
&=s_{2\theta}\qty(\frac{M_{H_2}^2-M_{H_1}^2}
{2v_{\Phi}v_{\chi}}).
\end{align}
As a result, we arrive at
\begin{align}
&\lambda_\Phi{v_\Phi}\qty(c_\theta+\frac{v_\Phi}{v_{\chi}}s_\theta)+2m_\Phi^2\frac{s_\theta}{v_\chi}=\lambda_\Phi{v_\Phi}\qty(c_\theta+\frac{v_\Phi}{v_{\chi}}s_\theta)-(\lambda_\Phi{v_\Phi^2}+\lambda_{\Phi\chi}v_\chi^2)\frac{s_\theta}{v_\chi} \notag\\
&=\lambda_\Phi{v_\Phi}c_\theta-\lambda_{\Phi\chi}{v_\chi}s_\theta=\frac{M_{H_2}^2}{2v_\Phi}(1-c_{2\theta})c_{\theta}+\frac{M_{H_1}^2}{2v_{\Phi}}(1+c_{2\theta})c_\theta-s_{2\theta}\qty(\frac{M_{H_2}^2-M_{H_1}^2}{2v_{\Phi}}){s_\theta} \notag\\
&=\frac{M_{H_2}^2}{2v_\Phi}[(1-c_{2\theta})c_{\theta}-s_{2\theta}s_{\theta}]+\frac{M_{H_1}^2}{2v_{\Phi}}[(1+c_{2\theta})c_\theta+s_{2\theta}{s_\theta}] \notag\\
&=\frac{M_{H_2}^2}{2v_\Phi}[2c_{\theta}s_{\theta}^2-2c_{\theta}s_{\theta}^2]+\frac{M_{H_1}^2}{2v_{\Phi}}[(1+c_{2\theta})c_\theta+s_{2\theta}{s_\theta}] \notag\\
    &=\frac{M_{H_1}^2}{2v_{\Phi}}[2c_{\theta}^2+2{s_\theta}^2]c_\theta=\frac{M_{H_1}^2}{v_{\Phi}}c_\theta.
\end{align}
%%%%%%%%%%%%%%%%%%%%%%%%%%%%%%%%%%%
\subsection*{Chiral $B-L$ Model} %%
%%%%%%%%%%%%%%%%%%%%%%%%%%%%%%%%%%%
We are going to calculate all the couplings 
related to the processes in chiral $B-L$ model.

The fermion Lagrangian will be expanded as follow
\begin{eqnarray}
\mathcal{L}_{Zff}
&\supset& 
\bar{u}^i\gamma_{\mu}
\Big[-\frac{1}{4}(1-\gamma_5)
(gc_W+(1-2Q_{u_L})g' s_W ) 
+\frac{1}{2}(1+\gamma_5)g'c_WY_{u_R}
\Big] c_{BL}' \; u^i \; Z^{\mu} 
\nonumber\\
&& 
+ \bar{d}^i \gamma_{\mu}
\Big[ -\frac{1}{4}(1-\gamma_5)
(-gc_W-(2Q_{d_L} + 1)g' s_W ) + 
\frac{1}{2}(1+\gamma_5)g'c_W 
Y_{d_R})c_{BL}'d^i \; Z^{\mu}
\nonumber\\
&& 
+ s_{BL}' g_X Y_X^{u^i}
\bar{u}^i\gamma_{\mu} u^i \; Z^{\mu} 
+s_{BL}'g_X Y_X^{d^i} 
\bar{d}^i\gamma_{\mu}d^i
\; Z^{\mu}
\nonumber\\
&& 
+
\bar{e}^i \gamma_{\mu} \Big[ 
\frac{1}{4}
(1-\gamma_5)(gc_Wc_{BL}'+(2Q_{e_L}+1)g' s_W )
+\frac{1}{2}
(1+\gamma_5)g' Y_{e_R}c_W]c_{BL}' e^i  \; Z^{\mu} 
\nonumber\\
&& + s_{BL}' g_X Y_X^{e^i} \bar{e}^i\gamma_{\mu}e^i \; Z^{\mu}. 
\end{eqnarray}
% %%%%%%%%%%%%%%%%%%%%%%%%%%%%%%%%%%
\begin{eqnarray}
\mathcal{L}_{Z'ff}
&\supset&
\bar{u}^i\gamma_{\mu}
\Big[\frac{-1}{4}(1-\gamma_5)(gc_W+(1-2Q_{u_L})g's_W)
+\frac{1}{2}(1+\gamma_5)g's_WY_{u_R}
\Big] s_{BL}'u^i\; Z'^{\mu} 
\nonumber\\
&&
+ \bar{d}^i\gamma_{\mu}
\Big[\frac{-1}{4}(1-\gamma_5)(-gc_W-(2Q_{d_L}+1)g's_W)
+\frac{1}{2}(1+\gamma_5)g's_WY_{d_R}\Big]
s_{BL}'d^i \; Z'^{\mu}
\nonumber\\
&&-c_{BL}'g_X Y_X^{u^i} 
\bar{u}^i\gamma_{\mu}u^i \; Z'^{\mu}
-c_{BL}' g_X Y_X^{d^i} \bar{d}^i\gamma_{\mu}d^i 
\; Z'^{\mu}
\nonumber\\
&&
+
\bar{e}^i\gamma_{\mu}\Big[\frac{1}{4} (1-\gamma_5)
(gc_Ws_{BL}'+(2Q_{e_L}+1)g's_W )
+\frac{1}{2} (1+\gamma_5)g'Y_{e_R} s_W  \Big]s_{BL}' 
e^i \; Z'^{\mu}
\nonumber\\
&&
-c_{BL}' g_X Y_X^{e^i}  \bar{e}^i\gamma_{\mu}e^i \; Z'^{\mu}. 
\end{eqnarray}
%%%%%%%%%%%%%%%%%%%%%%%%%%%%%%%%%%%
General formulas for the 
couplings of $Z$ ($Z'$) with fermion-pair
are given in Table~\ref{couplingchiralBL4}.

The components of Higgs kinetic for chiral 
$B-L$ model can be written as follows:
%%%%%%%%%%%%%%%%%%%%%%%%%%%%%%%%%%
\begin{eqnarray}
\mathcal{L}_{K} &\supset& 
(D_{\mu}\Phi)^{\dagger}(D_{\mu}\Phi)
+(D_{\mu}\varphi)^{\dagger}(D_{\mu}\varphi)
\\
&\supset&
\frac{g^2}{2} (v_{\Phi}c_{13}c_{12} 
- v_{\varphi}c_{13}s_{12})
\; hW^{\pm,\mu}W^{\mp}_{\mu}
\nonumber\\
&&
-\frac{gg's_W}{2}
(
c_{\alpha}v_\Phi 
+
s_{\alpha}v_\varphi
)
s_{BL}'
\; Z'^{\mu}W^{\pm}_{\mu}G^{\mp}  
\nonumber\\
&& 
- \frac{g g's_W }{2} 
(c_{\alpha}v_\Phi
+s_{\alpha}v_\varphi)
c_{BL}' 
\; 
Z^{\mu}W^{\pm}_{\mu}G^{\mp}
\nonumber\\
&& + 
\frac{i(gs_W+g'c_W)}{2}
\; 
(A^{\mu}G^-\partial_{\mu}G^+ 
- A^{\mu}G^+\partial_{\mu}G^-) 
\nonumber\\
&& 
+
\frac{i(gc_W-g's_W )c_{BL}'}{2}
\; 
(Z^{\mu}G^-\partial_{\mu}G^+ 
- Z^{\mu}G^+\partial_{\mu}G^-) 
\nonumber\\
&& 
+ \frac{i(gc_W-g's_W)s_{BL}'}{2}
\; ( Z'^{\mu}G^-\partial_{\mu}G^+ 
- Z'^{\mu}G^+\partial_{\mu}G^-) 
\nonumber\\
&& 
+ \frac{i(gs_W+g'c_W)}{2}
\; (A^{\mu}H^-\partial_{\mu}H^+ 
- A^{\mu}H^+\partial_{\mu}H^-) 
\nonumber\\
&& 
+ \frac{i(gc_W-g'c_W ) c_{BL}'}{2}
\; 
(Z^{\mu}H^-\partial_{\mu}H^+ 
- Z^{\mu}H^+\partial_{\mu}H^-) 
\nonumber\\
&& 
+ \frac{i(gc_W-g's_W)s_{BL}'}{2}
\; 
( Z'^{\mu}H^-\partial_{\mu}H^+ 
-  Z'^{\mu}H^+\partial_{\mu}H^-) 
\nonumber\\
&& 
+\frac{(gc_W + g' s_W )^2
(v_\Phi{c_{13}c_{12}} 
- v_{\varphi}c_{13}s_{12}) 
c_{BL}'^2 }{4}
\; hZ^{\mu}Z_{\mu} 
\nonumber\\
&& 
+ \frac{(gc_W + g's_W)^2
(v_\Phi{c_{13}c_{12}} 
-v_{\varphi}c_{13}s_{12}) 
s_{BL}'^2 }{4}
\; h{Z'}^{\mu}{Z'}_{\mu} 
\nonumber\\
&& 
+ \frac{(gc_W+g's_W)^2
(v_\Phi{c_{12}}-v_{\varphi}s_{12})
c_{13}
s_{BL}'c_{BL}' }{2}
\; 
hZ^{\mu}Z'_{\mu} 
\nonumber\\
&& 
+\frac{igc_{13}
(c_{12}c_{\alpha}
-s_{12}s_{\alpha})
}{2} 
\; 
(hW^-_{\mu}\partial^{\mu}G^+ 
- hW^+_{\mu}\partial^{\mu}G^- 
-\partial^{\mu}hW^-_{\mu}G^+
+\partial^{\mu}hW^+_{\mu}G^-). 
% \nonumber\\
% && 
% -\frac{igc_{13}
% (c_{12}c_{\alpha}
% -s_{12}s_{\alpha})
% }{2}
% \;
% (hW^+_{\mu}\partial^{\mu}G^- 
% - hW^-_{\mu}\partial^{\mu}G^+ 
% + \partial^{\mu}hW^-_{\mu}G^+ 
% -\partial^{\mu}hW^+_{\mu}G^-).
% \nonumber
\end{eqnarray}
%%%%%%%%%%%%%%%%%%%%%%%%%%%%%%%%%%
\begin{eqnarray}
\mathcal{L}_{K} 
&\supset&     
(D_{\mu}\varphi)^{\dagger}(D_{\mu}\varphi)
\\
&\supset&
g_X^2Y_X^2X^{\mu}X_{\mu}\qty[G^{\pm}_2G^{\mp}_2 
+ \frac{1}{2}(v_\varphi+R_2+iI_2)(v_\varphi+R_2-iI_2)] \nonumber\\
&\supset&
\frac{1}{2}g_X^2Y^2_Xv_{\sigma}(-s_{BL}'Z^\mu+c_{BL}'{Z'}^{\mu})
(-s_{BL}'Z_\mu+c_{BL}'{Z'}_{\mu})(v_{\varphi}+R_2)^2 \nonumber\\
&\supset& 
g_X^2Y^2_Xv_{\sigma}(-s_{BL}'Z^\mu+c_{BL}'{Z'}^{\mu})
(-s_{BL}'Z_\mu+c_{BL}'{Z'}_{\mu})v_{\varphi}(-c_{13}s_{12})h
\nonumber\\
&=& 
-g_X^2Y^2_Xv_{\sigma}c_{13}s_{12}h(s_{BL}'^2 
Z^{\mu}Z_{\mu} 
+ c_{BL}'^2{Z'}^{\mu}{Z'}_{\mu} 
- s_{BL}'c_{BL}'Z^{\mu}{Z'}_{\mu}-s_{BL}'c_{BL}' 
\;{Z'}_{\mu}{Z}^{\mu}) \nonumber\\
&=& 
-g_X^2Y^2_Xv_{\sigma}c_{13}s_{12}s_{BL}'^2
\; hZ^{\mu}Z_{\mu}-g_X^2Y^2_Xv_{\sigma}c_{13}s_{12}c_{BL}'^2
\; h{Z'}^{\mu}{Z'}_{\mu} \nonumber\\
&&+2g_X^2Y^2_Xv_{\sigma}c_{13}s_{12}s_{BL}'c_{BL}'
\; h{Z}^{\mu}{Z'}_{\mu}. 
\nonumber
\end{eqnarray}
%%%%%%%%%%%%%%%%%%%%%%%%%%%%%%%%%%
Another term in the kinematic part
of the Higgs sector is expressed
as:
\begin{eqnarray}
\mathcal{L}_{K} 
&\supset&
(D_{\mu}\sigma)^{\dagger}(D_{\mu}\sigma)
\\
&\supset&
\frac{1}{2}g_X^2 Y_X^2 X^{\mu}X_{\mu} (2v_{\sigma}) R_3
=g_X^2 Y^2_X v_{\sigma}(-s_{BL}'Z^\mu+c_{BL}'{Z'}^{\mu})
(-s_{BL}'Z_\mu+c_{BL}'{Z'}_{\mu})s_{13}h \nonumber\\
&=& 
g_X^2 Y^2_X v_{\sigma} s_{13}h(s_{BL}'^2 Z^{\mu}Z_{\mu} 
+ c_{BL}'^2{Z'}^{\mu}{Z'}_{\mu} 
- s_{BL}'c_{BL}' Z^{\mu}{Z'}_{\mu} 
-s_{BL}'c_{BL}' {Z'}^{\mu}{Z}_{\mu}) 
\nonumber\\
&=& g_X^2 Y^2_X v_{\sigma} s_{13} s_{BL}'^2 
\; hZ^{\mu}Z_{\mu} 
+ g_X^2 Y^2_X v_{\sigma} s_{13} c_{BL}'^2 
\; h{Z'}^{\mu}{Z'}_{\mu}
-2 g_X^2 Y^2_X v_{\sigma} s_{13} s_{BL}' c_{BL}'
\; h{Z}^{\mu}{Z'}_{\mu}. 
\nonumber
\end{eqnarray}
%%%%%%%%%%%%%%%%%%%%%%%%%%%%%%%%%%
The Higgs potential is
% %%%%%%%%%%%%%%%%%%%%%%%%%%%%%%%%%%
\begin{eqnarray}
\mathcal{V}(\Phi,\varphi,\sigma,\chi_d)
&\supset&
\Big\{
\lambda_{\Phi}v_{\Phi}c_{13}c_{12}c_{\alpha}^2 
-\lambda_{\varphi}v_{\varphi}c_{13}s_{12}s_{\alpha}^2
+\frac{\mu}{\sqrt{2}}s_{13}s_{\alpha}c_{\alpha}
+\frac{\mu}{\sqrt{2}}s_{13}s_{\alpha}c_{\alpha} 
\nonumber\\
&&
-\lambda_{\Phi\sigma}v_{\sigma}c_{\alpha}^2s_{13} 
-\lambda_{\varphi\sigma}v_{\sigma}s_{\alpha}^2s_{13} 
+ \lambda_{\Phi\varphi_1}\qty[-c_{13}s_{12}c_{\alpha}^2v_{\varphi} 
+ c_{13}c_{12}s_{\alpha}^2v_{\Phi}] \nonumber\\
&& 
+\lambda_{\Phi\varphi_2}\qty(-v_{\Phi}c_{13}s_{12}
+v_{\varphi}c_{13}c_{12})s_{\alpha}c_{\alpha}
\Big\}\; G^-G^+h \nonumber\\
&&
+\Big\{
\lambda_{\Phi}v_{\Phi}c_{13}c_{12}s_{\alpha}^2 
-\lambda_{\varphi}v_{\varphi}c_{13}s_{12}c_{\alpha}^2 
-\frac{\mu}{\sqrt{2}}s_{13}s_{\alpha}c_{\alpha}
-\frac{\mu}{\sqrt{2}}s_{13}s_{\alpha}c_{\alpha} 
\nonumber\\
&& 
-\lambda_{\Phi\sigma}v_{\sigma}s_{\alpha}^2s_{13} 
-\lambda_{\varphi\sigma}v_{\sigma}c_{\alpha}^2s_{13} 
+\lambda_{\Phi\varphi_1}\qty[-c_{13}s_{12}s_{\alpha}^2v_{\varphi}
+c_{13}c_{12}c_{\alpha}^2v_{\Phi}] \nonumber\\
&&
-\lambda_{\Phi\varphi_2}\qty(-v_{\Phi}c_{13}s_{12}
+v_{\varphi}c_{13}c_{12})s_{\alpha}c_{\alpha}
\Big\} H^-H^+h.
\end{eqnarray}

The Yukawa Lagrangian will be expanded as follow
\begin{align}
-\mathcal{L}_Y&=Y_e\bar{L}\Phi{e_R}+Y_u\bar{Q}\tilde{\Phi}{u_R}+Y_e\bar{Q}\Phi{d_R}+Y_\nu\bar{L}\tilde{\varphi}{\nu_R}+h.c \nonumber\\
&{\supset} Y_\nu 
\left(\begin{array}{cc}
       \bar{\nu}_L  &  \bar{e}_L 
\end{array}\right)\left(\begin{array}{c}
         \frac{1}{\sqrt{2}}(v_\varphi+R_2+iI_2)  \\
         G_2^{\pm}
\end{array}\right)\nu_R 
+Y_\nu\bar{\nu}_R\left(\begin{array}{cc}
\frac{1}{\sqrt{2}}(v_\varphi+R_2+iI_2) & G_2^{\mp}
\end{array}\right)\left(\begin{array}{cc}
       \nu_L  \\
       e_L 
\end{array}\right) \nonumber\\
%     &=\frac{Y_{\nu}}{\sqrt{2}}(v_\varphi+R_2+iI_2)\bar{\nu}_L\nu_R+Y_{\nu}\bar{e}_LG_{2}^{\pm}\nu_R+\frac{Y_\nu}{\sqrt{2}}\bar{\nu}_R(v_{\varphi}+R_2-iI_2)\nu_L+Y_{\nu}\bar{\nu}_RG_2^{\mp}e_L \nonumber\\
%     &{\supset}\frac{Y_{\nu}v_\varphi}{\sqrt{2}}\bar{\nu}_L\nu_R+\frac{Y_{\nu}v_\varphi}{\sqrt{2}}\bar{\nu}_R\nu_L+Y_{\nu}\bar{e}P_R(G^{\pm}s_\alpha+H^{\pm}c_\alpha)\nu_R+Y_{\nu}\bar{\nu}_R(G^{\mp}s_\alpha+H^{\mp}c_\alpha)P_Le \nonumber\\
&=\frac{Y_{\nu}v_\varphi}{\sqrt{2}}\bar{\nu}_L\nu_R+\frac{Y_{\nu}v_\varphi}{\sqrt{2}}\bar{\nu}_R\nu_L+\frac{s_\alpha{Y}_{\nu}(1+\gamma_5)}{2}\bar{e}G^{\pm}\nu_R+\frac{c_\alpha{Y}_{\nu}(1+\gamma_5)}{2}\bar{e}H^{\pm}\nu_R \nonumber\\
&+\frac{s_\alpha{Y}_{\nu}(1-\gamma_5)}{2}\bar{\nu}_RG^{\mp}e+\frac{c_\alpha{Y}_{\nu}(1-\gamma_5)}{2}\bar{\nu}_RH^{\mp}e
\end{align}
%%%%%%%%%%%%%%%%%%%%%%%%%%%%%%%%%%

In the limits of 
$c_{13}, c_{23} \rightarrow 1$, mixing angle between 
$H_1=h$ and $H_2$ is considered as $c_{\widehat{12}}$, 
the coupling of $hH^+H^-$ is then taken the form of
\begin{align}
g_{hH^{\pm}H^{\mp}}
&=\lambda_{\Phi}v_{\Phi}c_{\widehat{12}}s_{\alpha}^2-\lambda_{\phi}v_{\phi}s_{\widehat{12}}c_{\alpha}^2 +\lambda_{\Phi\phi_1}\qty(-s_{\widehat{12}}s_{\alpha}^2v_{\phi}+c_{\widehat{12}}c_{\alpha}^2v_{\Phi})-\lambda_{\Phi\phi_2}\qty(-v_{\Phi}s_{\widehat{12}}+v_{\phi}c_{\widehat{12}})s_{\alpha}c_{\alpha} 
\\
&=\frac{c_{2(\widehat{12})}}{v_{\Phi}}\qty[M_h^2\qty(c_{\widehat{12}}^2+{t_{2(\widehat{12})}}s_{\widehat{12}}c_{\widehat{12}})-M_H^2\qty(s_{\widehat{12}}^2-{t_{2(\widehat{12})}}s_{\widehat{12}}c_{\widehat{12}})]c_{\widehat{12}}s_{\alpha}^2-\frac{\mu{v_\phi{v_\sigma}}}{\sqrt{2}v_\Phi^2}c_{\widehat{12}}s_{\alpha}^2 
\notag\\
&\quad +\frac{c_{2(\widehat{12})}}{v_\phi}\qty[M_h^2\qty(s_{\widehat{12}}^2-{t_{2(\widehat{12})}}s_{\widehat{12}}c_{\widehat{12}})-M_H^2\qty(c_{\widehat{12}}^2+{t_{2(\widehat{12})}}s_{\widehat{12}}c_{\widehat{12}})]s_{\widehat{12}}c_{\alpha}^2+\frac{\mu{v_\Phi{v_\sigma}}}{\sqrt{2}v_\phi^2}s_{\widehat{12}}c_{\alpha}^2 \notag\\
&\quad +\qty[(M_H^2-M_h^2)\frac{s_{2(\widehat{12})}}{2v_{\Phi}v_{\phi}}-\frac{\mu{v_\sigma}}{\sqrt{2}v_{\Phi}v_{\phi}}+\frac{2M_{H^\pm}^2}{v^2}]\qty(-s_{\widehat{12}}s_{\alpha}^2v_{\phi}+c_{\widehat{12}}c_{\alpha}^2v_{\Phi}) \notag\\
&\quad -\qty(\frac{\sqrt{2}\mu{v_\sigma}}{v_\Phi{v_\phi}}-\frac{2M_{H^\pm}^2}{v^2})\qty(-v_{\Phi}s_{\widehat{12}}+v_{\phi}c_{\widehat{12}})s_{\alpha}c_{\alpha} 
\\
&=\frac{1}{v}\qty[\frac{\sqrt{2}{\mu}v_{\sigma}}{s_{\alpha}c_{\alpha}}\cot(2\alpha)s_{\alpha+\widehat{12}}+\frac{M_h^2}{s_{\alpha}c_{\alpha}}s_{\alpha-\widehat{12}}]+\frac{1}{v}\qty(2M_{H^\pm}^2-\frac{\sqrt{2}{\mu}v_{\sigma}}{s_{\alpha}c_{\alpha}}-M_h^2)c_{\alpha+\widehat{12}}.
\end{align}
%%%%%%%%%%%%%%%%%%%%%%%%%%%%%%%%%%%%%%%%%%%%%%%
Where we have applied the following relations:
%%%%%%%%%%%%%%%%%%%%%%%%%%%%%%%%%%%%%%%%%%%%%%%
\begin{align}
M_{h}^2&=\lambda_\Phi{v_\Phi^2}c_{\widehat{12}}^2+\frac{\mu{v_\phi{v_\sigma}}}{\sqrt{2}v_\Phi}c_{\widehat{12}}^2-\qty(\lambda_{12}v_\Phi{v_\phi}-\frac{\mu{v_\sigma}}{\sqrt{2}})s_{2(\widehat{12})}+\lambda_\phi{v_\phi^2}s_{\widehat{12}}^2+\frac{\mu{v_\Phi{v_\sigma}}}{\sqrt{2}v_\phi}s_{\widehat{12}}^2, 
\\
M_{H}^2
&=\lambda_\Phi{v_\Phi^2}s_{\widehat{12}}^2+\frac{\mu{v_\phi{v_\sigma}}}{\sqrt{2}v_\Phi}s_{\widehat{12}}^2+\qty(\lambda_{12}v_\Phi{v_\phi}-\frac{\mu{v_\sigma}}{\sqrt{2}})s_{2(\widehat{12})}+\lambda_\phi{v_\phi^2}c_{\widehat{12}}^2+\frac{\mu{v_\Phi{v_\sigma}}}{\sqrt{2}v_\phi}c_{\widehat{12}}^2
\end{align}
%%%%%%%%%%%%%%%%%%%%%%%%%%%%%%%%%%%%%%%%%%%%%
and 
\begin{align}
-v_{\phi}^2\lambda_{\phi} 
&={c_{2(\widehat{12})}}\qty[M_h^2\qty(s_{\widehat{12}}^2-{t_{2(\widehat{12})}}s_{\widehat{12}}c_{\widehat{12}})-M_H^2\qty(c_{\widehat{12}}^2+{t_{2(\widehat{12})}}s_{\widehat{12}}c_{\widehat{12}})]+
\frac{\mu{v_\Phi{v_\sigma}}}{\sqrt{2}v_\phi}, 
\\
v_{\Phi}^2\lambda_{\Phi}
&={c_{2(\widehat{12})}}\qty[M_h^2\qty(c_{\widehat{12}}^2+{t_{2(\widehat{12})}}s_{\widehat{12}}c_{\widehat{12}})-M_H^2\qty(s_{\widehat{12}}^2-{t_{2(\widehat{12})}}s_{\widehat{12}}c_{\widehat{12}})]
-\frac{\mu{v_\phi{v_\sigma}}}{\sqrt{2}v_\Phi}, 
\\
\lambda_{\Phi\phi_1}
&=\qty[M_H^2-M_h^2]\frac{s_{2(\widehat{12})}}{2v_{\Phi}v_{\phi}}-\frac{\mu{v_\sigma}}{\sqrt{2}v_{\Phi}v_{\phi}}+\frac{2M_{H^\pm}^2}{v^2}
.
\end{align}
%%%%%%%%%%%%%%%%%%%%%%%%%%%%%%

%%%%%%%%%%%%%%%%%%%%%%%%%%%%%%

\begin{thebibliography}{100}%%
%%%%%%%%%%%%%%%%%%%%%%%%%%%%%%
%%%%%%%%%%%%%%%%%%%%%%%%%%%%%%%%%%%%%%%%%%
%% Experimental data HL LHC
%%%%%%%%%%%%%%%%%%%%%%%%%%%%%%%%%%%%%%%%%%
%\cite{Liss:2013hbb}
\bibitem{Liss:2013hbb}
A.~Liss \textit{et al.} [ATLAS],
%``Physics at a High-Luminosity LHC with ATLAS,''
[arXiv:1307.7292 [hep-ex]].
%%%%%%%%%%%%%%%
%\cite{CMS:2013xfa}
\bibitem{CMS:2013xfa}
[CMS],
%``Projected Performance of an Upgraded CMS Detector at the LHC and HL-LHC: Contribution to the Snowmass Process,''
[arXiv:1307.7135 [hep-ex]].
%%%%%%%%%%%%%%%
%\cite{Baer:2013cma}
\bibitem{Baer:2013cma}
H.~Baer, T.~Barklow, K.~Fujii, Y.~Gao, A.~Hoang, S.~Kanemura, J.~List, H.~E.~Logan, A.~Nomerotski and M.~Perelstein, \textit{et al.}
%``The International Linear Collider Technical Design Report - Volume 2: Physics,''
[arXiv:1306.6352 [hep-ph]].
%%%%%%%%%%%%%%%%%%%%%%%%%%%%
%% Higgs couling
%\cite{ATLAS:2016neq}
\bibitem{ATLAS:2016neq}
G.~Aad \textit{et al.} [ATLAS and CMS],
%``Measurements of the Higgs boson production and decay rates and constraints on its couplings from a combined ATLAS and CMS analysis of the LHC pp collision data at $ \sqrt{s}=7 $ and 8 TeV,''
JHEP \textbf{08} (2016), 045
doi:10.1007/JHEP08(2016)045
[arXiv:1606.02266 [hep-ex]].

%\cite{ATLAS:2019nkf}
\bibitem{ATLAS:2019nkf}
G.~Aad \textit{et al.} [ATLAS],
%``Combined measurements of Higgs boson production and decay using up to $80$ fb$^{-1}$ of proton-proton collision data at $\sqrt{s}=$ 13 TeV collected with the ATLAS experiment,''
Phys. Rev. D \textbf{101} (2020) no.1, 012002
doi:10.1103/PhysRevD.101.012002
[arXiv:1909.02845 [hep-ex]].

%%%%%%%%%%%%%%%%%%%%%%%%%%%%%
%% LHC gamma gamma, Z gamma 
%%%%%%%%%%%%%%%%%%%%%%%%%%%%
%\cite{CMS:2014fzn}
\bibitem{CMS:2014fzn}
V.~Khachatryan \textit{et al.} [CMS],
%``Precise determination of the mass of the Higgs boson and tests of compatibility of its couplings with the standard model predictions using proton collisions at 7 and 8 $\,\text {TeV}$,''
Eur. Phys. J. C \textbf{75} (2015) no.5, 212
doi:10.1140/epjc/s10052-015-3351-7
[arXiv:1412.8662 [hep-ex]].
%%%%%%%%%%%%%%%%%
%\cite{ATLAS:2015egz}
\bibitem{ATLAS:2015egz}
G.~Aad \textit{et al.} [ATLAS],
%``Measurements of the Higgs boson production and decay rates and coupling strengths using pp collision data at $\sqrt{s}=7$ and 8 TeV in the ATLAS experiment,''
Eur. Phys. J. C \textbf{76} (2016) no.1, 6
doi:10.1140/epjc/s10052-015-3769-y
[arXiv:1507.04548 [hep-ex]].
%\cite{CMS:2021kom}
\bibitem{CMS:2021kom}
A.~M.~Sirunyan \textit{et al.} [CMS],
%``Measurements of Higgs boson production cross sections and couplings in the diphoton decay channel at $ \sqrt{\mathrm{s}} $ = 13 TeV,''
JHEP \textbf{07} (2021), 027
doi:10.1007/JHEP07(2021)027
[arXiv:2103.06956 [hep-ex]].
%%%%%%%%%%%
%\cite{D0:2008swt}
\bibitem{D0:2008swt}
V.~M.~Abazov \textit{et al.} [D0],
%``Search for a scalar or vector particle decaying into $Z \gamma$ in $p \bar{p}$ collisions at $\sqrt{s}$=1.96 TeV,''
Phys. Lett. B \textbf{671} (2009), 349-355
doi:10.1016/j.physletb.2008.12.009
[arXiv:0806.0611 [hep-ex]].
%%%%%%%
%\cite{CMS:2013rmy}
\bibitem{CMS:2013rmy}
S.~Chatrchyan \textit{et al.} [CMS],
%``Search for a Higgs Boson Decaying into a Z and a Photon in $pp$ Collisions at $\sqrt{s}$ = 7 and 8 TeV,''
Phys. Lett. B \textbf{726} (2013), 587-609
doi:10.1016/j.physletb.2013.09.057
[arXiv:1307.5515 [hep-ex]].
%%%%%%
%\cite{ATLAS:2017zdf}
\bibitem{ATLAS:2017zdf}
M.~Aaboud \textit{et al.} [ATLAS],
%``Searches for the $Z\gamma$ decay mode of the Higgs boson and for new high-mass resonances in $pp$ collisions at $\sqrt{s} = 13$ TeV with the ATLAS detector,''
JHEP \textbf{10} (2017), 112
doi:10.1007/JHEP10(2017)112
[arXiv:1708.00212 [hep-ex]].
%%%%%%%%
%\cite{ATLAS:2020qcv}
\bibitem{ATLAS:2020qcv}
G.~Aad \textit{et al.} [ATLAS],
%``A search for the $Z\gamma$ decay mode of the Higgs boson in $pp$ collisions at $\sqrt{s}$ = 13 TeV with the ATLAS detector,''
Phys. Lett. B \textbf{809} (2020), 135754
doi:10.1016/j.physletb.2020.135754
[arXiv:2005.05382 [hep-ex]].
%%%%%%%%%%%%%%%%%%%%%%%%%%
%%% ll gamma   %%%%%%%%%%%
%%%%%%%%%%%%%%%%%%%%%%%%%%
%\cite{CMS:2015tzs}
\bibitem{CMS:2015tzs}
V.~Khachatryan \textit{et al.} [CMS],
%``Search for a Higgs boson decaying into $\gamma^* \gamma \to \ell \ell \gamma$ with low dilepton mass in pp collisions at $\sqrt s = $ 8 TeV,''
Phys. Lett. B \textbf{753} (2016), 341-362
doi:10.1016/j.physletb.2015.12.039
[arXiv:1507.03031 [hep-ex]].
%%%%%%%%%%
%\cite{CMS:2017dyb}
\bibitem{CMS:2017dyb}
A.~M.~Sirunyan \textit{et al.} [CMS],
%``Search for Z$\gamma$ resonances using leptonic and hadronic final states in proton-proton collisions at $\sqrt{s}=$ 13 TeV,''
JHEP \textbf{09} (2018), 148
doi:10.1007/JHEP09(2018)148
[arXiv:1712.03143 [hep-ex]].
%%%%%%
%\cite{CMS:2018myz}
\bibitem{CMS:2018myz}
A.~M.~Sirunyan \textit{et al.} [CMS],
%``Search for the decay of a Higgs boson in the $\ell\ell\gamma$ channel in proton-proton collisions at $\sqrt{s} =$ 13 TeV,''
JHEP \textbf{11} (2018), 152
doi:10.1007/JHEP11(2018)152
[arXiv:1806.05996 [hep-ex]].
%%%%%%%
%\cite{ATLAS:2021wwb}
\bibitem{ATLAS:2021wwb}
G.~Aad \textit{et al.} [ATLAS],
%``Evidence for Higgs boson decays to a low-mass dilepton system and a photon in pp collisions at s=13 TeV with the ATLAS detector,''
Phys. Lett. B \textbf{819} (2021), 136412
doi:10.1016/j.physletb.2021.136412
[arXiv:2103.10322 [hep-ex]].

%%%%% Phenomenological %%%%%%%%%%%%%%%%%%%%%%
%\cite{Chen:2012ju}
\bibitem{Chen:2012ju}
L.~B.~Chen, C.~F.~Qiao and R.~L.~Zhu,
%``Reconstructing the 125 GeV SM Higgs Boson Through $\ell\bar{\ell}\gamma$,''
Phys. Lett. B \textbf{726} (2013), 306-311
[erratum: Phys. Lett. B \textbf{808} (2020), 135629]
doi:10.1016/j.physletb.2013.08.050
[arXiv:1211.6058 [hep-ph]].
%%%%%%%%%%%%
%\cite{Gainer:2011aa}
\bibitem{Gainer:2011aa}
J.~S.~Gainer, W.~Y.~Keung, I.~Low and P.~Schwaller,
%``Looking for a light Higgs boson in the $Z \gamma \to \ell \ell \gamma$ channel,''
Phys. Rev. D \textbf{86} (2012), 033010
doi:10.1103/PhysRevD.86.033010
[arXiv:1112.1405 [hep-ph]].
%%%%%%%%%%%%% 
%\cite{Korchin:2014kha}
\bibitem{Korchin:2014kha}
A.~Y.~Korchin and V.~A.~Kovalchuk,
%``Angular distribution and forward\textendash{}backward asymmetry of the Higgs-boson decay to photon and lepton pair,''
Eur. Phys. J. C \textbf{74} (2014) no.11, 3141
doi:10.1140/epjc/s10052-014-3141-7
[arXiv:1408.0342 [hep-ph]].
%%%%%%%%%%%%%%%%%%%%%%%%%%%%%%%%%%%%%%%%%%
%% Theory calculations in SM, BSM  
%%%%%%%%%%%%%%%%%%%%%%%%%%%%%%%%%%%%%%%%%%
%\cite{Abbasabadi:2000pb}
\bibitem{Abbasabadi:2000pb}
A.~Abbasabadi and W.~W.~Repko,
%``Higgs boson decay to muon anti-muon gamma,''
Phys. Rev. D \textbf{62} (2000), 054025
doi:10.1103/PhysRevD.62.054025
[arXiv:hep-ph/0004167 [hep-ph]].
%%%%%%%%%%%%%%%%
%%%%%%%%%
%\cite{Dicus:2013ycd}
\bibitem{Dicus:2013ycd}
D.~A.~Dicus and W.~W.~Repko,
%``Calculation of the decay $H\to e\bar{e}\gamma$,''
Phys. Rev. D \textbf{87} (2013) no.7, 077301
doi:10.1103/PhysRevD.87.077301
[arXiv:1302.2159 [hep-ph]].
%%%%%%%%% 
%\cite{Sun:2013rqa}
\bibitem{Sun:2013rqa}
Y.~Sun, H.~R.~Chang and D.~N.~Gao,
%``Higgs decays to gamma l+ l- in the standard model,''
JHEP \textbf{05} (2013), 061
doi:10.1007/JHEP05(2013)061
[arXiv:1303.2230 [hep-ph]].
%%%%%%%%%
%\cite{Passarino:2013nka}
\bibitem{Passarino:2013nka}
G.~Passarino,
%``Higgs Boson Production and Decay: Dalitz Sector,''
Phys. Lett. B \textbf{727} (2013), 424-431
doi:10.1016/j.physletb.2013.10.052
[arXiv:1308.0422 [hep-ph]].
%%%%%%%%% 
%\cite{Dicus:2013lta}
\bibitem{Dicus:2013lta}
D.~A.~Dicus, C.~Kao and W.~W.~Repko,
%``Comparison of $H\to\ell\bar{\ell}\gamma$ and $H\to\gamma\,Z, Z\to\ell\bar{\ell}$ including the ATLAS cuts,''
Phys. Rev. D \textbf{89} (2014) no.3, 033013
doi:10.1103/PhysRevD.89.033013
[arXiv:1310.4380 [hep-ph]].
%%%%%%%%%%
%\cite{Kachanovich:2020xyg}
\bibitem{Kachanovich:2020xyg}
A.~Kachanovich, U.~Nierste and I.~Ni\v{s}and\v{z}i\'c,
%``Higgs boson decay into a lepton pair and a photon revisited,''
Phys. Rev. D \textbf{101} (2020) no.7, 073003
doi:10.1103/PhysRevD.101.073003
[arXiv:2001.06516 [hep-ph]].
%\cite{Kachanovich:2021pvx}
\bibitem{Kachanovich:2021pvx}
A.~Kachanovich, U.~Nierste and I.~Ni\v{s}and\v{z}i\'c,
%``Higgs boson decay into a lepton pair and a photon: A roadmap to the discovery of H\textrightarrow{}Z\ensuremath{\gamma} and probes of new physics,''
Phys. Rev. D \textbf{105} (2022) no.1, 013007
doi:10.1103/PhysRevD.105.013007
[arXiv:2109.04426 [hep-ph]].
%%%%%%%%%%%%%%%%%%%%%%%%%%
%\cite{Ahmed:2023vyl}
\bibitem{Ahmed:2023vyl}
I.~Ahmed, U.~Hasan, S.~Iqbal, M.~Junaid, B.~Tariq and A.~Uzair,
%``Analysis of final state lepton polarization-dependent observables in $H\to \ell^{+}\ell^{-} \gamma$ in the SM at loop level,''
[arXiv:2309.07448 [hep-ph]].
%%%%%%%%%%%%%%%%%%%%%%%%%%%%%%%%%%%%%%%%%%%%%%%%%%
%% calculation in BSM   %%%%%%%%%%%%%%%%%%%%%%%%%%%
%%%%%%%%%%%%%%%%%%%%%%%%%%%%%%%%%%%%%%%%%%%%%%%%%%
%\cite{Li:1998rp}
\bibitem{Li:1998rp}
C.~S.~Li, C.~F.~Qiao and S.~H.~Zhu,
%``Radiative Higgs boson decays H ---\ensuremath{>} f anti-f gamma beyond the standard model,''
Phys. Rev. D \textbf{57} (1998), 6928-6933
doi:10.1103/PhysRevD.57.6928
[arXiv:hep-ph/9801334 [hep-ph]].
%%%%%%%
%\cite{Sasaki:2017fvk}
\bibitem{Sasaki:2017fvk}
K.~Sasaki and T.~Uematsu,
%``CP-odd Higgs boson production in $e\gamma$ collisions,''
Phys. Lett. B \textbf{781} (2018), 290-294
doi:10.1016/j.physletb.2018.04.005
[arXiv:1712.00197 [hep-ph]].

%%%%%%%%%
%\cite{Phan:2021xwc}
\bibitem{Phan:2021xwc}
K.~H.~Phan, L.~Hue and D.~T.~Tran,
%``General one-loop contributions to the decay $H\rightarrow \nu_l\bar{\nu}_l\gamma$,''
PTEP \textbf{2021} (2021) no.10, 103B07
doi:10.1093/ptep/ptab121
[arXiv:2106.14466 [hep-ph]].
%%%%%%%%%
%\cite{VanOn:2021myp}
\bibitem{VanOn:2021myp}
V.~Van On, D.~T.~Tran, C.~L.~Nguyen and K.~H.~Phan,
%``General one-loop formulas for $H\rightarrow f\bar{f}\gamma $ and its applications,''
Eur. Phys. J. C \textbf{82} (2022) no.3, 277
doi:10.1140/epjc/s10052-022-10225-z
[arXiv:2111.07708 [hep-ph]].
%%%%%%%%%
%\cite{Hue:2023tdz}
\bibitem{Hue:2023tdz}
L.~T.~Hue, D.~T.~Tran, T.~H.~Nguyen and K.~H.~Phan,
%``One-loop expressions for h \textrightarrow{} l\,l̅\ensuremath{\gamma} in Higgs extensions of the Standard Model,''
PTEP \textbf{2023} (2023) no.8, 083B06
doi:10.1093/ptep/ptad106
[arXiv:2305.04002 [hep-ph]]. 
%%%%%%%%%%%%%%%%%%%%%%%%%%%%%%%
%% other BSMs
%%%%%%%%%%%%%%%%%%%%%%%%%%%%%%%
%%%% Minimal B-L version %%%%%%
%%%%%%%%%%%%%%%%%%%%%%%%%%%%%%%
%\cite{Mandal:2023oyh}
\bibitem{Mandal:2023oyh}
S.~Mandal, H.~Prajapati and R.~Srivastava,
%``$B-L$ model in light of the CDF II result,''
[arXiv:2301.01522 [hep-ph]].
%\cite{Basso:2008iv}
\bibitem{Basso:2008iv}
L.~Basso, A.~Belyaev, S.~Moretti and C.~H.~Shepherd-Themistocleous,
%``Phenomenology of the minimal B-L extension of the Standard model: Z' and neutrinos,''
Phys. Rev. D \textbf{80} (2009), 055030
doi:10.1103/PhysRevD.80.055030
[arXiv:0812.4313 [hep-ph]].
%%%%%%%%%%%%%%%%%%%%%%%%%%%%%%%
%\cite{Basso:2009hf}
\bibitem{Basso:2009hf}
L.~Basso, A.~Belyaev, S.~Moretti and G.~M.~Pruna,
%``Probing the Z-prime sector of the minimal B-L model at future Linear Colliders in the e+ e- ---\ensuremath{>} mu+ mu- process,''
JHEP \textbf{10} (2009), 006
doi:10.1088/1126-6708/2009/10/006
[arXiv:0903.4777 [hep-ph]].
%%%%%%%%%%%%%%%%%
%\cite{Basso:2010jt}
\bibitem{Basso:2010jt}
L.~Basso, A.~Belyaev, S.~Moretti and G.~M.~Pruna,
%``Tree Level Unitarity Bounds for the Minimal B-L Model,''
Phys. Rev. D \textbf{81} (2010), 095018
doi:10.1103/PhysRevD.81.095018
[arXiv:1002.1939 [hep-ph]].
%%%%%%%%%%%%%%%%%%
%\cite{Basso:2010pe}
\bibitem{Basso:2010pe}
L.~Basso, A.~Belyaev, S.~Moretti, G.~M.~Pruna and C.~H.~Shepherd-Themistocleous,
%``$Z'$ discovery potential at the LHC in the minimal $B-L$ extension of the Standard Model,''
Eur. Phys. J. C \textbf{71} (2011), 1613
doi:10.1140/epjc/s10052-011-1613-6
[arXiv:1002.3586 [hep-ph]].
%%%%%%%%%%%%%%%%
%\cite{Basso:2010jm}
\bibitem{Basso:2010jm}
L.~Basso, S.~Moretti and G.~M.~Pruna,
%``A Renormalisation Group Equation Study of the Scalar Sector of the Minimal B-L Extension of the Standard Model,''
Phys. Rev. D \textbf{82} (2010), 055018
doi:10.1103/PhysRevD.82.055018
[arXiv:1004.3039 [hep-ph]].
%%%%%%%%%%%%%%%%%%%%
%\cite{Basso:2010hk}
\bibitem{Basso:2010hk}
L.~Basso, S.~Moretti and G.~M.~Pruna,
%``Constraining the $g'_1$ coupling in the minimal $B-L$ Model,''
J. Phys. G \textbf{39} (2012), 025004
doi:10.1088/0954-3899/39/2/025004
[arXiv:1009.4164 [hep-ph]].
%%%%%%%%%%%%%%%%%%%%%
%\cite{Basso:2010yz}
\bibitem{Basso:2010yz}
L.~Basso, S.~Moretti and G.~M.~Pruna,
%``Phenomenology of the minimal $B-L$ extension of the Standard Model: the Higgs sector,''
Phys. Rev. D \textbf{83} (2011), 055014
doi:10.1103/PhysRevD.83.055014
[arXiv:1011.2612 [hep-ph]].
%%%%%%%%%%%%%%%%%
%\cite{Basso:2010si}
\bibitem{Basso:2010si}
L.~Basso, S.~Moretti and G.~M.~Pruna,
%``The Higgs sector of the minimal $B-L$ model at future Linear Colliders,''
Eur. Phys. J. C \textbf{71} (2011), 1724
doi:10.1140/epjc/s10052-011-1724-0
[arXiv:1012.0167 [hep-ph]].
%%%%%%%%%%%%%%%%
%\cite{Basso:2011na}
\bibitem{Basso:2011na}
L.~Basso, S.~Moretti and G.~M.~Pruna,
%``Theoretical constraints on the couplings of non-exotic minimal $Z'$ bosons,''
JHEP \textbf{08} (2011), 122
doi:10.1007/JHEP08(2011)122
[arXiv:1106.4762 [hep-ph]].
%%%%%%%%%%%%%%%%%%
\bibitem{Pati:1974yy}
J.~C.~Pati and A.~Salam,
%``Lepton Number as the Fourth Color,''
Phys. Rev. D \textbf{10}, 275-289 (1974)
[erratum: Phys. Rev. D \textbf{11}, 703-703 (1975)].
%doi:10.1103/PhysRevD.10.275

\bibitem{Mohapatra:1974gc}
R.~N.~Mohapatra and J.~C.~Pati,
%``A Natural Left-Right Symmetry,''
Phys. Rev. D \textbf{11}, 2558 (1975).
%doi:10.1103/PhysRevD.11.2558
%%%%%%%%%%%%%%%%%

\bibitem{Senjanovic:1975rk}
G.~Senjanovic and R.~N.~Mohapatra,
%``Exact Left-Right Symmetry and Spontaneous Violation of Parity,''
Phys. Rev. D \textbf{12}, 1502 (1975).
%doi:10.1103/PhysRevD.12.1502

\bibitem{Singer:1980sw}
M.~Singer, J.~W.~F.~Valle and J.~Schechter,
%``Canonical Neutral Current Predictions From the Weak Electromagnetic Gauge Group SU(3) X $u$(1),''
Phys. Rev. D \textbf{22}, 738 (1980).
%doi:10.1103/PhysRevD.22.738

\bibitem{Valle:1983dk}
J.~W.~F.~Valle and M.~Singer,
%``Lepton Number Violation With Quasi Dirac Neutrinos,''
Phys. Rev. D \textbf{28}, 540 (1983).
%doi:10.1103/PhysRevD.28.540

\bibitem{Pisano:1991ee}
F.~Pisano and V.~Pleitez,
%``An SU(3) x U(1) model for electroweak interactions,''
Phys. Rev. D \textbf{46}, 410-417 (1992).
%doi:10.1103/PhysRevD.46.410

\bibitem{Frampton:1992wt}
P.~H.~Frampton,
%``Chiral dilepton model and the flavor question,''
Phys. Rev. Lett. \textbf{69}, 2889-2891 (1992).
%doi:10.1103/PhysRevLett.69.2889
\bibitem{Diaz:2004fs}
R.~A.~Diaz, R.~Martinez and F.~Ochoa,
%``SU(3)(c) x SU(3)(L) x U(1)(X) models for beta arbitrary and families with mirror fermions,''
Phys. Rev. D \textbf{72}, 035018 (2005).
%
\bibitem{Fonseca:2016tbn}
R.~M.~Fonseca and M.~Hirsch,
%``A flipped 331 model,''
JHEP \textbf{08} (2016), 003
%doi:10.1007/JHEP08(2016)003
%[arXiv:1606.01109 [hep-ph]].
%%%%%%%%%%%
\bibitem{Foot:1994ym}
R.~Foot, H.~N.~Long and T.~A.~Tran,
%``$SU(3)_L \otimes U(1)_N$ and $SU(4)_L \otimes U(1)_N$ gauge models with right-handed neutrinos,''
Phys. Rev. D \textbf{50}, no.1, R34-R38 (1994).
%--------------
%
\bibitem{Sanchez:2004uf}
L.~A.~Sanchez, F.~A.~Perez and W.~A.~Ponce,
%``SU(3) ($c$) x SU(4) (L) x U(1) ($X$) model for three families,''
Eur. Phys. J. C \textbf{35} (2004), 259-265
[arXiv:hep-ph/0404005 [hep-ph]].
%%%%%%%%%%%%%%%%%%%%%%
\bibitem{Ponce:2006vw}
W.~A.~Ponce and L.~A.~Sanchez,
%``Systematic study of the SU(3)(c) X SU(4)(L) X U(1)(X) gauge symmetry,''
Mod. Phys. Lett. A \textbf{22} (2007), 435-448
%doi:10.1142/S0217732307021275
[arXiv:hep-ph/0607175 [hep-ph]].
%%%%%%%%%%%%%%%%%%%
\bibitem{Riazuddin:2008yx}
Riazuddin and Fayyazuddin,
%``SU(L)(4) x U(1) model for electroweak unification and sterile neutrinos,''
Eur. Phys. J. C \textbf{56} (2008), 389-394
%doi:10.1140/epjc/s10052-008-0665-8
[arXiv:0803.4267 [hep-ph]].
%%%%%%%%%%%%%%%%%%
\bibitem{Jaramillo:2011qu}
A.~Jaramillo and L.~A.~Sanchez,
%``FCNC, CP violation and implications for some rare decays in an $SU(4)_L \otimes U(1)_x$ extension of the standard model,''
Phys. Rev. D \textbf{84} (2011), 115001
%doi:10.1103/PhysRevD.84.115001
[arXiv:1110.3363 [hep-ph]].
%%%%%%%%%%%%%%%%%%%
\bibitem{Long:2016lmj}
H.~N.~Long, L.~T.~Hue and D.~V.~Loi,
%``Electroweak theory based on SU(4)$_L\otimes$U(1)$_X$ gauge group,''
Phys. Rev. D \textbf{94} (2016) no.1, 015007
%doi:10.1103/PhysRevD.94.015007
[arXiv:1605.07835 [hep-ph]].
%%%%%%%%%%%%%%%%%%%%%%%%%
%%% H\gamnma at e-e+   %%
%%%%%%%%%%%%%%%%%%%%%%%%%
%% SM
%\cite{Abbasabadi:1995rc}
\bibitem{Abbasabadi:1995rc}
A.~Abbasabadi, D.~Bowser-Chao, D.~A.~Dicus and W.~W.~Repko,
%``Higgs photon associated production at $e \bar{e}$ colliders,''
Phys. Rev. D \textbf{52} (1995), 3919-3928
doi:10.1103/PhysRevD.52.3919
[arXiv:hep-ph/9507463 [hep-ph]].
%%%%%%%%%%%%%%%%
%\cite{Djouadi:1996ws}
\bibitem{Djouadi:1996ws}
A.~Djouadi, V.~Driesen, W.~Hollik and J.~Rosiek,
%``Associated production of Higgs bosons and a photon in high-energy e+ e- collisions,''
Nucl. Phys. B \textbf{491} (1997), 68-102
doi:10.1016/S0550-3213(96)00711-0
[arXiv:hep-ph/9609420 [hep-ph]].
%%%%%%%%%%%%%%%
%\cite{Abbasabadi:1997zr}
\bibitem{Abbasabadi:1997zr}
A.~Abbasabadi, D.~Bowser-Chao, D.~A.~Dicus and W.~W.~Repko,
%``Higgs - photon associated production at hadron colliders,''
Phys. Rev. D \textbf{58} (1998), 057301
doi:10.1103/PhysRevD.58.057301
[arXiv:hep-ph/9706335 [hep-ph]].
%%% HESM
%\cite{Arhrib:2014pva}
\bibitem{Arhrib:2014pva}
A.~Arhrib, R.~Benbrik and T.~C.~Yuan,
%``Associated Production of Higgs at Linear Collider in the Inert Higgs Doublet Model,''
Eur. Phys. J. C \textbf{74} (2014), 2892
doi:10.1140/epjc/s10052-014-2892-5
[arXiv:1401.6698 [hep-ph]].
%%%%%%%%%%
%\cite{Rahili:2019ixf}
\bibitem{Rahili:2019ixf}
L.~Rahili, A.~Arhrib and R.~Benbrik,
%``Associated production of SM Higgs with a photon in type-II seesaw models at the ILC,''
Eur. Phys. J. C \textbf{79} (2019) no.11, 940
doi:10.1140/epjc/s10052-019-7471-3
[arXiv:1909.07793 [hep-ph]].
%%%%%%%%%%%%%%
%\cite{Kanemura:2018esc}
\bibitem{Kanemura:2018esc}
S.~Kanemura, K.~Mawatari and K.~Sakurai,
%``Single Higgs production in association with a photon at electron-positron colliders in extended Higgs models,''
Phys. Rev. D \textbf{99} (2019) no.3, 035023
doi:10.1103/PhysRevD.99.035023
[arXiv:1808.10268 [hep-ph]].
%% MSSM
%\cite{Demirci:2019ush}
\bibitem{Demirci:2019ush}
M.~Demirci,
%``Associated production of Higgs boson with a photon at electron-positron colliders,''
Phys. Rev. D \textbf{100} (2019) no.7, 075006
doi:10.1103/PhysRevD.100.075006
[arXiv:1905.09363 [hep-ph]].
%%%%%%%%%%%%%%%%%%%%%%
%% Tools%%%%%%%%%%%%%%
%\cite{Hahn:2000kx}
\bibitem{Hahn:2000kx}
T.~Hahn,
%``Generating Feynman diagrams and amplitudes with FeynArts 3,''
Comput. Phys. Commun. \textbf{140} (2001), 418-431
doi:10.1016/S0010-4655(01)00290-9
[arXiv:hep-ph/0012260 [hep-ph]].

%\cite{Mertig:1990an}
\bibitem{Mertig:1990an}
R.~Mertig, M.~Bohm and A.~Denner,
%``FEYN CALC: Computer algebraic calculation of Feynman amplitudes,''
Comput. Phys. Commun. \textbf{64} (1991), 345-359
doi:10.1016/0010-4655(91)90130-D

%\cite{Hahn:1998yk}
\bibitem{Hahn:1998yk}
T.~Hahn and M.~Perez-Victoria,
%``Automatized one loop calculations in four-dimensions and D-dimensions,''
Comput. Phys. Commun. \textbf{118} (1999), 153-165
doi:10.1016/S0010-4655(98)00173-8
[arXiv:hep-ph/9807565 [hep-ph]].

%%%%%%%%%%%%%%%%%%%%%
%% contrainst B-L %%%
%%%%%%%%%%%%%%%%%%%%%
%\cite{Deppisch:2019kvs}
\bibitem{Deppisch:2019kvs}
F.~Deppisch, S.~Kulkarni and W.~Liu,
%``Heavy neutrino production via $Z'$ at the lifetime frontier,''
Phys. Rev. D \textbf{100} (2019) no.3, 035005
doi:10.1103/PhysRevD.100.035005
[arXiv:1905.11889 [hep-ph]].
%%%%%%%%%%%
%\cite{Carena:2004xs}
\bibitem{Carena:2004xs}
M.~Carena, A.~Daleo, B.~A.~Dobrescu and T.~M.~P.~Tait,
%``$Z^\prime$ gauge bosons at the Tevatron,''
Phys. Rev. D \textbf{70} (2004), 093009
doi:10.1103/PhysRevD.70.093009
[arXiv:hep-ph/0408098 [hep-ph]].
%%%%%%%%%%
%\cite{Cacciapaglia:2006pk}
\bibitem{Cacciapaglia:2006pk}
G.~Cacciapaglia, C.~Csaki, G.~Marandella and A.~Strumia,
%``The Minimal Set of Electroweak Precision Parameters,''
Phys. Rev. D \textbf{74} (2006), 033011
doi:10.1103/PhysRevD.74.033011
[arXiv:hep-ph/0604111 [hep-ph]].
%%%%%%%%%%
%\cite{Electroweak:2003ram}
\bibitem{Electroweak:2003ram}
t.~Electroweak [LEP, ALEPH, DELPHI, L3, OPAL, LEP Electroweak Working Group, SLD Electroweak Group and SLD Heavy Flavor Group],
%``A Combination of preliminary electroweak measurements and constraints on the standard model,''
[arXiv:hep-ex/0312023 [hep-ex]].
%%%%%%%%%%%%%
%\cite{ATLAS:2019erb}
\bibitem{ATLAS:2019erb}
G.~Aad \textit{et al.} [ATLAS],
%``Search for high-mass dilepton resonances using 139 fb$^{-1}$ of $pp$ collision data collected at $\sqrt{s}=$13 TeV with the ATLAS detector,''
Phys. Lett. B \textbf{796} (2019), 68-87
doi:10.1016/j.physletb.2019.07.016
[arXiv:1903.06248 [hep-ex]].
%%%%%%%%%%%
%\cite{CMS:2019kaf}
\bibitem{CMS:2019kaf}
A.~M.~Sirunyan \textit{et al.} [CMS],
%``Combination of CMS searches for heavy resonances decaying to pairs of bosons or leptons,''
Phys. Lett. B \textbf{798} (2019), 134952
doi:10.1016/j.physletb.2019.134952
[arXiv:1906.00057 [hep-ex]].
%%%%%%%%%%%%%%
% \bibitem{LHCb:2017trq}
% R.~Aaij \textit{et al.} [LHCb],
% %``Search for Dark Photons Produced in 13 TeV $pp$ Collisions,''
% Phys. Rev. Lett. \textbf{120} (2018) no.6, 061801
% doi:10.1103/PhysRevLett.120.061801
% [arXiv:1710.02867 [hep-ex]].
% %%%%%%%%%%%%%%
% %\cite{Bauer:2018onh}
% \bibitem{Bauer:2018onh}
% M.~Bauer, P.~Foldenauer and J.~Jaeckel,
% %``Hunting All the Hidden Photons,''
% JHEP \textbf{07} (2018), 094
% doi:10.1007/JHEP07(2018)094
% [arXiv:1803.05466 [hep-ph]].
% %%%%%%%%%%%%%
% %\cite{CHARM-II:1993phx}
% \bibitem{CHARM-II:1993phx}
% P.~Vilain \textit{et al.} [CHARM-II],
% %``Measurement of differential cross-sections for muon-neutrino electron scattering,''
% Phys. Lett. B \textbf{302} (1993), 351-355
% doi:10.1016/0370-2693(93)90408-A.
% %%%%%%%%%%%%%
%\cite{Bechtle:2015pma}
\bibitem{Bechtle:2015pma}
P.~Bechtle, S.~Heinemeyer, O.~Stal, T.~Stefaniak and G.~Weiglein,
%``Applying Exclusion Likelihoods from LHC Searches to Extended Higgs Sectors,''
Eur. Phys. J. C \textbf{75} (2015) no.9, 421
doi:10.1140/epjc/s10052-015-3650-z
[arXiv:1507.06706 [hep-ph]].
%%%%%%%%%%%%%%%%%%%%%%%%%%
%\cite{Robens:2016xkb}
\bibitem{Robens:2016xkb}
T.~Robens and T.~Stefaniak,
%``LHC Benchmark Scenarios for the Real Higgs Singlet Extension of the Standard Model,''
Eur. Phys. J. C \textbf{76} (2016) no.5, 268
doi:10.1140/epjc/s10052-016-4115-8
[arXiv:1601.07880 [hep-ph]].
%%%%%%%%%%%%%%%%%%%%%%%
%\cite{Chalons:2016jeu}
\bibitem{Chalons:2016jeu}
G.~Chalons, D.~Lopez-Val, T.~Robens and T.~Stefaniak,
%``The Higgs singlet extension at LHC Run 2,''
PoS \textbf{ICHEP2016} (2016), 1180
doi:10.22323/1.282.1180
[arXiv:1611.03007 [hep-ph]].
%%%%%%%%%%%%%%%%%%%%%%%
%\cite{Ilnicka:2018def}
\bibitem{Ilnicka:2018def}
A.~Ilnicka, T.~Robens and T.~Stefaniak,
%``Constraining Extended Scalar Sectors at the LHC and beyond,''
Mod. Phys. Lett. A \textbf{33} (2018) no.10n11, 1830007
doi:10.1142/S0217732318300070
[arXiv:1803.03594 [hep-ph]].
%%%%%%%%%%%%%%%%%%%%%%%%%%
%\cite{L3:1992xaz}
\bibitem{L3:1992xaz}
O.~Adriani \textit{et al.} [L3],
%``Search for isosinglet neutral heavy leptons in Z0 decays,''
Phys. Lett. B \textbf{295} (1992), 371-382
doi:10.1016/0370-2693(92)91579-X
%%%%%%%%%%%%%%%%%%%%%%%%%%
%\cite{DELPHI:1996qcc}
\bibitem{DELPHI:1996qcc}
P.~Abreu \textit{et al.} [DELPHI],
%``Search for neutral heavy leptons produced in Z decays,''
Z. Phys. C \textbf{74} (1997), 57-71
[erratum: Z. Phys. C \textbf{75} (1997), 580]
doi:10.1007/s002880050370
%%%%%%%%%%%%%%%%%%%%%%%%%%%
%\cite{CMS:2012wqj}
\bibitem{CMS:2012wqj}
S.~Chatrchyan \textit{et al.} [CMS],
%``Search for heavy Majorana Neutrinos in $\mu^{\pm}\mu^{\pm} +$ Jets and $e^{\pm}e^{\pm} +$ Jets Events in pp Collisions at $\sqrt{s} =$ 7 TeV,''
Phys. Lett. B \textbf{717} (2012), 109-128
doi:10.1016/j.physletb.2012.09.012
[arXiv:1207.6079 [hep-ex]].
%\cite{ATLAS:2012yoa}
\bibitem{ATLAS:2012yoa}
 [ATLAS],
%``Search for Majorana neutrino production in pp collisions at sqrt(s)=7 TeV in dimuon final states with the ATLAS detector,''
ATLAS-CONF-2012-139.

%\cite{CMS:2015qur}
\bibitem{CMS:2015qur}
V.~Khachatryan \textit{et al.} [CMS],
%``Search for heavy Majorana neutrinos in $\mu^\pm \mu^\pm+$ jets events in proton-proton collisions at $\sqrt{s}$ = 8 TeV,''
Phys. Lett. B \textbf{748} (2015), 144-166
doi:10.1016/j.physletb.2015.06.070
[arXiv:1501.05566 [hep-ex]]. 

%\cite{CMS:2018iaf}
\bibitem{CMS:2018iaf}
A.~M.~Sirunyan \textit{et al.} [CMS],
%``Search for heavy neutral leptons in events with three charged leptons in proton-proton collisions at $\sqrt{s} =$ 13 TeV,''
Phys. Rev. Lett. \textbf{120} (2018) no.22, 221801
doi:10.1103/PhysRevLett.120.221801
[arXiv:1802.02965 [hep-ex]].

%\cite{CMS:2018jxx}
\bibitem{CMS:2018jxx}
A.~M.~Sirunyan \textit{et al.} [CMS],
%``Search for heavy Majorana neutrinos in same-sign dilepton channels in proton-proton collisions at $ \sqrt{s}=13 $ TeV,''
JHEP \textbf{01} (2019), 122
doi:10.1007/JHEP01(2019)122
[arXiv:1806.10905 [hep-ex]].

%\cite{CMS:2017ybg}
\bibitem{CMS:2017ybg}
A.~M.~Sirunyan \textit{et al.} [CMS],
%``Search for Evidence of the Type-III Seesaw Mechanism in Multilepton Final States in Proton-Proton Collisions at $\sqrt{s}=13\text{ }\text{ }\mathrm{TeV}$,''
Phys. Rev. Lett. \textbf{119} (2017) no.22, 221802
doi:10.1103/PhysRevLett.119.221802
[arXiv:1708.07962 [hep-ex]].
%%%%%%%%%%%%%%%%%%%%%%%%%%%%%%
%\cite{Aoki:2019sht}
\bibitem{Aoki:2019sht}
Y.~Aoki, K.~Fujii, S.~Jung, J.~Lee, J.~Tian and H.~Yokoya,
%``Study of the $h\gamma Z$ coupling using $e^+e^-\to h \gamma$ at the ILC,''
[arXiv:1902.06029 [hep-ex]].
%%%%%%%%%%%%%%%%%%%%%%%%%%%%%
%%%%%%%%%%%%%%%%%%%%%%%%%%%%%
%\cite{Denner:2005nn}
\bibitem{Denner:2005nn}
A.~Denner and S.~Dittmaier,
%``Reduction schemes for one-loop tensor integrals,''
Nucl. Phys. B \textbf{734} (2006), 62-115
%%%%%%%%%%%%%%%%%%%%%%%%%%%%%%
\end{thebibliography}
\end{document}